\newcommand{\onlinecite}[1]{\hspace{-1 ex} \nocite{#1}\citenum{#1}}
\documentclass[a4paper,aps,twocolumn]{revtex4-1}

\usepackage[english]{babel}
\usepackage{dcolumn}
\usepackage{amsmath}
\usepackage{graphicx}
\usepackage{tabularx}
\usepackage[center]{subfigure}
\usepackage{color}
\usepackage{booktabs}
\usepackage{calc}
\begin{document}
\title{
Accuracy and Resource Estimations for Quantum Chemistry on 
a Near-term Quantum Computer 
}

\author{Michael K\"uhn}
\email{michael.b.kuehn@basf.com}
\affiliation{BASF SE $-$ Quantum Chemistry Materials, Carl-Bosch-Str. 38, 67063 Ludwigshafen, Germany}

\author{Sebastian Zanker}
\affiliation{HQS Quantum Simulations, Haid-und-Neu-Str. 18, 76131 Karlsruhe, Germany}

\author{Peter Deglmann}
\affiliation{BASF SE $-$ Quantum Chemistry Materials, Carl-Bosch-Str. 38, 67063 Ludwigshafen, Germany}

\author{Michael Marthaler}
\email{michael.marthaler@quantumsimulations.de}
\affiliation{HQS Quantum Simulations, Haid-und-Neu-Str. 18, 76131 Karlsruhe, Germany}

\author{Horst Wei{\ss}}
\affiliation{BASF SE $-$ Quantum Chemistry Materials, Carl-Bosch-Str. 38, 67063 Ludwigshafen, Germany}

\date{\today}

\begin{abstract}
The study and prediction of chemical reactivity is one of the most important application areas of molecular quantum chemistry.
Large-scale, fully error-tolerant quantum computers could provide exact or near-exact solutions to the underlying electronic structure problem
with exponentially less effort than a classical computer thus enabling highly accurate predictions for comparably large molecular systems.
In the nearer future, however, only ``noisy'' devices with a limited number of qubits that are subject to decoherence will be available.  
For such near-term quantum computers the hybrid quantum-classical variational quantum eigensolver algorithm in combination with the unitary coupled-cluster {\it ansatz} (UCCSD-VQE)\cite{Peruzzo14,McClean15} 
has become an intensively discussed approach
that could provide accurate results before the dawn of error-tolerant quantum computing.
In this work we present 
an implementation of UCCSD-VQE that allows for the first time to treat both open- and closed-shell molecules. 
We study the accuracy of the obtained energies for nine small molecular systems as well as for four exemplary chemical reactions 
by comparing to well-established electronic structure methods like (non-unitary) coupled-cluster and density functional theory.
Finally, we roughly estimate the required 
quantum hardware resources to obtain ``useful'' results for practical purposes.
\end{abstract}

\maketitle

\section{Introduction}
One of the most important application areas of molecular electronic structure calculations in both industry and academia 
is the investigation and prediction of chemical reactivity.\cite{Deglmann15} 
A typical quantum-chemical study of chemical reactions usually starts with the calculation of relative electronic energies 
of reactants, products and transition states. 
From the obtained energies, thermodynamics (enthalpies, entropies and Gibbs free energies) as well as kinetics (activation barriers and reaction rates via Eyring's equation\cite{Eyring35}) can be computed. This then in principle allows a prediction of the course of a chemical reaction.    
As for many other molecular properties the quality of the prediction significantly depends on the accuracy of the calculated electronic energies $-$ especially in the case of reaction rates which depend exponentially on the energy difference between transition state and reactants. 

Several different methods for calculating molecular energies have been implemented in a large number of quantum-chemical program packages and made available as standard tools. 
Density functional theory (DFT) is a widely used method that can also be applied to relatively large systems due to its moderate computational requirements compared to high-level correlated electronic structure methods.
However, due to the known deficiencies of DFT the obtained accuracy might be not good enough for some practically relevant systems.
High accuracy in electronic energies usually can be achieved by computationally more demanding 
coupled-cluster (CC) methods which exhibit a systematic convergence towards the exact full configuration-interaction (FCI) result.
As a compromise between efficiency and accuracy the CC expansion is most commonly truncated at the doubles term yielding the coupled-cluster singles and doubles (CCSD)\cite{Purvis82} method. 
Additionally augmenting this method by a perturbative treatment of the triple
excitations (CCSD(T))\cite{Raghavachari89} yields the ``gold standard'' method in quantum chemistry, which exhibits high accuracy for most practical purposes (with the exception of multireference systems / strong static correlation) and a computational effort that formally scales with the system size to the seventh power whereas the exact solution (FCI) would require exponential (factorial) resources on a classical computer.

\textcolor{black}{Quantum computers instead could provide exact or near-exact results for quantum systems without exponential scaling. It has been shown that a quantum computer can efficiently simulate the time 
evolution of a quantum system.\cite{Lloyd96} 
At the same time it is also clear that it is not possible to efficiently find the ground state of an arbitrary (non-natural) Hamiltonian.\cite{Jordan10}  
If a universal (error-free) quantum computer with a large number
of qubits would be available the phase estimation algorithm\cite{Abrams97,Abrams99} should provide access to the exact ground-state energy, as long as it is possible to construct an initial guess with a non-vanishing overlap with 
the real ground state. }

In the near term, however, we can only expect a rather small number of qubits, and even more importantly we can only use a small number of gates since the prepared quantum state is subject to decoherence. This is also a result of the fact that currently the number of qubits is 
too small to accommodate the overhead for quantum error correction, therefore every operation has a finite fidelity. 
Nonetheless, it should be possible to efficiently approximate the ground state of a natural system, as would be the case for a chemical electronic structure problem. 
For near-term quantum computers it has become popular to use algorithms based on the so-called variational quantum eigensolver (VQE),\cite{Peruzzo14,McClean15} which combines a classical algorithm that tries to solve a variational problem with a quantum algorithm that provides the parameters for the classical approach. 
One particular example is the unitary coupled-cluster {\it ansatz} (UCC)\cite{Romero17} which is the quantum computing equivalent to the classical CC methods. It reduces the problem of finding the ground state to a variational problem,
which in turn reduces the number of gates needed to be carried out on the quantum computer. 
However, at the same time the algorithm has to be repeated more often, and therefore it is not necessarily shorter in total
as compared to phase estimation. Since UCC is similar to the well established CC approach it has generated a substantial amount of interest and will be the quantum algorithm tested in this paper.

Until now the following molecular systems have been studied on quantum hardware or simulators using different algorithms and different experimental realizations:
(i) molecular hydrogen ($\rm H_{2}$) using a NMR-based quantum hardware\cite{Du10} and on a superconducting chip applying UCCSD-VQE\cite{Malley16}
as well as on simulators including excited-state calculations\cite{McClean17,Colless18};
(ii) dimer of $\rm H_{2}$ using UCCSD-VQE on a simulator\cite{Romero17};
(iii) helium hydride cation ($\rm HeH^{+}$) in a solid-state spin register\cite{Wang15} as well as using UCCSD-VQE on a quantum photonic chip\cite{Peruzzo14}; 
(iv) lithium hydride ($\rm LiH$) together with $\rm H_{2}$ on a trapped-ion system applying UCCSD-VQE\cite{Hempel18};
(v) beryllium dihydride ($\rm BeH_2$) together with LiH and $\rm H_{2}$ on a superconducting quantum processor with VQE\cite{Kandala17};
(vi) water molecule ($\rm H_{2}O$) on a simulator comparing VQE to other methods.\cite{Bian18,Barkoutsos18}
Additionally, it was argued by Reiher {\it et al.} that a \textcolor{black}{(long-term)} error-corrected quantum computer could in principle 
be used to elucidate reaction mechanisms as complicated as the open question of biological nitrogen fixation 
in nitrogenase enzymes.\cite{Reiher17}   

To the best of our knowledge, in the previous studies
molecular energies of closed-shell systems were mainly compared to the numerically exact (FCI) results and in most cases only small basis sets (mostly minimal basis sets) were applied.
In this study we 
present an implementation of UCCSD-VQE that $-$ in addition to closed-shell species $-$ allows for the first time to treat open-shell molecules (e.g. systems with an odd number of electrons). We  
compare molecular energies of small systems 
not only to FCI but also to CCSD and CCSD(T), methods that are routinely used in quantum chemistry.
\textcolor{black}{We note that in the context of classical computing there are several studies investigating the accuracy of UCCSD and its extensions to higher excitations for basis sets of up to valence double- and triple-$\zeta$ quality.\cite{Cooper10,Evangelista11,Harsha18}}
Furthermore, in comparison to previous studies we do not only consider molecular energies but also reaction energies and compare those to DFT results. 
We note that in order to obtain ``useful'' results for practical applications $-$ like the study of chemical reactions $-$ not only an accurate, highly-correlated method is needed but also a sufficiently large basis set of at least triple- or quadruple-$\zeta$ quality. Only by combining both these prerequisites the major portion of the correlation energy can be captured and a high accuracy is reached.
Therefore, we give a rough estimation of the quantum resources (number of qubits and two-qubit gates which are even more critical) 
needed to obtain results for small molecules with an accuracy that is comparable to today's ``gold standard'' method CCSD(T) in combination with a sufficiently large basis set.
The following molecules are subject to our study: 
water ($\rm H_{2}O$), hydroxyl radical ($\rm OH$), lithium hydride ($\rm LiH$), atomic lithium ($\rm Li$), 
molecular nitrogen ($\rm N_2$), molecular hydrogen ($\rm H_2$), ammonia ($\rm NH_3$), 
triplet methylene ($\rm {:}CH_2$) and 
singlet methylene ($\rm CH_2$).
With these, we investigate the reaction energies of four exemplary chemical reactions:
\\
(i) homolytic $\rm O-H$-bond dissociation in the water molecule, $\rm H_{2}O \, \rightarrow \, \, OH \, + \, H$;
\\
(ii) homolytic dissociation of lithium hydride, $\rm LiH \, \rightarrow \, \, Li \, + \, H$;
\\
(iii) Haber-Bosch process, $\rm N_2 \, + \, 3\,H_2 \,  \rightarrow \, \, 2\, NH_3$;
\\
(iv) energy gap between triplet- and singlet-methylene, $\rm {:}CH_2 \, \rightarrow \, \, CH_2$.
\\

The paper is organized as follows:
In Section \ref{sec:theory} we briefly summarize the theory of 
the UCCSD-VQE approach and give some details specific to our implementation for reducing the number of required gates.
In Section \ref{sec:appl} we study the accuracy of molecular energies (Section \ref{sec:appl:molener}) as well as
reaction energies (Section \ref{sec:appl:reactener}) by comparing to different methods. We furthermore roughly estimate the required quantum resources
to obtain ``useful'' results for practical applications.

%

\section{Theory and Implementation}
\label{sec:theory}

\subsection{The UCCSD-VQE approach}
In the VQE paradigm it is the goal to prepare a parameter-dependent quantum state in a register of qubits,
measure a particular observable and find the parameters that optimize the value of the observable. In our case
we are interested to find the energetic ground state of the electronic structure problem for a molecule and therefore we measure the average energy of 
the prepared states. \textcolor{black}{For more details on the measurement process see for example Refs. \onlinecite{McClean15} and \onlinecite{Izmaylov19}}.

When it comes to preparing a reasonable parameter-dependent state we will use an approach based on the CC operators, which are well established
in quantum chemistry. The CC operator, including single and double excitations is given by
\begin{eqnarray}\label{eq_Coupled_Cluster_operators_definition}
 T &=& T_1+T_2\\
 T_1 &=& 
 \!\!\!\!\!\!\!\!
\, \sum\limits_{\tiny
\begin{array}{l}
i \in {\rm occ} \\\label{eq_Coupled_Cluster_single_operator_definition}
a\in {\rm virt}
\end{array}
} 
\!\!\!\!\!\!\!
\, t_a^i  c_a^{\dag}c_i \\\label{eq_Coupled_Cluster_double_operator_definition}
T_2 &=& \!\!\!\!\!\!\!\!\!\!
\sum\limits_{\tiny
\begin{array}{l}
i > j\in {\rm occ} \\
a > b\in {\rm virt}
\end{array}
} \!\!\!\!\!\!\!\!\!\!
\, t_{ab}^{ij}  c_a^{\dag} c_b^{\dag} c_i c_j\,, 
\end{eqnarray}
where $c_{i}$ and $c_{a}^{\dag}$ are the fermionic annihilation and creation operators of the orbitals $i$ and $a$, respectively, obeying the according commutation relations. 
 In this {\it ansatz} we assume that the CC operator is applied to a trivial initial state
with a certain number of occupied orbitals, $N_{\text{occ}}$, and a certain number of unoccupied or virtual orbitals, $N_{\text{virt}}$.
The indices $i,j$ refer to occupied orbitals, and $a,b$ refer to unoccupied or virtual orbitals. 
\textcolor{black}{To reduce the computational requirements the summations in Equations \eqref{eq_Coupled_Cluster_single_operator_definition} and \eqref{eq_Coupled_Cluster_double_operator_definition} might be restricted to a chemically relevant subset of orbital indices, which 
is equivalent to choosing an ``active space'' that includes a selection of orbitals that should enter the correlation treatment.}
As the term ``active space'' could be misinterpreted because it is at the same time used in multireference calculations to
describe the space of orbitals from which a basis of Slater determinants is formed to enter the variational optimization of the multireference function, we
use in the following the term ``non-frozen orbital space''.  
The variational parameters
are given by $t_a^i$ and $t_{ab}^{ij}$ which we assume to be real. 
For later purposes it is sensible to summarize all terms in the coupled cluster operator with a single
index $T=\sum_x T_x$.

Using the unitary exponential of the CC operator in Equation \eqref{eq_Coupled_Cluster_operators_definition} we prepare the state
\begin{equation}
 |\psi\rangle= e^{T-T^{\dag}} |\psi_{\rm HF}\rangle\,, 
\end{equation}
where the initial state is the Hartree-Fock (HF) state which is trivially given by a state of the form
\begin{equation}
 |\psi_{\rm HF}\rangle=|0000\ldots 1111\rangle\,,
\end{equation}
where $|0/1\rangle$ refers to a single qubit being in the state $0$ or $1$. In the multi-qubit register we assign each qubit an orbital, with orbital energies from high to low. 
The state $|0\rangle$ refers to an unoccupied orbital, while $|1\rangle$ refers to an occupied orbital. The fermionic nature of the state is preserved by the anti-commutation relations 
of the annihilation and creation operators. 

The unitary operator which is to be applied to the HF state can be disassembled 
into smaller parts using the Trotter expansion \cite{Trotter59}
\begin{equation}\label{eq_UCCSD_Unitary_operator_Trotter_expansion}
 e^{T-T^{\dag}} =
\left(e^{(T-T^{\dag})/n}\right)^n
=
\left(\prod_x e^{(T_x-T_x^{\dag})/n}\right)^n\,,
\end{equation}
where for \textcolor{black}{VQE $n=1$ is usually sufficient}.
A single term in the Trotter expansion contains an Operator $T_x$ which consists of a certain number of fermionic annihilation and creation operators. 

\begin{figure}[t]
\begin{center}
 \includegraphics[width=9cm]{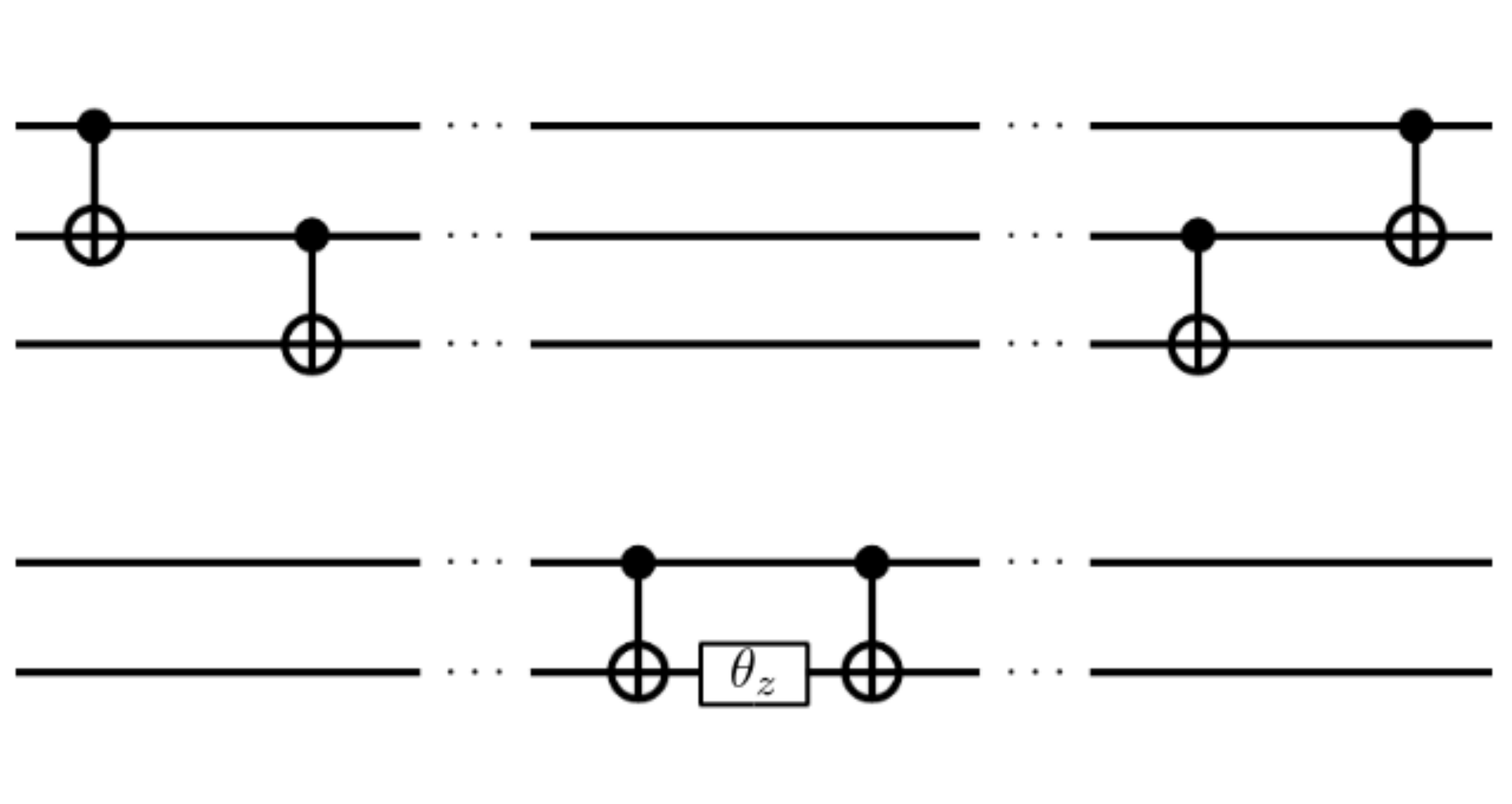}
 \caption{{\bf Gate sequence to create arbitrary Pauli strings. }}\label{fig_Pauli_string_unitary}
 \end{center}
\end{figure}

We translate the fermionic operators to spin operators using the Jordan Wigner transformation.\cite{Jordan28,Somma02} As an example, a single excitation term transforms as
\begin{equation}
 c_a^{\dag}c_i = \sigma_+^a \sigma_z^{a+1} \ldots \sigma_z^{i-1}\sigma_-^i\,,
\end{equation}
where $\sigma_i$ are the Pauli operators acting on the $i$-th qubit.
This means that any operator $T_x$ can be written as a string of Pauli operators. 
In Figure \ref{fig_Pauli_string_unitary} we show a gate sequence which implements a unitary operator
of the form
\begin{equation}
e^{i \sigma_z^a \sigma_z^{a+1} \ldots \sigma_z^{i-1}\sigma_z^i }\,.            
\end{equation}
Any Pauli string can be generated from this expression in combination with single qubit rotations. 
Out of the ingredients presented in this subsection we can construct the full UCCSD-VQE 
routine, which allows us to access energies close to the ground-state energy of an
electronic structure problem using a quantum computer. 

Our implementation is partly based on the ProjectQ framework \cite{Steiger16,Haner16} and the OpenFermion library\cite{openfermion} and can be applied to closed- as well as open-shell systems.
The HF reference state was prepared using the open-source quantum chemistry package pySCF.\cite{pyscf}
``Noise'' originating from decoherence of the prepared quantum states was not considered in our simulations. We use the (classical) optimization method L-BFGS-B
to find the optimal values of the parameters $t_a^i $ and $t_{ab}^{ij}$.
\textcolor{black}{We tested various methods available in the Python package \texttt{scipy.optimize} and found L-BFGS-B to perform best. } 

\subsection{MP2 pre-screening}
\label{MP2}

When implementing the unitary operator from Equation \eqref{eq_UCCSD_Unitary_operator_Trotter_expansion} a majority of 
operations is spent on the implementation of $T_2$. However, we know that probably not all $T_2$-amplitudes 
$t_{ab}^{ij}$ will be relevant for the final result. We can use second-order M\o{}ller-Plesset perturbation theory (MP2)\cite{Moller34} to calculate an initial estimate for the
order of magnitude of $t_{ab}^{ij}$. If the resulting estimate of $t_{ab}^{ij}$ is very small, we can use this result to 
directly choose $t_{ab}^{ij}=0$ and therefore reduce the number of terms in $T_2$ and finally the number of gates to be carried out. However it is obviously important to test
what cut-off value for $t_{ab}^{ij}$ should be chosen. 
For this purpose single-point molecular energy calculations were carried out applying our UCCSD-VQE implementation exploiting potential cancellation of gates. 

\begin{figure}[t]
\begin{center}
\includegraphics[width=9cm ,angle=0,clip=]{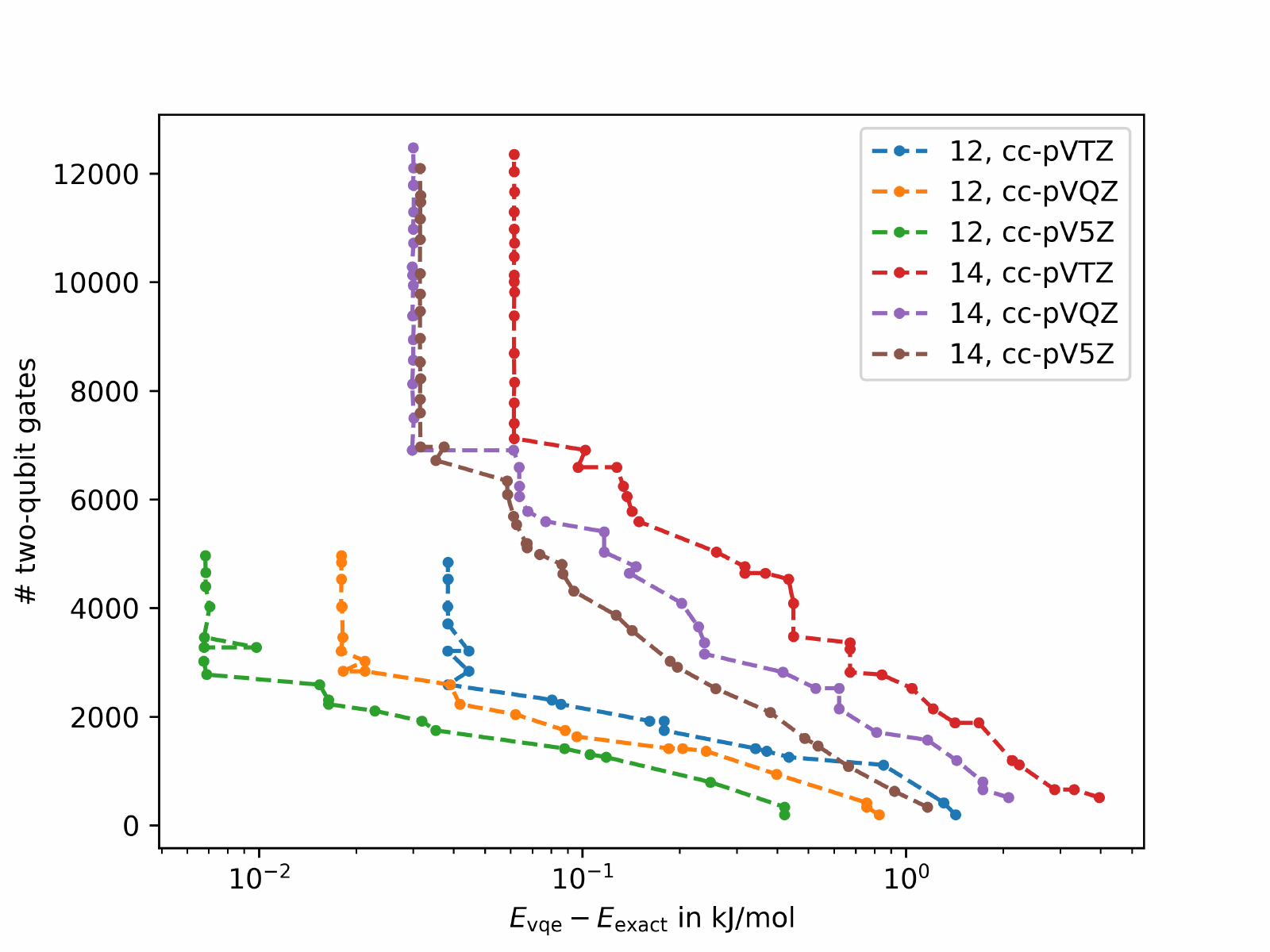}
\caption{{\bf Precision of the ground-state energy compared to the number of two-qubit gates for $\rm{H}_2 \rm{O}$.} Different sizes of the ``non-frozen orbital space'' (12 and 14 spin orbitals) in combination with different basis sets are investigated. We show the number of two-qubit gates needed to 
implement the unitary operator from Equation \ref{eq_UCCSD_Unitary_operator_Trotter_expansion} as a function of the difference between the VQE energy and the exact ground-state energy. 
}
\label{fig_MP2_prescreening_energy_accuracy_and_CNOT_reduction}
 \end{center}
\end{figure}

We test MP2 pre-screening on a small molecule, in this case $\rm{H}_2 \rm{O}$.
We use a ``non-frozen orbital space'' of either $12$ spin orbitals (which corresponds to the orbitals in the minimal basis set with the energetically lowest occupied orbital being excluded) 
or $14$ spin orbitals (obtained by adding the energetically next higher-lying two unoccupied orbitals in order to simulate the effect of adding further states), while the initial HF state is calculated using different basis sets. For more computational details we refer to Section \ref{sec:appl:details}.
In Figure \ref{fig_MP2_prescreening_energy_accuracy_and_CNOT_reduction} we show the number of two-qubit (e.g. CNOT) gates needed to implement the unitary operator from Equation
\eqref{eq_UCCSD_Unitary_operator_Trotter_expansion} as a function of the difference between the VQE energy and the exact ground-state energy
at the given method and basis set. 
We count the two-qubit gates, since we assume that the algorithm is supposed to be implemented on a 
quantum computer without quantum error correction. In that case the fidelity of the two-qubit gates
is most probably the limiting factor. We find that with virtually no changes in the VQE energy the number of two-qubit gates can
roughly be reduced by 
more than an order of magnitude.

Finally we have checked for all molecules within this study (in the minimal basis set)
that the UCCSD-VQE procedure converges to the same final energy both when using MP2 to initialize $t_{ab}^{ij}$ ($t_{a}^{i}=0$)
and CCSD to initialize $t_{ab}^{ij}$ and $t_{a}^{i}$ (without pre-screening). Since CCSD in most cases should provide an initial state close to the final UCCSD-VQE
state we therefore conclude that also MP2 provides a sufficiently accurate initial guess.

\begin{figure}[t]
\begin{center}
  \includegraphics[width=9cm ,angle=0,clip=]{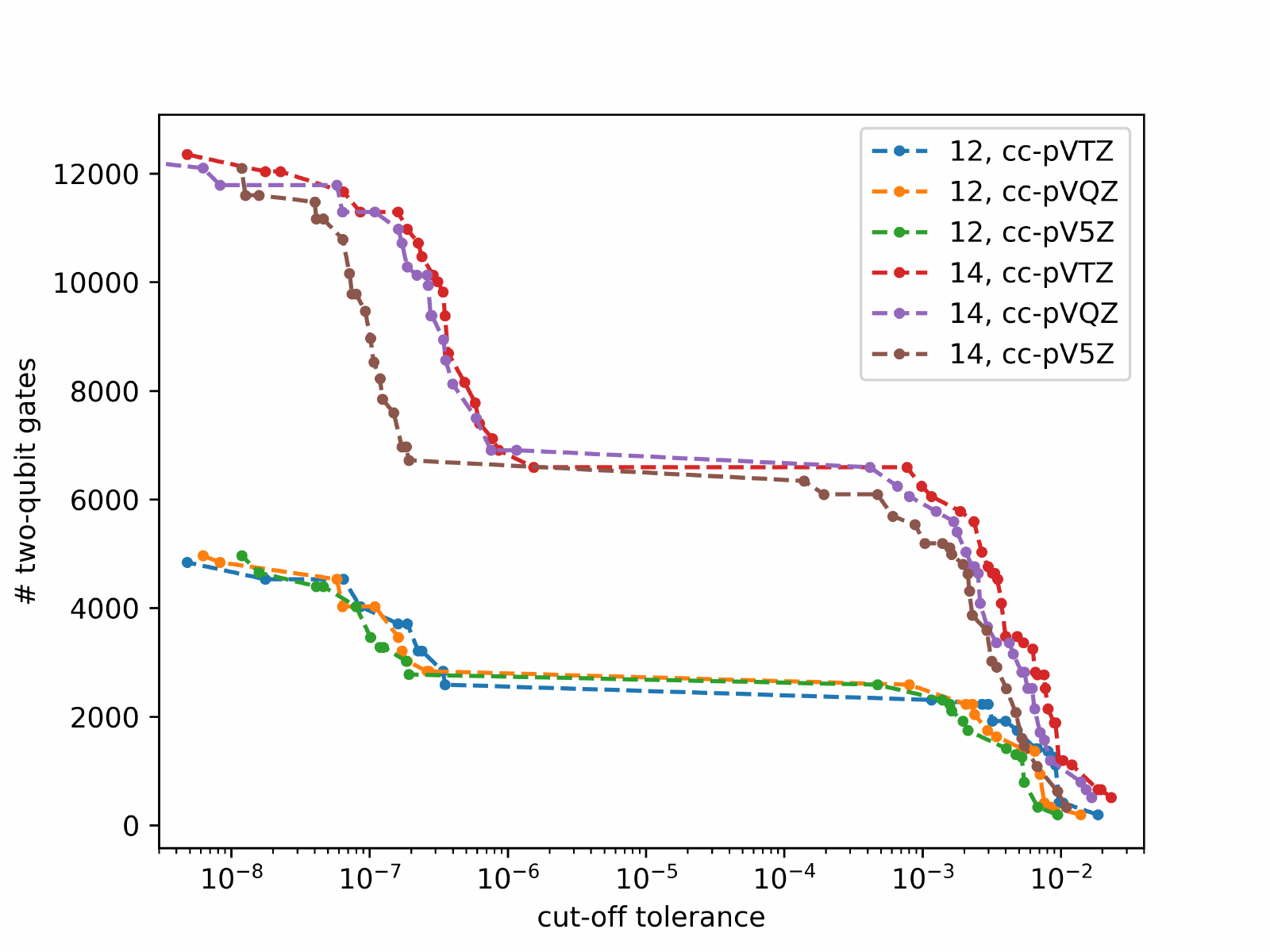}
 \caption{{\bf Number of two-qubit gates as a function of the cut-off tolerance for $\rm{H}_2 \rm{O}$.} If our pre-screening indicates that $t_{ab}^{ij}$ is below the cut-off tolerance then we choose $t_{ab}^{ij}=0$. For further information see Figure \ref{fig_MP2_prescreening_energy_accuracy_and_CNOT_reduction}}
 \label{fig_MP2_prescreening_cut_off_tolerance_and_CNOT_reduction}
 \end{center}
\end{figure}

In Figure \ref{fig_MP2_prescreening_cut_off_tolerance_and_CNOT_reduction} we show how the number of two-qubit gates depends on the cut-off tolerance. 
We find that the reduction in the number of two-qubit gates is rather large below a threshold of $10^{-5}$ while at the same time we see from 
Figure \ref{fig_MP2_prescreening_energy_accuracy_and_CNOT_reduction} that the energy results are hardly affected up to this point. 
Therefore we choose only to include $T_2$-amplitudes larger than $10^{-5}$ obtained from MP2 for all further UCCSD-VQE calculations.

\subsection{Gate Cancellations}
\label{cancel}

In Figure \ref{fig_Pauli_string_unitary} we show the gate sequence we used to implement the Jordan Wigner transformation. The depicted gate sequence 
has to be repeated for each set of fermionic operators.
Additionally, before and after each of these gate sets we apply single-qubit gates to create the correct Pauli string. 
From the structure of the gates it seems clear that
commuting the application of the two-qubit gates $-$ here CNOT gates $-$ with the application of the single-qubit gates instead of applying 
one gate sequence after another could allow for a substantial amount
of gate cancellations, since $\rm CNOT \, CNOT = {\bf 1}$.

This is also the underlying idea of the improved implementation of the gate sequence discussed in 
Ref. \onlinecite{Hastings15}. \textcolor{black}{Similar ideas have been proposed in Ref. \onlinecite{Ryabinkin18}}.
The discussed gate sequence in principle allows for the implementation of arbitrary Pauli strings, however, the application of the different terms 
in the CC operator need to be sorted
in a particular way to make use of the gate cancellations. 
The gate sequence described in Ref. \onlinecite{Hastings15} has the disadvantage that it needs two-qubit gates different from the CNOT gate.
Even more profound is the need for very well-connected qubits. 
We note that an algorithm has been published that allows for the implementation of fermionic operators on a one-dimensional
qubit geometry.\cite{Kivlichan18} 
Only very recently
this algorithm has been extended
to potentially allow for the full implementation of CC operators, such as shown in Equation \ref{eq_Coupled_Cluster_operators_definition}.\cite{Motta18}

\begin{figure}[t]
\begin{center}
  \includegraphics[width=9cm ,angle=0,clip=]{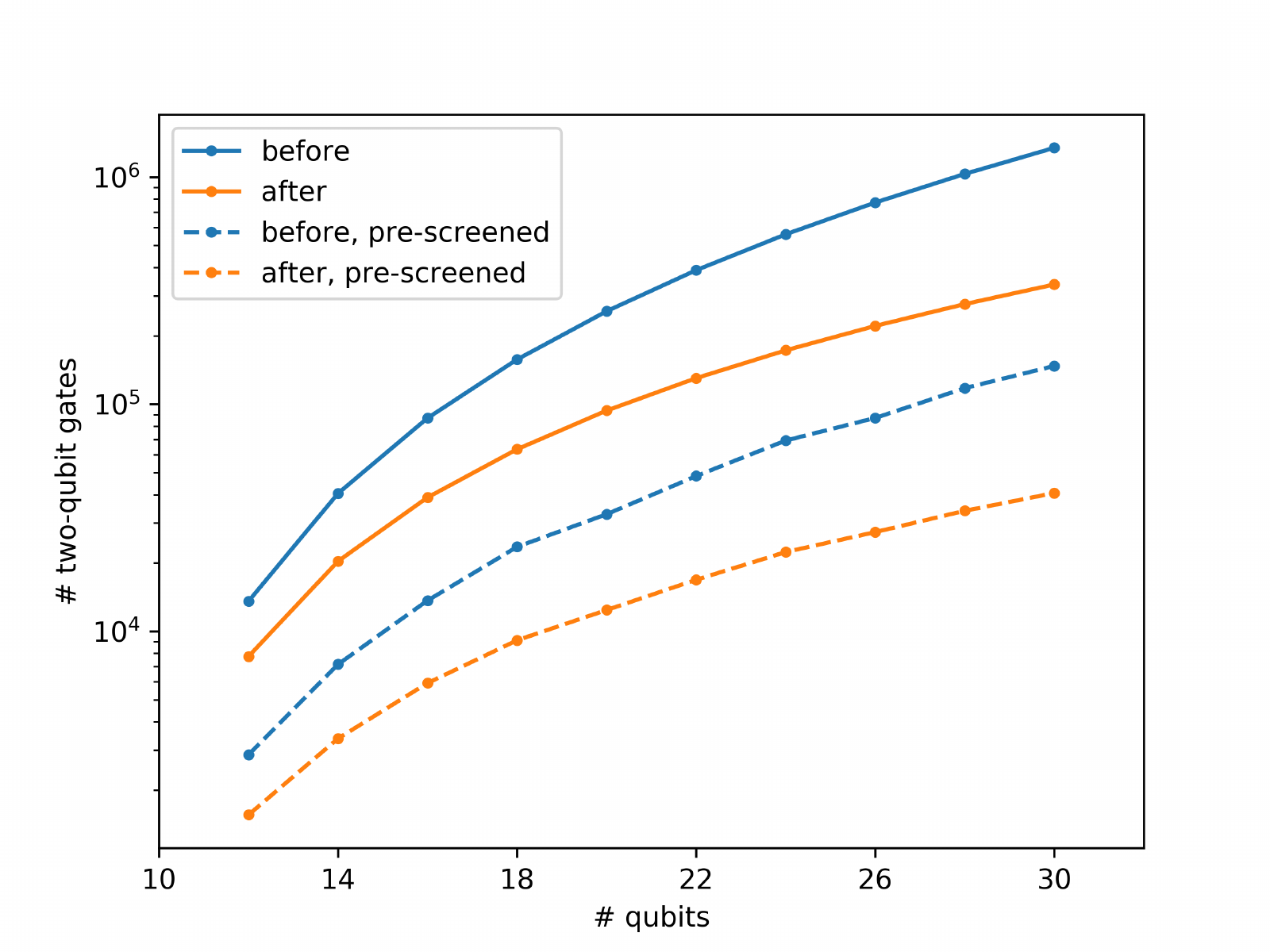}
 \caption{{\bf The number of two-qubit gates 
before and after exploiting gate cancellations as well as with and without MP2 pre-screening for $\rm H_{2}O$}. The cc-pV5Z basis set is used.} 
 \label{optimized_gate_counts}
 \end{center}
\end{figure}

In this work we implemented the gate cancellations as described in Ref. \onlinecite{Hastings15} to get a first impression of the impact
optimized algorithms can have on the overall number of two-qubit gates. 
For a large number of qubits we reduce the number of two-qubit gates by a factor of four as shown in Figure \ref{optimized_gate_counts}, 
which could be useful under certain special experimental conditions.
However, considering the need for increased connectivity and the fact that MP2 pre-screening alone creates a much larger reduction in the gate count
future work should also consider other gate reduction schemes like the ones discussed above.

%
\section{Estimation of Accuracy and Required Quantum Resources for Small Quantum Chemistry Problems}
\label{sec:appl}


\subsection{Computational Details}
\label{sec:appl:details}

The ground-state structures of all molecules were optimized using DFT in combination 
with Becke's three-parameter hybrid functional with Lee-Yang-Parr correlation (B3LYP)\cite{met:Becke93a}
and doubly polarized quadruple-$\zeta$ valence basis sets (def2-QZVPP). \cite{bas:Weigend05a}
All structure optimizations were done using the TURBOMOLE program suite.\cite{turbomole}

All UCCSD-VQE calculations were done exploiting gate cancellations (see Section \ref{cancel}) as well as $T_2$-amplitude
pre-screening based on MP2 (see Section \ref{MP2}) to reduce gate counts. 
For comparison of the resulting energies additionally CCSD and CCSD(T) calculations were carried out using TURBOMOLE.
The numerically exact results (identical to FCI) were obtained by exact diagonalization of the Hamiltonian in the given basis set using the OpenFermion implementation.
In $\rm H_{2}O$, $\rm OH$, $\rm NH_3$, $\rm {:}CH_2$ and $\rm CH_2$ the energetically lowest and in $\rm N_2$ the two energetically lowest molecular orbital(s) was(were) always excluded from all correlation treatments, 
corresponding to a freezing of the energetically well-separated 1s atomic orbitals in oxygen, nitrogen and carbon. 
In case of reaction energies, additional comparison to DFT was made using the generalized gradient approximation functional of Becke and Perdew (BP86),\cite{met:Becke88a,met:Perdew86}
the hybrid functional B3LYP as well as Truhlar's hybrid meta exchange-correlation functional M06-2X\cite{Zhao08} designed for main-group thermochemistry applications.
DFT-quadrature grids of size m3\cite{Treutler95} were employed.

The minimal basis set STO-3G\cite{Hehre69} was used as well as Dunning's correlation-consistent basis sets of valence
$X$-tuple-$\zeta$ quality (cc-pV$X$Z)\cite{Dunning89} with $X \rm = D,\,T,\,Q,\,5$.
For Li and H additionally split-valence (SV), double-$\zeta$ (DZ) and triple-$\zeta$ (TZ) basis sets were employed.\cite{Schaefer92}
In all TURBOMOLE calculations corresponding CC auxiliary basis sets\cite{Weigend98,Weigend01,Haettig05b}
were used.
All calculations were performed on the BASF supercomputer ``Quriosity''.\cite{quriosity}

\subsection{Molecular Energies}
\label{sec:appl:molener}

In this section we present a detailed investigation of the accuracy of molecular energies obtained with our UCCSD-VQE implementation.
We compare UCCSD-VQE molecular energies to FCI as well as CCSD and the ``gold standard'' method CCSD(T).
Furthermore, we roughly estimate the required quantum resources (number of qubits and number of two-qubit gates) that are needed to obtain results that are comparable to CCSD/CCSD(T) results when applying a sufficiently large basis set.
Here we focus on $\rm H_{2}O$, $\rm N_2$ as well as the open-shell molecules $\rm OH$ (doublet ground state, i.e. one unpaired electron) and $\rm {:}CH_2$ (triplet ground state, i.e. two unpaired electrons).
Results for additional systems $-$ $\rm LiH$, $\rm Li$ (doublet ground state), $\rm H_2$, $\rm NH_3$ and $\rm CH_2$ $-$
are in line with the discussion below and
can be found in the Supporting Information (SI) in Tables I-II and Figures 1-4. 
The molecular energies discussed in this section will later be used to study reaction energies (Section \ref{sec:appl:reactener}) which are simply given as 
stoichiometrically weighted differences between different molecular energies (partial error compensation possible).

\begin{table*}[htbp]
\centering
\caption[]{
{\bf Accuracy of molecular energies obtained with UCCSD-VQE for ${\rm H_{2}O}$ and ${\rm N_{2}}$
as well as for the open-shell species ${\rm OH}$ and ${\rm {:}CH_2}$.}
The HF total energy as well as the CCSD, CCSD(T), FCI and UCCSD-VQE correlation energies are given together with the respective differences to the FCI energy ($\Delta$FCI).
Additionally, the number of required qubits and two-qubit gates is shown.
All results were obtained using the minimal basis set STO-3G. Energies are in kJ/mol.
 }
\label{tab:appl:molener}
\tiny
\begin{tabular}{lcccccccccccc}
 & \phantom{x} & \multicolumn{2}{c}{\bf{$\boldsymbol{\rm H_{2}O}$}} &  \phantom{x}      &  \multicolumn{2}{c}{\bf{$\boldsymbol{\rm N_{2}}$}} &  \phantom{x} & \multicolumn{2}{c}{\bf{$\boldsymbol{\rm OH}$}} & \phantom{x} & \multicolumn{2}{c}{\bf{$\boldsymbol{\rm {:}CH_{2}}$}} \\
 &              &  & $\Delta$FCI                                             &               &      & $\Delta$FCI                           &              &       & $\Delta$FCI                      &             &      & $\Delta$FCI    \\
\hline
$E_{\text{total}}$({\text{HF}}) & &                 -196815.7& $-$ &       &-282224.5&$-$&&       -195240.2&$-$&&    -100909.4&$-$      \\
\hline
$E_{\text{corr}}$(CCSD) & &          -130.159&0.314     &   &-395.526&9.890&&      -64.900&0.001&&    -97.590&0.448       \\
$E_{\text{corr}}$(CCSD(T)) & &       -130.340&0.132     &   &-399.948&5.469&&      -64.900&0.001&&    -97.843&0.196       \\
$E_{\text{corr}}$(FCI) & &           -130.473& 0        &   &-405.417&0&&          -64.901&0&&        -98.039&0          \\
\hline
$E_{\text{corr}}$(UCCSD-VQE) & &     -130.205&0.267     &   &-399.910&5.507&&      -64.899&0.002&&    -97.601&0.438       \\
\# qubits & &                         12&$-$&                &16&$-$&&              10&$-$&&           12&$-$              \\
\# two-qubit gates & &                1198& $-$        &     &2510&$-$&        &    538&$-$&         & 1542&$-$            \\

\end{tabular}
\end{table*}

\subsubsection{Method Comparison in the Minimal Basis Set}
\label{sec:appl:molener:method}

In Table \ref{tab:appl:molener} the HF total energies as well as the CCSD, CCSD(T), FCI and UCCSD-VQE correlation energies using the minimal basis set (STO-3G) are shown for four exemplary molecules. 
In general, 
for any correlated method (FCI, CCSD, CCSD(T), UCCSD-VQE, etc.) the total energy can be decomposed
into a HF (mean-field) and a correlation contribution.
In the minimal basis set the correlation energy of the presented systems is roughly of the order of a hundred kJ/mol, which is only $0.1\,\%$ of the HF energy. 
In an adequate basis set for correlated methods, such as for example the cc-pVQZ basis set, the correlation energy increases by almost one order of magnitude but still remains significantly smaller than the HF energy 
(see Table \ref{tab:appl:reactener_extrapol_H2O} for the example of $\rm H_{2}O$ and $\rm OH$).  
Nevertheless, an accurate description of the correlation energy is of utmost importance, since it usually
is of similar size or even larger than typical reaction energies, which range from a few kJ/mol to a few hundred kJ/mol (few thousand kJ/mol for combustion processes of large molecules), and can decide whether a chemical reaction will happen or not. 
Therefore, the task of accurately describing the effect of electron correlation
lies at the heart of quantum chemistry.

Comparing CCSD and CCSD(T) to the exact (FCI) correlation energy in the minimal basis set
reveals that errors are small compared to the absolute value of the correlation energy
amounting to less than $0.5\, \rm kJ/mol$ or $0.5\,\%$ for CCSD and less than $0.2\, \rm kJ/mol$ or $0.2\,\%$ for CCSD(T) in case of $\rm H_{2}O$, $\rm OH$ and $\rm {:}CH_2$. 
In the case of $\rm N_2$,
which is known to be a challenge to approximate electronic structure methods, 
the errors increase to up to $10\, \rm kJ/mol$ or $2.5\,\%$ for CCSD and $5.5\, \rm kJ/mol$ or $1.4\,\%$ for CCSD(T).

For all molecules studied within this work UCCSD-VQE is between CCSD and CCSD(T). \textcolor{black}{Together with the findings in previous studies on the accuracy of UCCSD\cite{Cooper10,Evangelista11,Harsha18} this indicates that the CCSD energy can be used as a worst-case and the CCSD(T) energy as a best-case estimate for the UCCSD-VQE energy, see also 
Tables I-II in the SI.}
In detail, 
for $\rm {:}CH_2$ (and $\rm LiH$) UCCSD-VQE is very close
to CCSD and in case of $\rm H_{2}O$ (and $\rm NH_{3}$) closer to CCSD than to CCSD(T). In case of $\rm CH_2$ UCCSD-VQE is halfway between CCSD and CCSD(T) and for $\rm N_2$ UCCSD-VQE is very close to CCSD(T).
In case of $\rm OH$ (and $\rm Li$, $\rm H_2$) 
there is virtually no difference between all studied correlated methods (UCCSD-VQE, CCSD, CCSD(T) and FCI are formally identical for two-electron systems such as $\rm H_2$).
We note that the error introduced by the choice of the small basis set is by orders of magnitude larger than the errors due to the applied approximate electronic structure methods, see for example Table \ref{tab:appl:reactener_extrapol_H2O}. 

We roughly estimate the required quantum resources for a full UCCSD-VQE treatment of those example molecules on quantum hardware assuming that the number of qubits is equal to the number of spin orbitals in the given minimal basis set. This assumption leads to 10 qubits for $\rm OH$, 12 for 
$\rm H_{2}O$ and $\rm {:}CH_2$ and 16 for $\rm N_2$. 
The number of required two-qubit gates $-$ which are known to be the most critical gates $-$ within our gate-reduced UCCSD-VQE code 
exploiting T$_2$-amplitude pre-screening 
was counted to be 500 for $\rm OH$, 1200 for $\rm H_{2}O$, 1500 for $\rm {:}CH_2$ (as the number of occupied and unoccupied orbitals is more balanced than in $\rm H_{2}O$) and 2500 for $\rm N_2$.
For resource estimations of additional molecules see Table I in the SI.

\begin{figure*}
\subfigure[]{
\includegraphics[width=7cm ,angle=0,clip=]{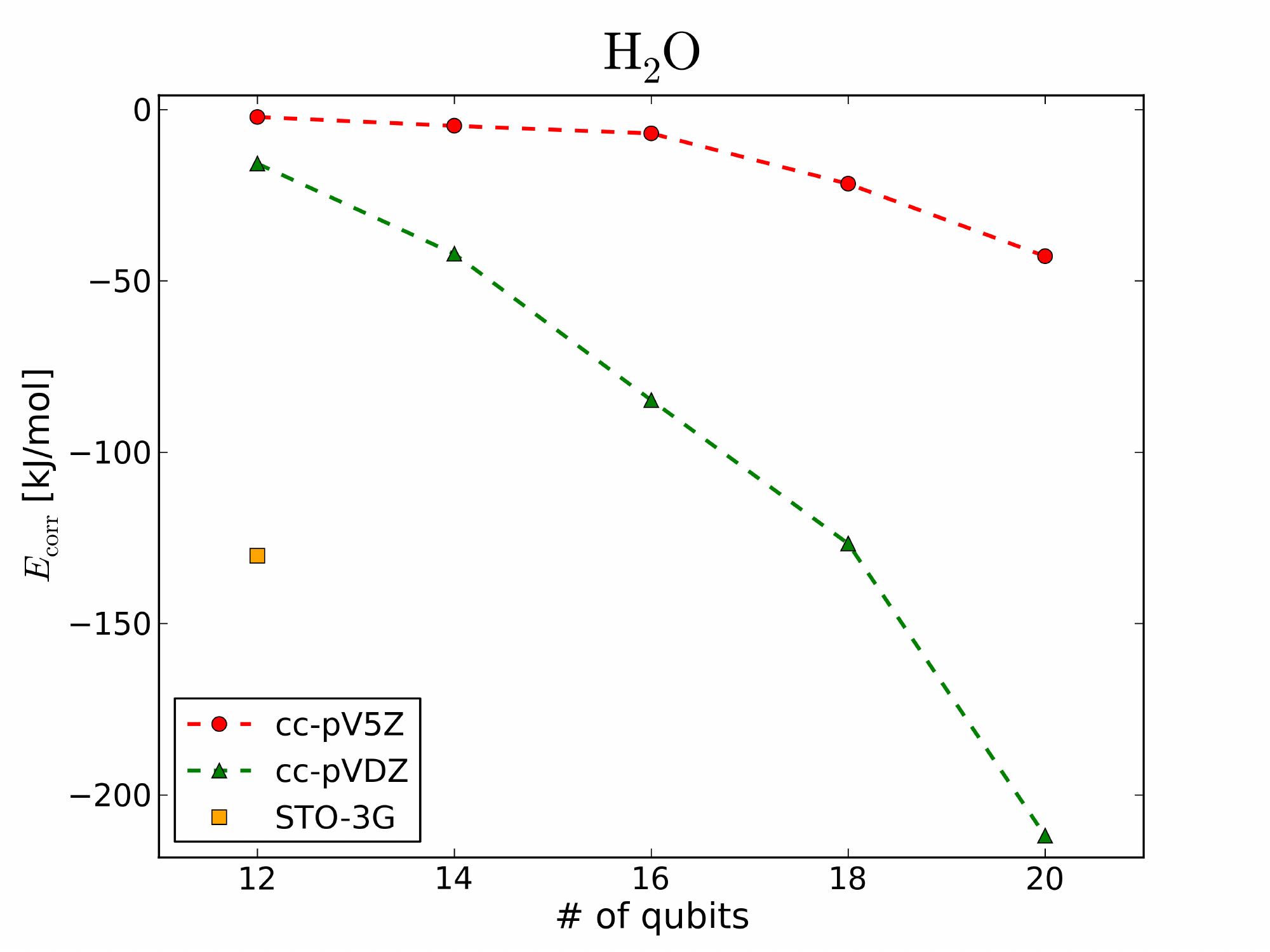}
}
\hspace{1cm}
\subfigure[]{
\includegraphics[width=7cm ,angle=0,clip=]{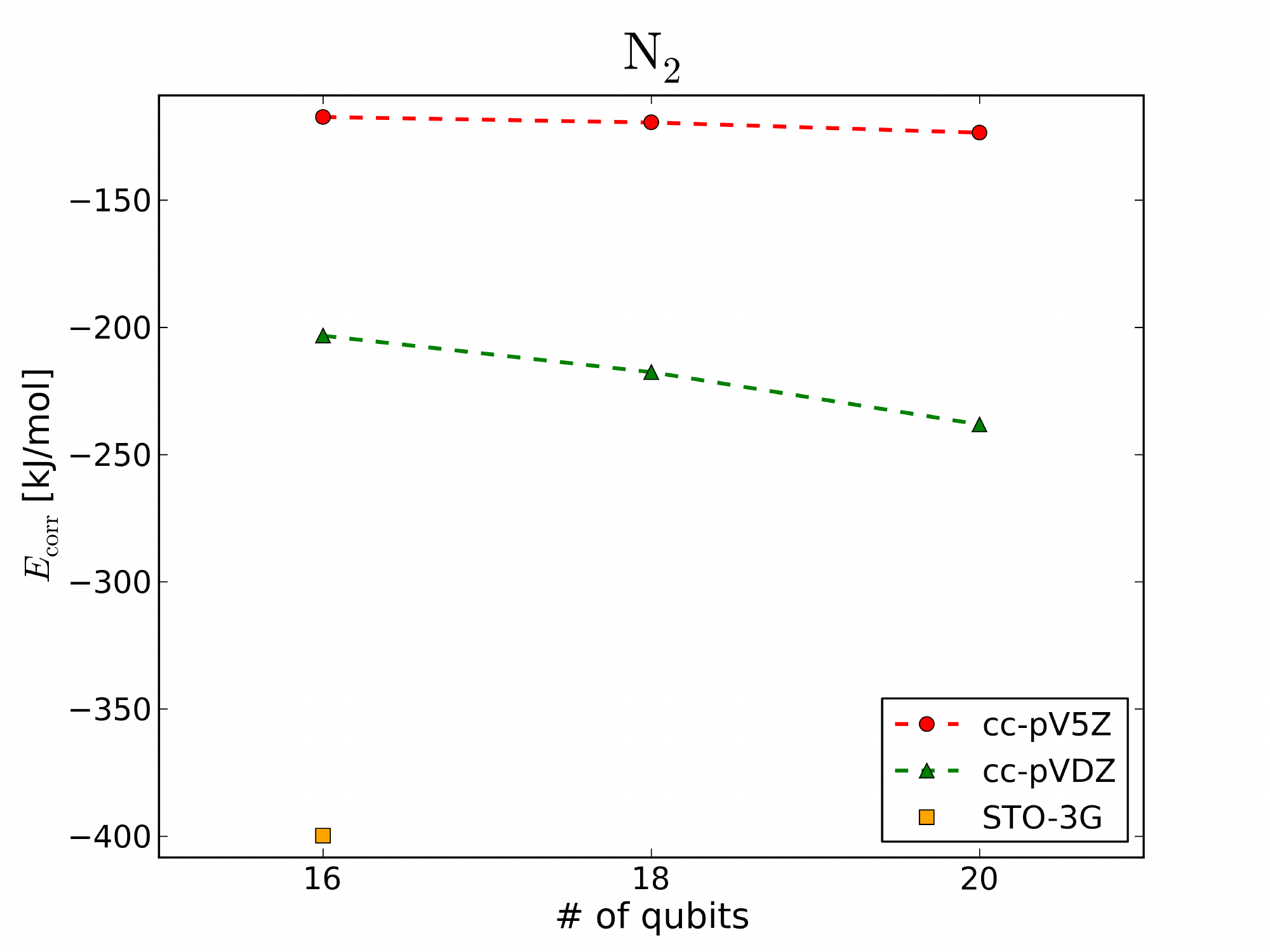}
} \\
\subfigure[]{
\includegraphics[width=7cm ,angle=0,clip=]{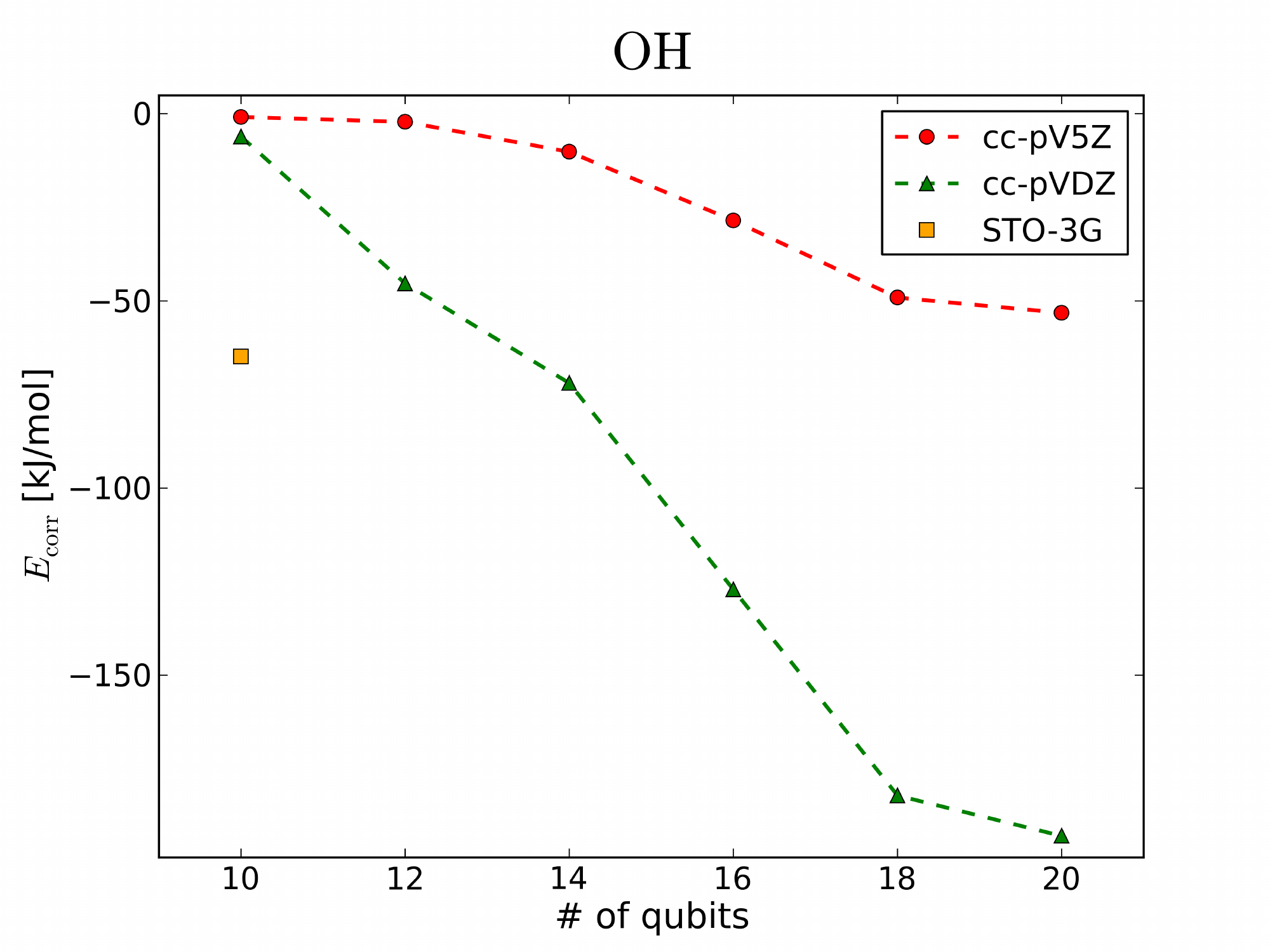}
}
\hspace{1cm}
\subfigure[]{
\includegraphics[width=7cm ,angle=0,clip=]{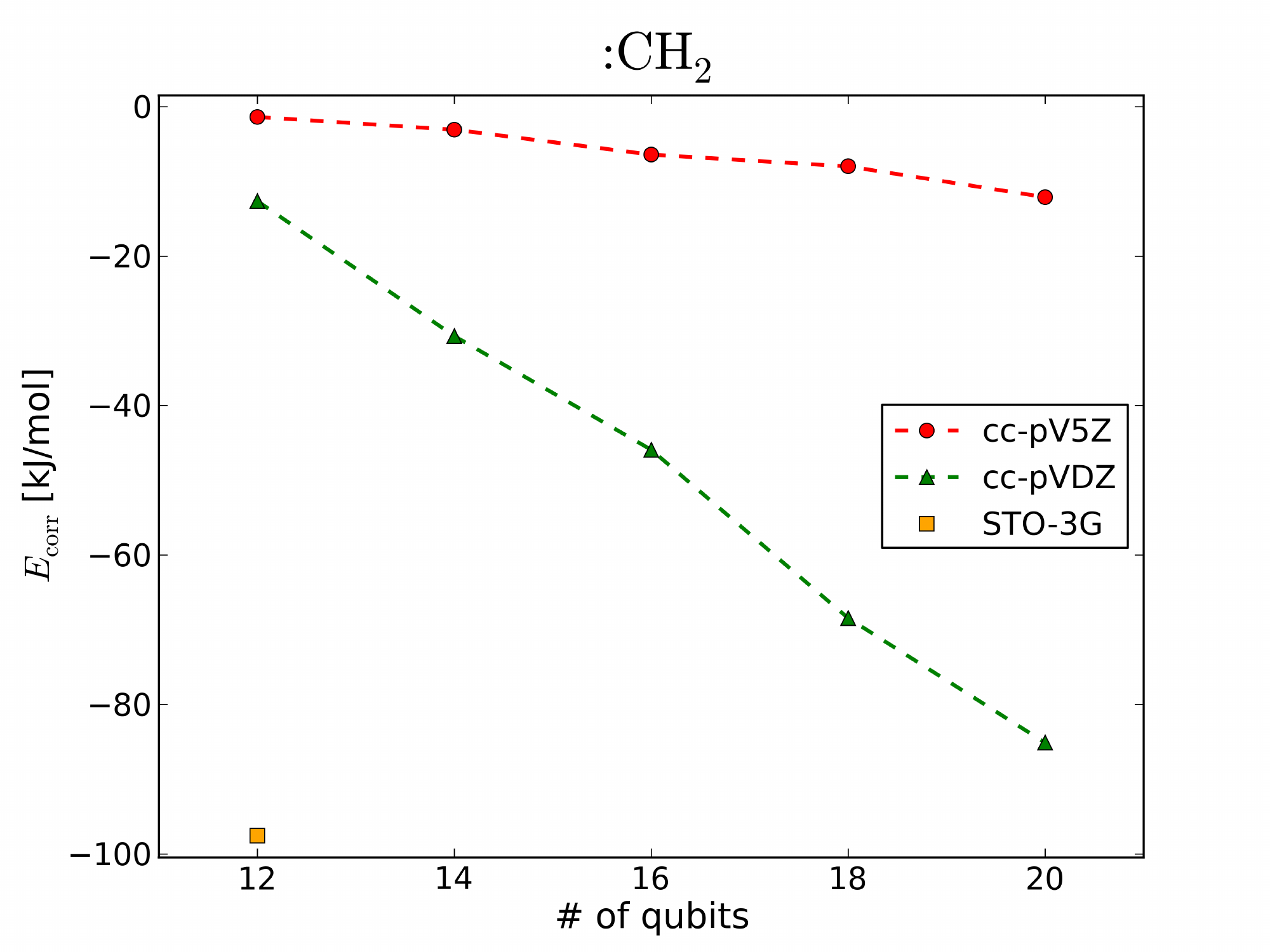}
}
\caption{
{\bf UCCSD-VQE correlation energies for up to 20 simulated qubits.}
Results are shown for molecules ${\rm H_{2}O}$ (a), ${\rm N_{2}}$ (b), ${\rm OH}$ (c) and ${\rm {:}CH_2}$ (d) in combination with
different basis sets (STO-3G, cc-pVDZ and cc-pV5Z).
}
\label{fig:appl:ecorr}
\end{figure*}

\subsubsection{UCCSD-VQE Calculations and Method Comparison in Larger Basis Sets}
\label{sec:appl:molener:basisuccsd}
As mentioned previously, in order to 
obtain highly accurate results usually large basis sets need to be applied in combination with correlated methods. 
For example, FCI will always give the exact result in the given basis set but only in the limit of a complete basis set this result will correspond to the exact solution of the electronic Schr{\"o}dinger equation.
It is known that closely approaching this basis set limit requires the use of very large basis sets, e.g. of valence 6-tuple-$\zeta$ quality or even larger, which dramatically increases the number of virtual orbitals to be included in the correlation treatment and therefore the computational cost (both on classical and quantum hardware). 
We note that as a more economic alternative to straightforwardly applying very large basis sets 
so-called explicitly correlated variants of CC have been developed.\cite{Haettig12} However, those methods exhibit a more complicated structure than conventional CC methods and require the introduction of an additional basis set and evaluation of additional integrals.
Therefore, in the following we will not put further focus on them. Nevertheless, we believe that they might also be of 
great value in the field of quantum chemistry on quantum computers in the longer term.

For most practical purposes, e.g. the study of chemical reactions, we assume that we are in general close 
enough to the basis set limit when applying a valence quintuple-$\zeta$ basis set, such as cc-pV5Z, and ``useful'', reliable predictions might already be made with smaller basis sets of valence triple- or quadruple-$\zeta$ quality, such as cc-pVTZ or cc-pVQZ.
Thus, in addition to the calculations in the minimal basis set discussed in the previous Section  
we also carried out UCCSD-VQE calculations using our ``reference'' basis set cc-pV5Z as well as an intermediate-sized basis set of valence double-$\zeta$ quality (cc-pVDZ).

\textcolor{black}{In Figure \ref{fig:appl:ecorr} the UCCSD-VQE correlation energy obtained with both basis sets is plotted against the number of simulated qubits for four exemplary molecules.
To check whether it is possible to reduce the required quantum resources by simple truncation of the space of canonical (HF) virtual orbitals, in a first step we only correlated the same orbitals as in the minimal basis set (e.g. for $\rm H_{2}O$ eight non-frozen occupied and four virtual spin orbitals yielding a total of 12 
simulated qubits). Then we gradually increased the number of simulated qubits by including more and more of the energetically lowest-lying virtual orbitals.
In total we ran the full UCCSD-VQE procedure for up to 20 qubits (also checking the robustness of our implementation).} 
Resulting correlation energies for additional molecules are given in 
Figure 1 in the SI.
Due to the exponential increase in the UCCSD-VQE computation cost with the number of qubits
we were not able to correlate all virtual orbitals present in the cc-pVDZ 
(for $\rm H_{2}O$ 8 occupied plus 38 virtual orbitals requiring a total of 46 qubits) and the cc-pV5Z basis sets (400 qubits for $\rm H_{2}O$).
However, for each molecule we counted the number of required two-qubit gates for up to 64 qubits.
The corresponding results for four exemplary molecules 
when exploiting gate cancellations as well as $T_2$-amplitude
pre-screening to reduce the overall gate counts
are visualized 
in Figure \ref{fig:appl:gates} (see Figures \ref{fig:appl:react1}-\ref{fig:appl:react4} for gate counts of additional molecules). 

As 
expected, 
the absolute value of the correlation energy increases with the number of simulated qubits (see Figure \ref{fig:appl:ecorr})
\textcolor{black}{while the number of required two-qubit gates also increases (quadratically, see Figure
\ref{fig:appl:gates}, which will be discussed later)}. 
Furthermore, it can be seen that for a fixed, small number of qubits (e.g. 12 available qubits for simulating the $\rm H_{2}O$ molecule) 
a larger portion of the correlation energy can be captured when 
applying a smaller basis set that ideally just requires all of the available qubits (e.g. 12) for inclusion of all canonical virtual orbitals in the correlation treatment. 
\textcolor{black}{The slower convergence of the correlation energy for the cc-pV5Z basis set compared to cc-pVDZ 
can be traced back to the fact that for large basis sets 
our simple truncation of the virtual orbital space in combination with
energetically low-lying
diffuse basis functions leads to inclusion of virtual orbitals that contribute only little to the correlation energy.
Therefore, it seems disadvantageous to apply large basis sets and then reduce the required quantum resources by only considering the energetically lowest-lying canonical virtual orbitals in the correlation treatment.
We note that this problem can at least partly be solved by rotating the canonical orbital basis to an improved orbital basis, e.g. to MP2 natural orbitals. The frozen natural orbital (FNO) approach is known to 
typically allow for significant truncation (of up to about $50\%$) of the virtual orbital space for large basis sets with minimal loss in accuracy for classical CC methods.\cite{Sosa89,Taube08,DePrince13} 
This therefore might 
also be a promising approach to reduce the required resources (number of qubits and thus also number of two-qubit gates) 
in case of UCCSD-VQE on quantum computers.}
The number of required two-qubit gates 
turns out to be virtually independent of the chosen basis set as long as the numbers of both occupied and virtual orbitals included in the correlation treatment are identical (i.e. for a given ``non-frozen orbital space'' which leads to a given number of qubits).

When comparing the UCCSD-VQE results for different basis sets and different numbers of qubits to the respective FCI, CCSD and CCSD(T) correlation energies, differences are very small compared to the absolute value of the correlation energy as 
already discussed in Section \ref{sec:appl:molener:method} for the minimal basis set. Thus, in Figure \ref{fig:appl:ecorr} all of the other methods would be on spot 
with the UCCSD-VQE results. As the absolute value of the correlation energy increases the differences between these different methods
also increase but the relative error (of up to $2.5\,\%$ in case of $\rm N_2$) remains nearly constant (slight increase).

We note that in case of $\rm LiH$ we were able to additionally carry out 
UCCSD-VQE calculations in combination with the SV, DZ and TZ basis sets correlating all available orbitals,
see Table II in the SI,
with all findings being in line with the above discussion.

\begin{figure*}
\subfigure[]{
\includegraphics[width=7cm ,angle=0,clip=]{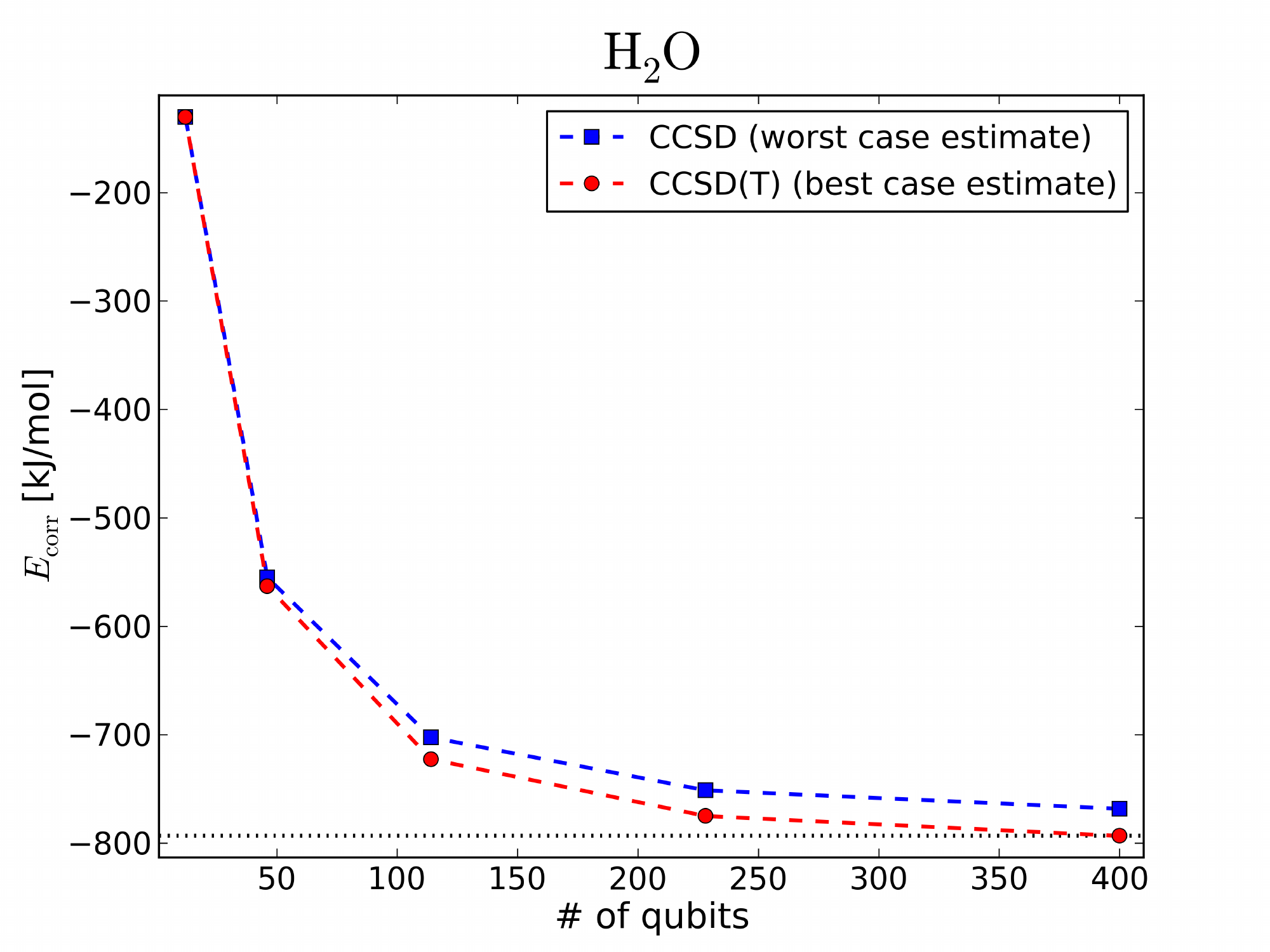}
}
\hspace{1cm}
\subfigure[]{
\includegraphics[width=7cm ,angle=0,clip=]{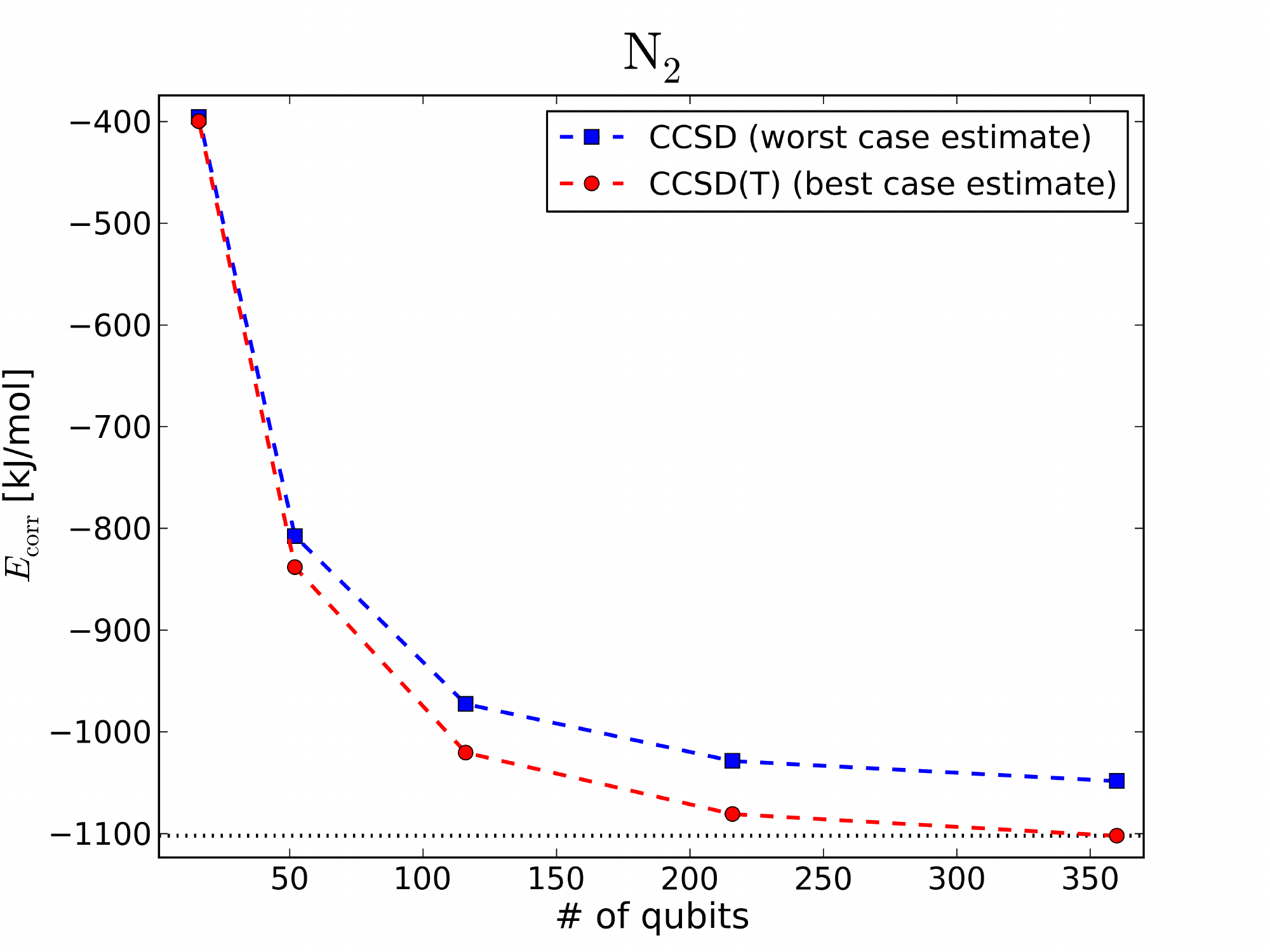}
} \\
\subfigure[]{
\includegraphics[width=7cm ,angle=0,clip=]{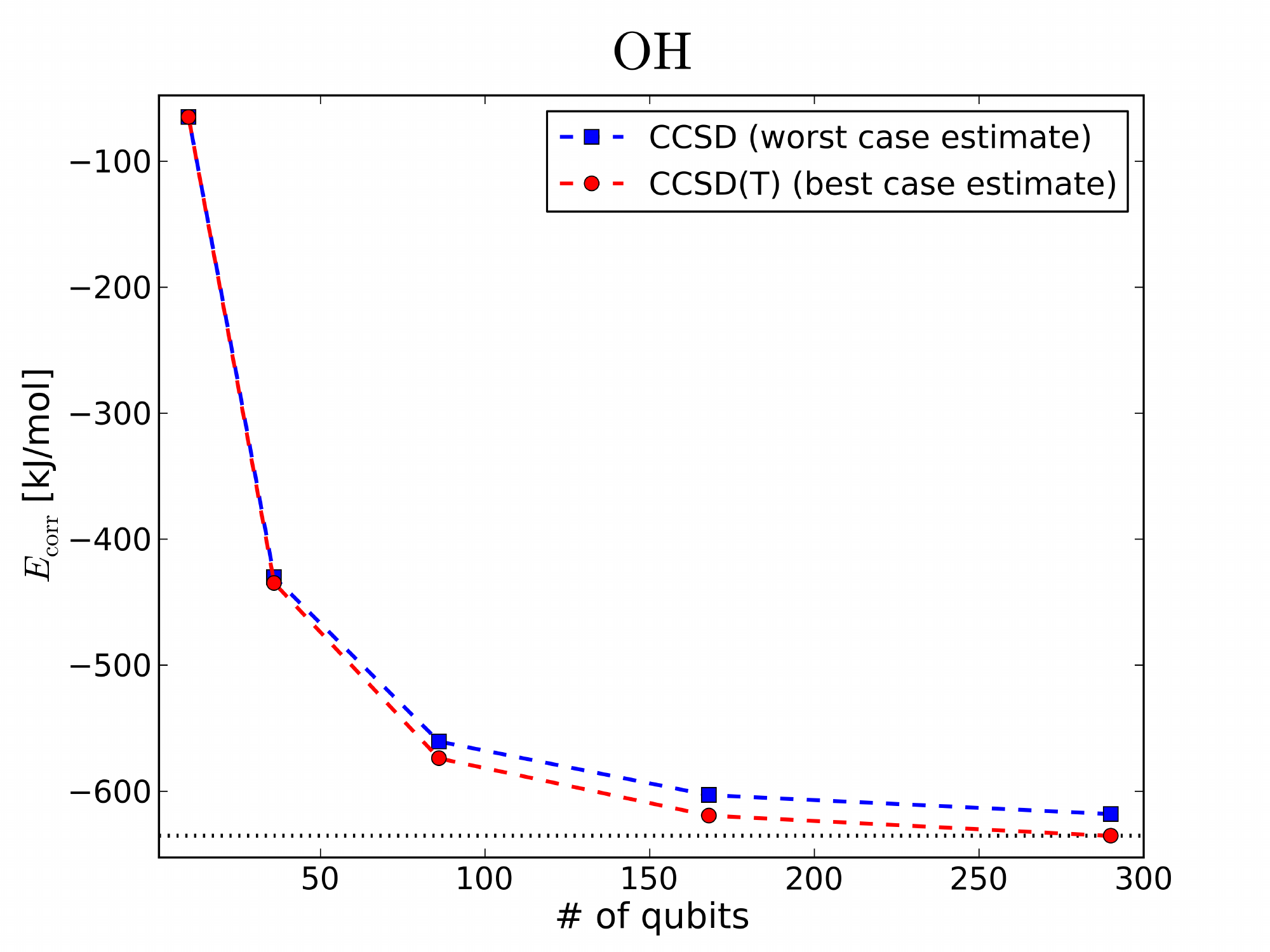}
}
\hspace{1cm}
\subfigure[]{
\includegraphics[width=7cm ,angle=0,clip=]{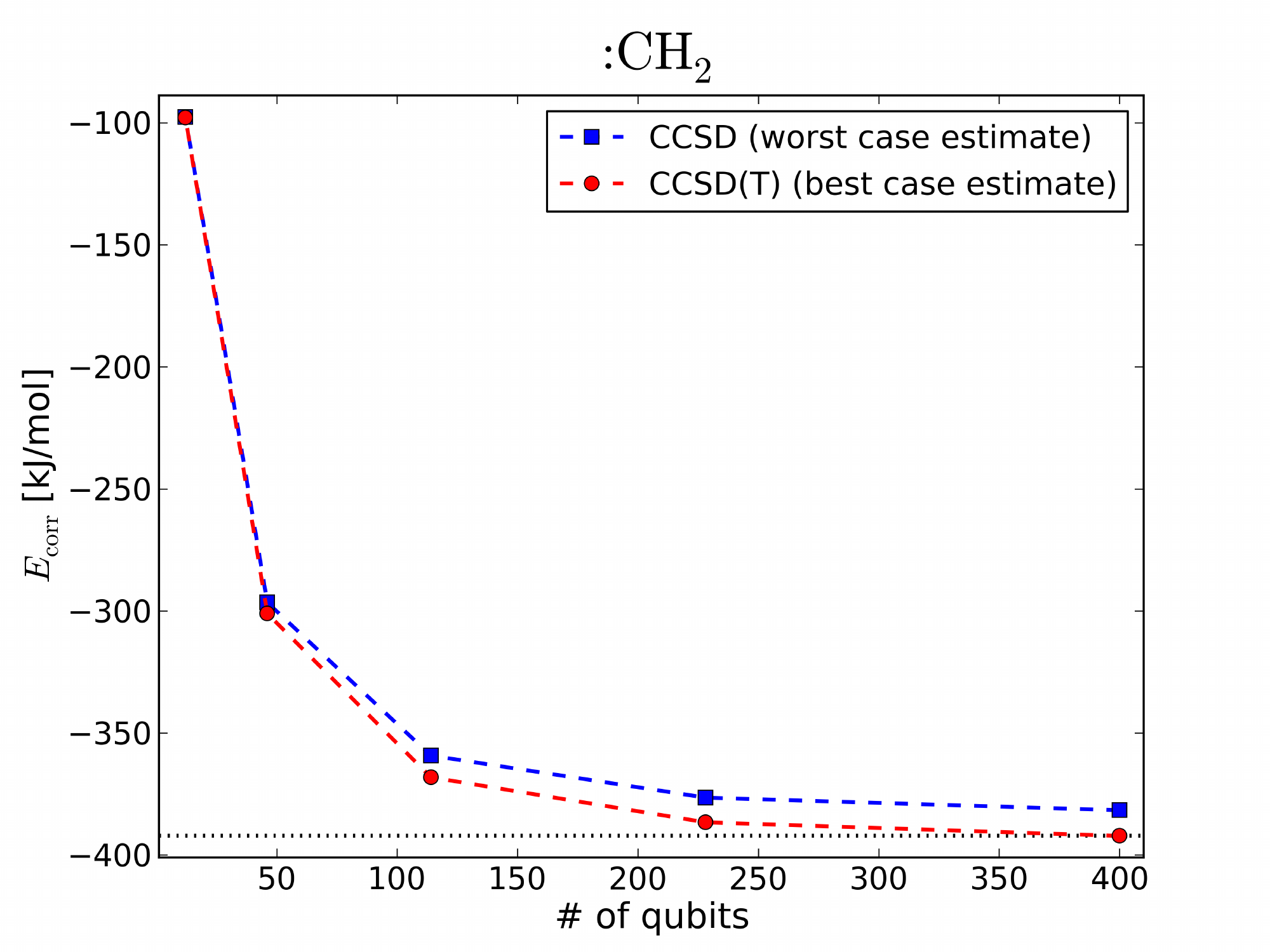}
}
\caption{
{\bf Extrapolated UCCSD-VQE correlation energies for a larger number of qubits.}
The extrapolation is done by using CCSD (worst-case estimate) and CCSD(T) (best-case estimate) correlation energies in accordance with the results shown in Table \ref{tab:appl:molener}.
Results are depicted for molecules ${\rm H_{2}O}$ (a), ${\rm N_{2}}$ (b), ${\rm OH}$ (c) and ${\rm {:}CH_2}$ (d) in combination with
basis sets of increasing size (STO-3G, cc-pVDZ, cc-pVTZ, cc-pVQZ and cc-pV5Z).
The dotted horizontal line represents the CCSD(T)/cc-pV5Z result which is chosen to be the target/reference energy.
}
\label{fig:appl:extrapolbest}
\end{figure*}

\subsubsection{Extrapolation to Large Basis Sets}
\label{sec:appl:molener:basis}

In order to estimate the accuracy and required quantum resources for UCCSD-VQE when consistently correlating all 
available orbitals within basis sets of increasing size up to our ``reference'' basis set cc-pV5Z,
we resort to CCSD and CCSD(T) which in contrast to UCCSD-VQE do not exhibit exponential scaling on a classical hardware (instead formally $N^6$- and $N^7$-scaling, respectively, with $N$ being a measure for the system size). Thus, those methods are applicable to much larger systems, i.e. a larger number of spin orbitals (number of qubits) can be correlated.
This ``extrapolation'' procedure to larger basis sets is based on our previous findings, that for all studied systems the UCCSD-VQE energy is always between the CCSD and the CCSD(T) energy as shown for the minimal basis set and for larger basis sets for up to 20 qubits (see also Section \ref{sec:appl:molener:method} and Table \ref{tab:appl:molener} as well as Tables I-II in the SI). 
\textcolor{black}{
Furthermore and in agreement with our observations, in previous
studies using purely classical UCCSD implementations\cite{Cooper10,Evangelista11,Harsha18} it was found that in cases where static correlation is weak the difference between UCCSD and CCSD is typically very small with UCCSD being slightly more accurate than CCSD but still inferior to CCSD(T). 
When static correlation is strong, e.g. in bond-breaking situations, 
UCCSD is typically also found to be somewhat more accurate and robust than CCSD. However, in this case both methods usually exhibit a significant error with respect to ''true`` multireference CC approaches and the exact FCI solution. 
The findings in those previous studies, in which comparably large basis sets of up to valence double- and triple-$\zeta$ quality were employed, additionally support our rough ``extrapolation'' technique:
in the following we will simply use  
the CCSD energy as a worst-case and the CCSD(T) energy as a best-case estimate for the UCCSD-VQE energy.
}

In Figure \ref{fig:appl:extrapolbest} the aforementioned UCCSD-VQE energy extrapolation by CCSD(T) and CCSD is shown for 
basis sets of increasing size (STO-3G, cc-pVDZ, cc-pVTZ, cc-pVQZ, cc-pV5Z) and
four exemplary molecules.
As the target/reference correlation energy for practical applications we choose the CCSD(T)/cc-pV5Z result, i.e. the quantum-chemical ``gold standard''
result close to the basis set limit, which is depicted as a dotted horizontal line.
We note that this ``gold standard'' method in a large basis set
should not be seen as the ``true'' solution but rather as a computationally still affordable accuracy measure for all other practically
relevant electronic structure methods.
The exact solution of the electronic Schr{\"o}dinger equation given by FCI in the basis set limit would in most cases further increase the absolute value of the correlation energy.
Therefore, one should keep in mind that in general (but not for two-electron systems) there is an additional energy gap between the exact result and CCSD(T)/cc-pV5Z. 
Furthermore, we note that even exactly solving the Schr{\"o}dinger equation
might not be sufficient to describe special phenomena such as spin-orbit coupling, non-Born-Oppenheimer effects, etc.
depending on the choice of the underlying Hamiltonian (usually electronic, non-relativistic and within the Born-Oppenheimer approximation).

In the SI (Figures 2-3 therein) we additionally show extrapolated UCCSD-VQE energies when not all virtual orbitals are included in the correlation treatment in order to reduce the required resources as done in Section \ref{sec:appl:molener:basisuccsd}.
As mentioned before, since this procedure in general leads to losses in the correlation energy $-$ and thus possibly to inferior results $-$ we will only focus on the results obtained by consistently correlating all virtual orbitals
present in the chosen basis set, which leads to an optimal and balanced correlation energy for each basis set. 

\begin{figure}
\includegraphics[width=9cm ,angle=0,clip=]{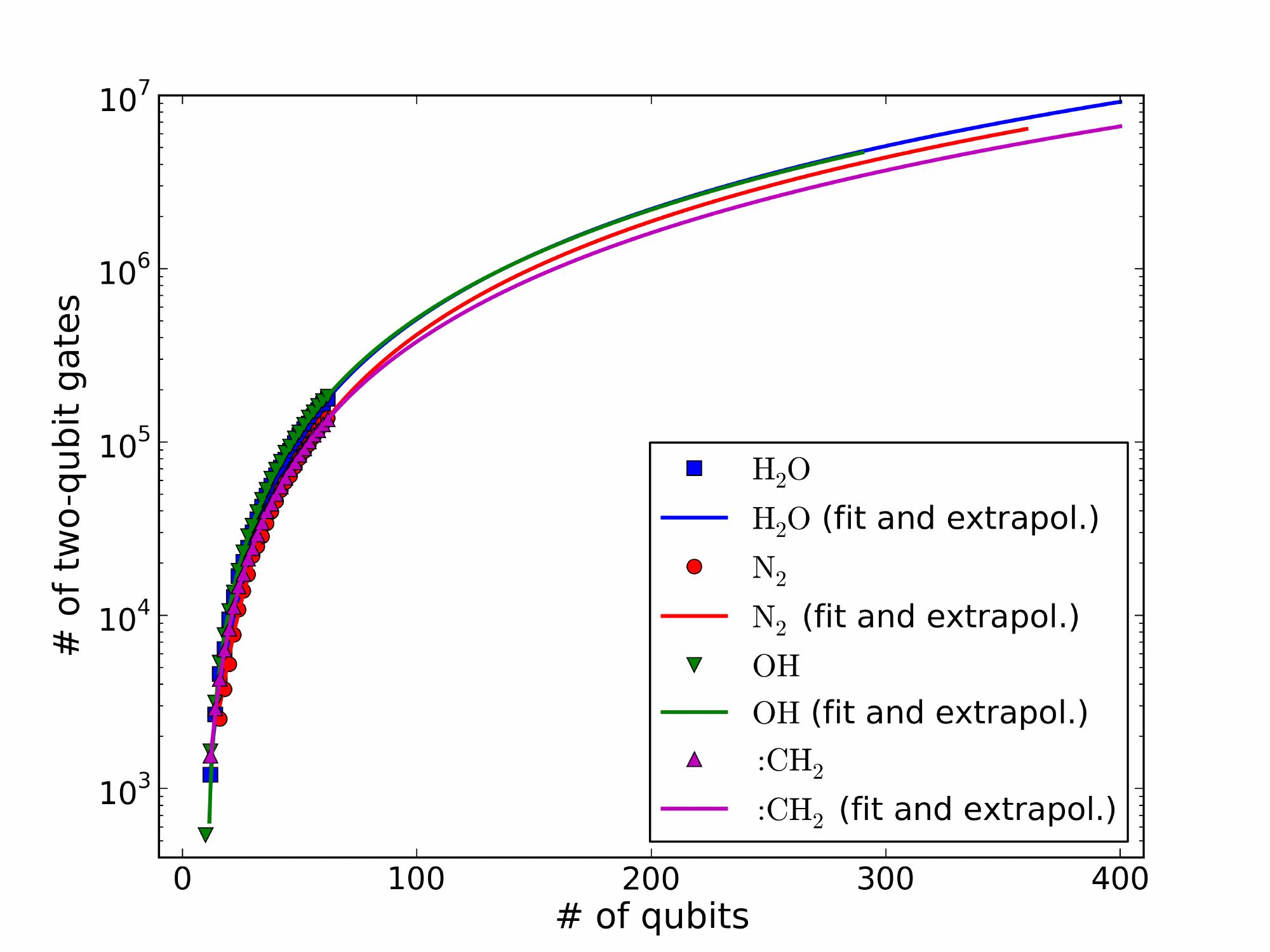}
\caption{
\textcolor{black}{
{\bf Estimated number of required two-qubit gates for the
UCCSD-VQE treatment of $\rm H_{2}O$, $\rm N_2$, OH and $\rm {:}CH_2$.}
We exploit gate cancellations as well as $T_2$-amplitude pre-screening based on MP2 to reduce gate counts.
The number of required two-qubit gates is evaluated for up to 64 qubits applying the cc-pV5Z basis set.
A quadratic polynomial is fitted to the calculated data and then used for extrapolation to a larger number of qubits, see also Section \ref{sec:appl:molener:basis}.
For further information see Figures \ref{fig:appl:react1}-\ref{fig:appl:react4}.
}
}
\label{fig:appl:gates}
\end{figure}

In order to obtain the corresponding number of required two-qubit gates for the larger basis sets we extrapolated our energies to,
we first fit a quadratic polynomial to the calculated gate counts (for up to 64 qubits) in Figure 
\ref{fig:appl:gates}. 
The high quality of the fit is expressed by very large coefficients of determination ($R^{2}$) which are always larger than 0.999 except for the smallest systems in our study (for LiH, Li and $\rm H_2$ larger than 0.98). 
For each molecule the fitted quadratic polynomial is then used for extrapolation to a larger number of qubits (larger basis sets). 

The finding that the number of two-qubit gates depends quadratically on the number of simulated qubits can be rationalized as follows. The number of two-qubit gates is proportional to the amount of amplitudes in the UCCSD
ansatz, Equation \eqref{eq_Coupled_Cluster_operators_definition}.
According to Equations \eqref{eq_Coupled_Cluster_single_operator_definition} and
\eqref{eq_Coupled_Cluster_double_operator_definition} the number of singles amplitudes scales as
$N_\mathrm{virt}N_\mathrm{occ}$ while the number of doubles amplitudes scales as
$N_\mathrm{virt}^2N_\mathrm{occ}^2$.
Consequently, the number of two-qubit gates can be expressed as
\begin{equation}
  N_\mathrm{two-qubit} = c_{\text{s}} N_\mathrm{occ} N_\mathrm{virt} + c_{\text{d}} N_\mathrm{occ}^2 N_\mathrm{virt}^2,
\label{eqn:max_number_two_qubit_gates}
\end{equation}
where $c_{\text{s}}$ and $c_{\text{d}}$ can be interpreted as the average number of two-qubit gates per singles and doubles excitation, which can be reduced by means of several strategies, e.g. introducing MP2 pre-screening and/or exploiting gate cancellations. 
Thus, from Equation \eqref{eqn:max_number_two_qubit_gates} it becomes obvious that the number of two-qubit gates increases
quadratically with $N_\mathrm{virt}$, which is the quantity we varied when increasing the number of simulated qubits.

From Figures \ref{fig:appl:extrapolbest} and \ref{fig:appl:gates} it can be seen that 
the absolute value of the correlation energy increases with the number of simulated qubits 
(i.e. the size of the basis set). The reference result is asymptotically approached 
while the number of required two-qubit gates increase quadratically.
For example, in case of $\rm H_{2}O$ 
applying the minimal basis set 
will introduce a large basis set incompleteness error of $\rm +660\, kJ/mol$ (which is $83\,\%$ of the reference correlation energy of $\rm -793.3\, kJ/mol$) while
requiring 12 qubits as well as $1.2\cdot10^{3}$ two-qubit gates. 
From this it also becomes evident that UCCSD-VQE calculations in the minimal basis set will most likely not have any practical chemical relevance $-$ but nevertheless are of great importance for the understanding and advancement of quantum computing.
By our definition, the reference correlation energy is finally reached (or at least very well approximated depending on whether UCCSD-VQE is actually closer to CCSD(T) or CCSD) by a UCCSD-VQE calculation applying the cc-pV5Z basis set requiring 400 qubits and $9.2\cdot10^{6}$ two-qubit gates.
We note in passing that the dependence of the correlation energy on the size of the basis set has always been subject to intense research in quantum chemistry. 
Besides the aforementioned explicitly correlated variants of CC, there are also techniques which extract (approximate) 
results close to the basis set limit by extrapolating correlation energies obtained with a series of basis sets of increasing size as
suggested by the similar shape of all curves in Figure \ref{fig:appl:extrapolbest}.\cite{Haettig12}

For most other molecules studied in this work estimations are similar. 
Roughly speaking, approaching the ``basis set limit'' requires
around 300 to 400 qubits as well as $10^{6}$ to $10^{7}$ two-qubit gates for the small molecules presented in Figure \ref{fig:appl:extrapolbest} 
(less for $\rm H_2$, somewhat more in case of $\rm NH_3$, see Figures 2-4 in the SI). 
As mentioned before, ``useful'' predictions for practical purposes $-$ especially for chemical reactions where only energy differences matter (thus 
possibility of partial error compensation) $-$ 
might already be made with smaller basis sets like cc-pVQZ or cc-pVTZ, which would roughly halve the number of required qubits as well as reduce the
number of required two-qubit gates by up to one order of magnitude (for cc-pVTZ)
while introducing an average loss in the correlation energy of roughly around $\rm 3\, \%$ (cc-pVQZ) and $\rm 8\, \%$ (cc-pVTZ) for each individual species participating in the reaction. This loss is 
then expected to be of similar size as the error introduced by the approximate correlation treatment due to the choice of the method, which will be discussed in the following.
 
Besides the error due to the incompleteness of the basis set the error introduced by the approximate correlated electronic structure method itself 
also influences the accuracy of the result.
Thus, we will discuss
the difference between the CCSD and CCSD(T) correlation energies $-$ serving as our worst- and best-case estimates for UCCSD-VQE, respectively.
As the 
size of the basis set increases, 
this difference also increases roughly up to the cc-pVTZ level (for the example of $\rm H_{2}O$ then amounting to $\rm 20.1\, kJ/mol$ or $\rm 2.8\,\%$). Further increasing the  size of the basis set then only leads to a comparably small increase in the difference between
CCSD and CCSD(T) due to the fact that the correlation energy converges.
The same behavior is observed for all other molecules in our study with the special case of $\rm N_2$ where the 
difference between CCSD and CCSD(T) is comparably large for all basis sets increasing from $\rm 4.4\, kJ/mol$ ($\rm 1.1\,\%$) for STO-3G up to $\rm 52.2\, kJ/mol$ ($\rm 4.8\,\%$) for cc-pV5Z. 
As expected, in case of $\rm H_2$, Li and LiH there is (virtually) no difference between CCSD and CCSD(T) (as well as presumably UCCSD-VQE) for any of the applied basis sets.


In summary,
the finding that the difference between the CCSD and CCSD(T) correlation energy increases
with the size of the basis set
up to cc-pVTZ (from where it then remains within $\rm 4\, \%$ on average) indicates that at least a valence triple-$\zeta$ quality basis set is needed to approximately converge 
the perturbative triples contribution to CCSD(T).
As discussed above, at this point it is still possible to gain a significant amount 
of correlation energy when going
to basis sets of valence quadruple-$\zeta$ quality (on average reducing the basis set incompleteness error from
$\rm 8\,\%$ to $\rm 3\, \%$ with the exception of LiH and Li), which can be attributed to the known fact that the doubles contribution to CCSD(T)
converges slower with the size of the basis set than the perturbative triples contribution.
From our comparison we finally estimate that 
when combining CCSD/CCSD(T) with basis sets of valence triple- or quadruple-$\zeta$ quality 
the error introduced by the 
finite basis set is of similar size as the error introduced by the approximate correlated method itself. 
Thus, as a rule of thumb, 
to obtain ``useful'' results for practical
applications at least basis sets of valence triple- or quadruple-$\zeta$ quality should be used.
Since UCCSD-VQE is expected to be between CCSD and CCSD(T) 
we assume 
that the same basis set requirements apply,
i.e. for the calculation of small molecules $-$ as done in this study $-$ at least basis sets of 
valence triple- or quadruple-$\zeta$ quality with all available virtual orbitals correlated should be used roughly requiring between  
100 and 300 qubits and between $10^{5}$ and $10^{7}$ two-qubit gates. In the next section we will check and further refine those estimations by studying the accuracy of UCCSD-VQE reaction energies.

\subsection{Reaction Energies}
\label{sec:appl:reactener}

In this section we study the accuracy of relative energies obtained with UCCSD-VQE for four exemplary
chemical reactions by comparing to FCI, CCSD and CCSD(T) as well as to additional results obtained with DFT $-$ 
the most commonly used method for practically relevant quantum chemistry problems. 
Furthermore, we again roughly estimate the required quantum resources to obtain 
``meaningful'' results, i.e. reaction energies that are close to the CCSD/CCSD(T) results when
applying a sufficiently large basis set.
Subject to our study are the following chemical reactions:
$\rm H_{2}O$-dissociation, LiH-dissociation, Haber-Bosch process and the triplet-singlet transition in $\rm CH_{2}$.
The corresponding reaction energies are simply computed as the stoichiometrically weighted energy differences between 
all products and reactants meaning that they are just the weighted sums and differences of the molecular energies discussed in Section
\ref{sec:appl:molener}.
Thus, in practice reaction energies are obtained by separate calculation of all species participating in the reaction
and the overall hardware
requirements for treating a certain
reaction on a quantum computer would typically be determined by the computationally most demanding molecule.
We note that in practical applications such reaction energies build the foundation for further 
calculation of (experimentally accessible) thermodynamical and kinetical quantities, e.g. enthalpies, entropies, Gibbs free energies as well as reaction rates.

\begin{table*}[htbp]
\centering
\caption[]{
{\bf Accuracy of reaction energies obtained with UCCSD-VQE for four exemplary chemical reactions.}
The HF, CCSD, CCSD(T), FCI and UCCSD-VQE reaction energies are given together with the respective differences to the FCI reaction energy ($\Delta$FCI).
All results were obtained using the minimal basis set STO-3G. All values are in kJ/mol.
 }
\label{tab:appl:reactener}
\tiny
\begin{tabular}{lcccccccccccc}
 & \phantom{x} & \multicolumn{2}{c}{\bf{$\boldsymbol{\rm H_{2}O \, \rightarrow \, \, OH \, + \, H}$}} &  \phantom{x}      &  \multicolumn{2}{c}{\bf{$\boldsymbol{\rm LiH \, \rightarrow \, \, Li \, + \, H}$}} &  \phantom{x} & \multicolumn{2}{c}{\bf{$\boldsymbol{\rm N_2 \, + \, 3\,H_2 \,  \rightarrow \, \, 2\, NH_3}$}} & \phantom{x} & \multicolumn{2}{c}{\bf{$\boldsymbol{\rm {:}CH_2 \, \rightarrow \, \, CH_2}$}} \\
 &              &  & $\Delta$FCI                                             &               &      & $\Delta$FCI                           &              &       & $\Delta$FCI                      &             &      & $\Delta$FCI    \\
\hline
$E_{\text{react}}$(HF) & &                 350.569& -65.572 &       &210.111&-52.532&&       -168.523&-227.940&&    163.733&60.379      \\
\hline
$E_{\text{react}}$(CCSD) & &          415.828&-0.313     &   &262.616&-0.028&&      50.660&-8.757&&    104.112&0.757       \\
$E_{\text{react}}$(CCSD(T)) & &       416.008&-0.132     &   &262.638&-0.006&&      54.443&-4.974&&    103.817&0.462       \\
$E_{\text{react}}$(FCI) & &           416.140& 0        &   &262.644&0&&          59.417&0&&        103.355&0          \\
\hline
$E_{\text{react}}$(UCCSD-VQE) & &     415.875&-0.265     &   &262.616&-0.028&&      54.843&-4.574&&    103.843&0.488       \\

\end{tabular}
\end{table*}

\subsubsection{Method Comparison in the Minimal Basis Set}
\label{sec:appl:reactener:method}

In Table \ref{tab:appl:reactener}
the reaction energies obtained with different methods (HF, CCSD, CCSD(T), FCI and UCCSD-VQE) 
in combination with the minimal basis set are given.
Furthermore, the respective differences to the FCI reaction energies $-$ where electron correlation is fully 
taken into account in the given basis set $-$ are also shown.
All FCI reaction energies are positive and range between $\rm 59\, kJ/mol$ for the Haber-Bosch process 
and $\rm 416\, kJ/mol$ in case of the $\rm H_{2}O$-dissociation meaning that {\it at this level of theory}
for each reaction the products are higher in energy than the reactants and thus all reactions are expected to consume energy.
In case of the $\rm H_{2}O$- and LiH-dissociation
as well as the Haber-Bosch process all other methods underestimate those FCI reaction energies whereas 
for the triplet-singlet transition in $\rm CH_{2}$ an overestimation is observed.
Largest errors with respect to FCI are given in case of the HF method 
ranging up to $\rm -227.9\, kJ/mol$ for the Haber-Bosch process which even leads to a sign change of the corresponding reaction energy.

Comparing CCSD and CCSD(T) to the FCI result shows
that in case of   
the $\rm H_{2}O$- and LiH-dissociation
errors are relatively small (up to $\rm -0.3 \, kJ/mol$ or $\rm -0.1\, \%$) compared to 
the absolute values of the reaction energies with CCSD(T) being closer to FCI than CCSD.
For both reactions the CCSD/CCSD(T) errors in reaction energies are virtually identical to the errors
in the $\rm H_{2}O$ and LiH molecular energies, respectively,
which can be attributed to the fact
that in case of all other participating molecules/atoms, namely OH, Li and H, 
all applied correlated methods lead to virtually identical results. We note that in case of the hydrogen atom, which is a single-electron system, there is no electron correlation and thus already HF is exact within the given basis set.
For the triplet-singlet transition in $\rm CH_{2}$
CCSD/CCSD(T) errors are somewhat larger (up to $\rm +0.8 \, kJ/mol$ or $\rm 0.7\, \%$).
By comparing to the errors in the corresponding individual molecular energies of $\rm {:}CH_{2}$ and $\rm CH_{2}$
(Table \ref{tab:appl:molener} and Table I in the SI) it can be concluded that there is partial error compensation when calculating the transition energy.
In case of the Haber-Bosch process CCSD and CCSD(T) errors are largest
due to the involvement of the challenging $\rm N_2$ (see Section \ref{sec:appl:molener:method}) amounting to  
$\rm -8.8 \, kJ/mol$ ($\rm -14.7\, \%$)
and $\rm -5.0 \, kJ/mol$ ($\rm -8.4\, \%$), respectively,
despite 
partial error compensation.

UCCSD-VQE reaction energies
are usually between CCSD and CCSD(T) reaction energies $-$ as expected from the molecular energies discussed in 
Section \ref{sec:appl:molener:method}.
In detail, 
for the LiH-dissociation all methods are very close with UCCSD-VQE being virtually identical to CCSD. 
In case of the $\rm H_{2}O$-dissociation UCCSD-VQE is closer to CCSD than to CCSD(T) and for 
the triplet-singlet transition in $\rm CH_{2}$
UCCSD-VQE is very close to CCSD(T). In case of the Haber-Bosch process the UCCSD-VQE reaction energy
is even slightly closer (error of $\rm -4.6 \, kJ/mol$) to FCI than CCSD(T)
due to better error compensation.
Roughly speaking,  
these results again indicate that also for reaction energies CCSD can be used as a worst-case 
and CCSD(T) as a best-case estimate for UCCSD-VQE.
We note that $-$ as for molecular energies $-$ the error introduced by the choice
of the small basis set is by orders of magnitude larger than errors due to the approximate correlated method itself,
see for example Table 
\ref{tab:appl:reactener_extrapol_H2O}.

As mentioned above, the overall required quantum resources for a UCCSD-VQE calculation
of a chemical reaction are determined by the molecule requiring the largest number of qubits and/or two-qubit gates. They have already been discussed in Section \ref{sec:appl:molener:method} in case the minimal basis set is applied. 
This leads to a requirement of 12 qubits and around 1200$-$1500 two-qubit gates for the $\rm H_{2}O$- and LiH-dissociation as well as for the triplet-singlet transition in $\rm CH_{2}$ (computationally most demanding molecules for each reaction being $\rm H_{2}O$, LiH and $\rm {:}CH_{2}$, respectively). 
In case of the Haber-Bosch process 16 qubits (for $\rm N_2$) and around 6000 two-qubit gates (for $\rm NH_3$) are roughly needed. 
The fact that in our estimations $\rm NH_3$ requires more two-qubit gates than $\rm N_2$ (6000 vs. 2500) despite a smaller number of qubits (14 vs. 16) is mainly due to a more efficient $T_2$-amplitude pre-screening 
in the case of $\rm N_2$. We note that for larger basis sets of the cc-pV$X$Z family 
$\rm NH_3$ always is computationally more demanding than $\rm N_2$ both with respect to the number of required qubits and two-qubit gates. 

\begin{table*}[htbp]
\centering
\caption[]{
{\bf Extrapolated UCCSD-VQE reaction energies for the $\rm H_{2}O$-dissociation, ${\rm H_{2}O \, \rightarrow \, \, OH \, + \, H}$, 
along with corresponding individual molecular energies for basis sets of increasing size.}
The extrapolation is carried out by using CCSD energies (worst-case estimates) in accordance with the results shown in Table \ref{tab:appl:molener}.
The obtained CCSD total energies are splitted into a HF total energy contribution and a CCSD correlation energy contribution. For comparison, DFT reaction energies using the functionals BP86, B3LYP and M06-2X are also shown.
Additionally, the number of required qubits as well as the number of two-qubit gates are estimated for the
computationally most demanding system at a certain basis set quality (here ${\rm H_{2}O}$),
see also \textcolor{black}{the extrapolation in Figure \ref{fig:appl:react1}}.
All energies are in kJ/mol.
 }
\label{tab:appl:reactener_extrapol_H2O}
\tiny
\begin{tabular}{llcccccc}
 &   & \phantom{x} & STO-3G & cc-pVDZ & cc-pVTZ & cc-pVQZ & cc-pV5Z  \\
\hline
                                    & $E_{\text{total}}$(HF)                     &&     -196815.7       &                                       -199607.9       &       -199687.5               &               -199707.7       &       -199713.6            \\
{\bf{$\boldsymbol{\rm H_{2}O}$}}    & $E_{\text{corr}}$(CCSD)                    &&     -130.2  &                                       -554.9  &       -702.4          &               -751.2  &       -768.2                                               \\
                                    & $E_{\text{total}}$(CCSD)                   &&     -196945.9       &                                       -200162.8       &       -200389.9               &               -200458.9       &       -200481.8            \\
\hline
                                    & $E_{\text{total}}$(HF)                     &&     -195240.2       &                                       -197946.3       &       -198012.9               &               -198030.2       &       -198035.1             \\
{\bf{$\boldsymbol{\rm OH}$}}        & $E_{\text{corr}}$(CCSD)                    &&     -64.9   &                                       -430.2  &       -560.5          &               -603.1  &       -618.1                                                \\
                                    & $E_{\text{total}}$(CCSD)                   &&     -195305.1       &                                       -198376.4       &       -198573.4               &               -198633.3       &       -198653.2              \\
\hline
{\bf{$\boldsymbol{\rm H}$}}         & $E_{\text{total}}$                         &&     -1225.0 &                                       -1310.9 &       -1312.3         &               -1312.6 &       -1312.7                                                \\
\hline
                                    & $E_{\text{react,\,total}}$(HF)              &&    350.6   &                                       350.8   &       362.4           &               364.8   &       365.8                                                  \\
                                    & $E_{\text{react,\,corr}}$(CCSD)             &&    65.3    &                                       124.7   &       141.8           &               148.1   &       150.1                                                   \\
\bf Reaction                            & $E_{\text{react,\,total}}$(CCSD)            &&        415.8   &                                       475.5   &       504.3           &               512.9   &       \bf 515.9                                                   \\
                                    & \# qubits                                  &&     12      &                                       46      &       114             &               228     &       400                                                     \\
                                    & \textcolor{black}{\# two-qubit gates}        &&   $1.2 \cdot 10^{3}$      &                                       $8.9 \cdot 10^{4}$      &       $6.8 \cdot 10^{5}$              &               $2.9 \cdot 10^{6}$      &       $9.2 \cdot 10^{6}$                \\
\hline
                                    & $E_{\text{react,\,total}}$(BP86)            &&    504.2   &                                       507.0   &       523.5           &               527.0   &       528.5                                                    \\
\bf Reaction (DFT)                      & $E_{\text{react,\,total}}$(B3LYP)           &&        488.9   &                                       490.3   &       507.1           &               510.9   &       512.5                                                    \\
                                    & $E_{\text{react,\,total}}$(M06-2X)          &&    499.2   &                                       500.2   &       515.9           &               518.2   &       519.6                                                       \\
\end{tabular}
\end{table*}

\subsubsection{Extrapolation to Large Basis Sets}
\label{sec:appl:reactener:basis}

In the following
we roughly estimate the required quantum resources
to obtain ``useful'' reaction energies
by extrapolating UCCSD-VQE to sufficiently large basis sets
as done in Section \ref{sec:appl:molener:basis}
for molecular energies.
In Table \ref{tab:appl:reactener_extrapol_H2O}
extrapolated UCCSD-VQE reaction energies as well as quantum resource estimations 
for the example of the 
$\rm H_{2}O$-dissociation are shown 
along with the corresponding molecular energies
for basis sets of increasing size (always correlating all virtual orbitals within the chosen basis set).
\textcolor{black}{CCSD energies are again used as worst-case and CCSD(T) energies as best-case estimates for UCCSD-VQE energies in accordance with all previous findings.\cite{Cooper10,Evangelista11,Harsha18}}
For comparison, we also present DFT reaction energies obtained with three commonly used exchange-correlation functionals: 
generalized gradient approximation functional BP86, hybrid functional B3LYP,
hybrid meta functional M06-2X designed for main-group thermochemistry.
In Figure  
\ref{fig:appl:react1}
the main results from Table \ref{tab:appl:reactener_extrapol_H2O}, i.e. the basis set convergence of CCSD and CCSD(T) reaction energies,
are visualized. In Figure \ref{fig:appl:react1} (a) the errors of the 
extrapolated UCCSD-VQE reaction energies 
with respect to the CCSD(T)/cc-pV5Z reference reaction energy are plotted
against the number of simulated qubits.
Additionally, errors in B3LYP/cc-pV5Z and M06-2X/cc-pV5Z reaction energies are shown.
\textcolor{black}{
In Figure \ref{fig:appl:react1} (b) the estimated number of required two-qubit gates for our gate-reduced UCCSD-VQE treatment 
is plotted against the number of qubits for all
many-electron species participating in the reaction, see also Figure \ref{fig:appl:gates} for more details. 
}
Respective tables for all other reactions can be found in the SI, Tables IV-VI. 
Corresponding results are visualized in Figures
\ref{fig:appl:react2}-\ref{fig:appl:react4}.

\subsubsection*{Resource Estimations for Obtaining Practically Relevant Results}

For each chemical reaction we now roughly estimate the required quantum resources
to obtain UCCSD-VQE results within $\rm 4\, kJ/mol$ (approximately $\rm 1\, kcal/mol$ or $\rm 1\, mH$, a commonly used accuracy threshold in quantum chemistry also termed as ``chemical accuracy'') 
of the respective 
CCSD(T)/cc-pV5Z reference reaction energy, which is approached by successively increasing the number of 
qubits (size of the basis set) as well as the number of two-qubit gates. 
We note that the above-mentioned ``chemical accuracy'' threshold 
only serves as a rough guideline, since
in cases where UCCSD-VQE is closer to CCSD than to CCSD(T), see Section 
\ref{sec:appl:reactener:method}, this accuracy criterion might actually never be fulfilled $-$ even in the limit of a complete basis set $-$ if the difference between CCSD and CCSD(T) is too large. 

\begin{figure*}[htb]
\subfigure[]{
\includegraphics[width=7cm ,angle=0,clip=]{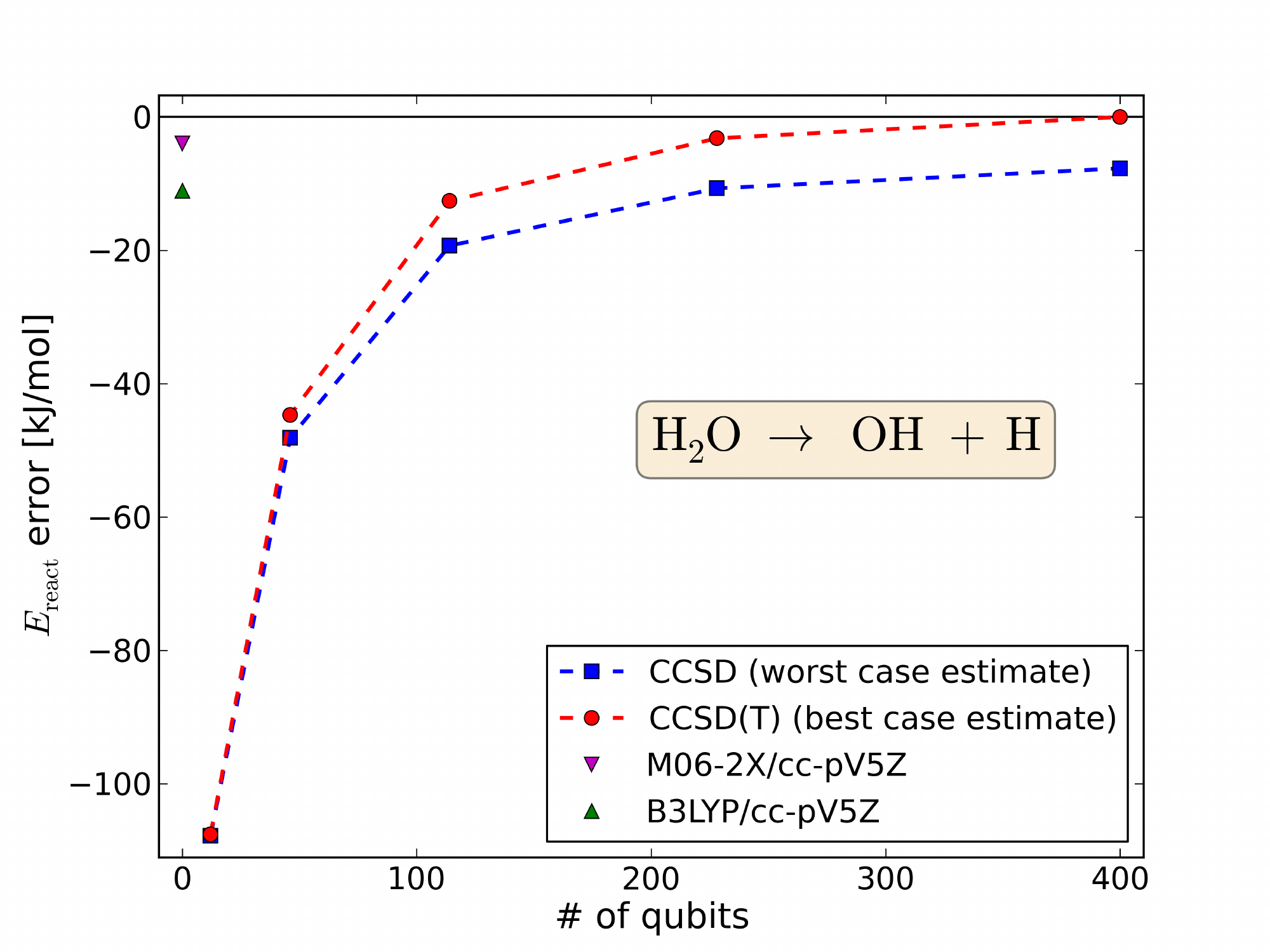}
}
\hspace{1cm}
\subfigure[]{
\includegraphics[width=7cm ,angle=0,clip=]{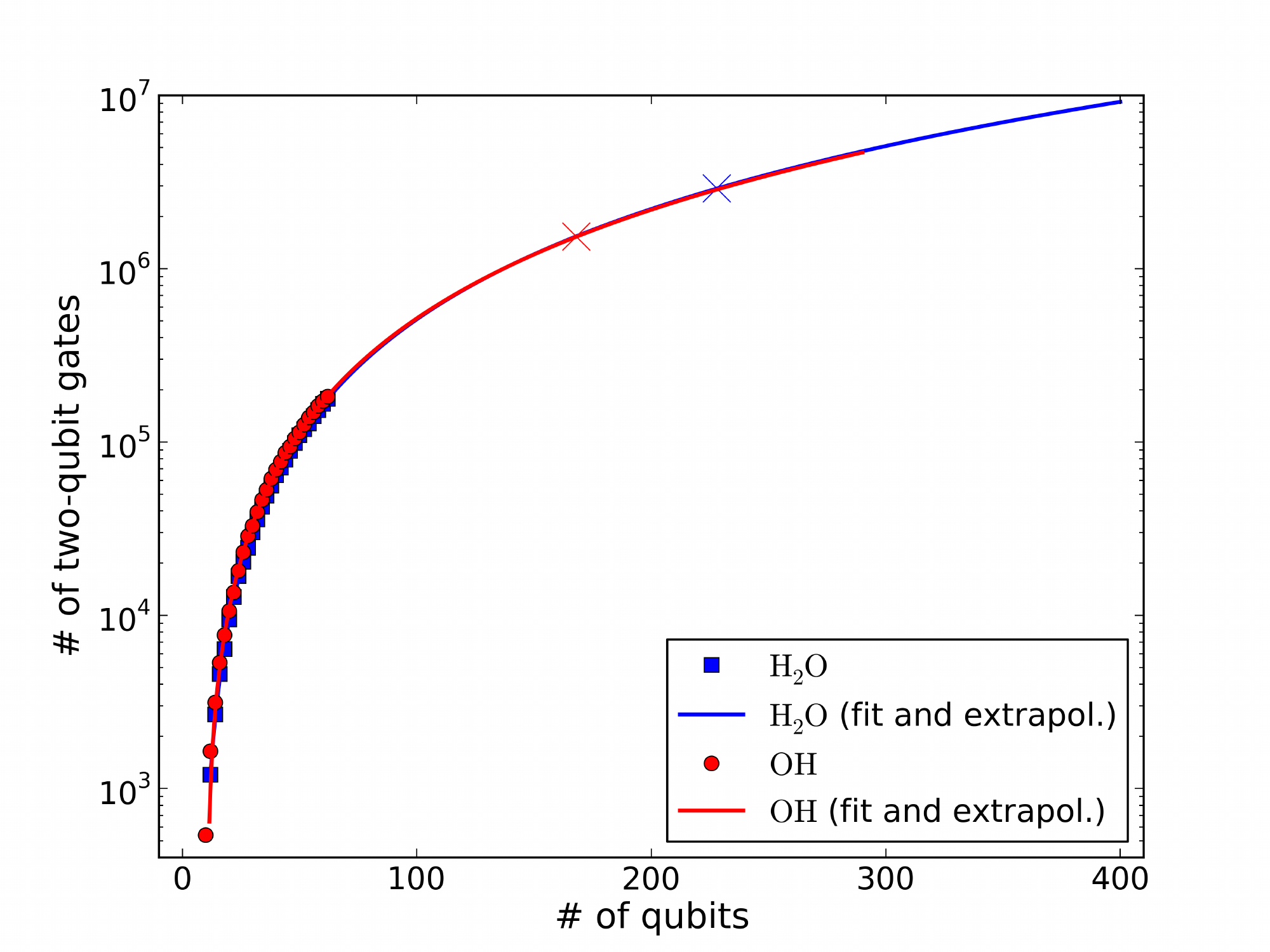}
}
\caption{
{\bf Estimation of the required quantum resources for obtaining reaction energies of certain accuracy for the
$\rm H_{2}O$-dissociation.}
\\
(a) Error in (extrapolated) UCCSD-VQE reaction energies for a larger number of qubits with respect to the CCSD(T)/cc-pV5Z target/reference reaction energy (523.6 kJ/mol). The extrapolated UCCSD-VQE results are obtained by using CCSD (worst-case estimate) and CCSD(T) (best-case estimate) reaction energies. For further information see Table \ref{tab:appl:reactener_extrapol_H2O}.
For comparison, errors in DFT reaction energies using the functionals B3LYP and M06-2X in combination with the cc-pV5Z basis set are also depicted.
\\
\textcolor{black}{
(b) Estimated number of required two-qubit gates for an optimized UCCSD-VQE treatment.
A quadratic polynomial ($an_{\text{qubits}}^2+bn_{\text{qubits}}+c$) is fitted to the calculated data
($a_{\rm H_{2}O} = 59.6$, $b_{\rm H_{2}O} = -885.2$, $c_{\rm H_{2}O} = 3258.5$, $a_{\rm OH} = 57.7$, $b_{\rm OH} = -597.4$, $c_{\rm OH} = -117.1$) and then used for extrapolation to a larger number of qubits.
For orientation, the required quantum resources for a calculation
using the cc-pVQZ basis set (needed to obtain a reaction energy within or close to ``chemical accuracy'' of the CCSD(T)/cc-pV5Z reference)
are visualized by an ``x'' for each molecule.
For further information see also Figure \ref{fig:appl:gates}.
}
}
\label{fig:appl:react1}
\end{figure*}

In case of the $\rm H_{2}O$-dissociation, see Figure \ref{fig:appl:react1} and Table \ref{tab:appl:reactener_extrapol_H2O},
increasing basis set from STO-3G to cc-pVDZ and cc-pVTZ roughly reduces the error in the UCCSD-VQE reaction energy 
(with respect
to the reference of $\rm 523.6\, kJ/mol$ which is 
close to the experimental result of $\rm 492\, kJ/mol$\cite{Maksyutenko06} if the comparably large zero-point vibrational energy contribution of $\rm -33.8\, kJ/mol$ is also considered)
from $\rm -110\, kJ/mol$ to $\rm -45\, kJ/mol$ 
and $\rm -20\, kJ/mol$ while increasing the number of qubits from 12 to 46 and 114 as well as the number of two-qubit gates from $\rm 1.2\cdot 10^{3}$ to $\rm 8.9\cdot 10^{4}$ and $\rm 6.8\cdot 10^{5}$, respectively.
At the cc-pVQZ level requiring 228 qubits and $\rm 2.9\cdot 10^{6}$ two-qubit gates the UCCSD-VQE reaction energy is finally within 
``chemical accuracy'' or
a few kJ/mol
of the reference cc-pV5Z result (which is approached using 400 qubits and $\rm 9.2\cdot 10^{6}$ two-qubit gates).
The required quantum resources for obtaining the desired ``chemical accuracy'' (cc-pVQZ basis set needed)
are visualized by an ``x'' for each molecule in Figure \ref{fig:appl:react1} (b).
When comparing those requirements for $\rm H_{2}O$ and OH it is obvious that $\rm H_{2}O$ is the more demanding system both with respect 
to the number of qubits and number of two-qubit gates, thus dictating the overall quantum hardware requirements. However, we note that for a fixed number of qubits (e.g. 150) OH would require slightly
more two-qubit gates than $\rm H_{2}O$.
For all basis sets the above-discussed errors in reaction energies are by a factor of five smaller than the errors in $\rm H_{2}O$ and OH 
molecular energies due to partial error compensation.
As for molecular energies
the difference between the CCSD and CCSD(T)
reaction energy increases
with the size of the basis set 
from $\rm 0.2\, kJ/mol$ for STO-3G 
up to $\rm 6.7\, kJ/mol$ for cc-pVTZ with the 
CCSD reaction energy always being below the CCSD(T) reaction energy.
Further increasing the size of the basis set only leads to comparably small increases in those differences up to 
$\rm 7.7\, kJ/mol$ for cc-pV5Z. 
This again indicates that at least basis sets of valence triple-$\zeta$ quality are needed to approximately converge the
perturbative triples contribution in CCSD(T) to the reaction energy.
Comparing to $\rm H_{2}O$ and OH molecular energies, 
differences between CCSD and CCSD(T)  
are approximately by a factor of three smaller when calculating the dissociation reaction energy due to partial error compensation.  
In summary, applying the cc-pVQZ basis set requiring 228 qubits and $\rm 2.9\cdot 10^{6}$ two-qubit gates leads to a basis set incompleteness 
error in the reaction energy which is of similar size as the difference between CCSD and CCSD(T) (a few kJ/mol).

\begin{figure*}[htb]
\subfigure[]{
\includegraphics[width=7cm ,angle=0,clip=]{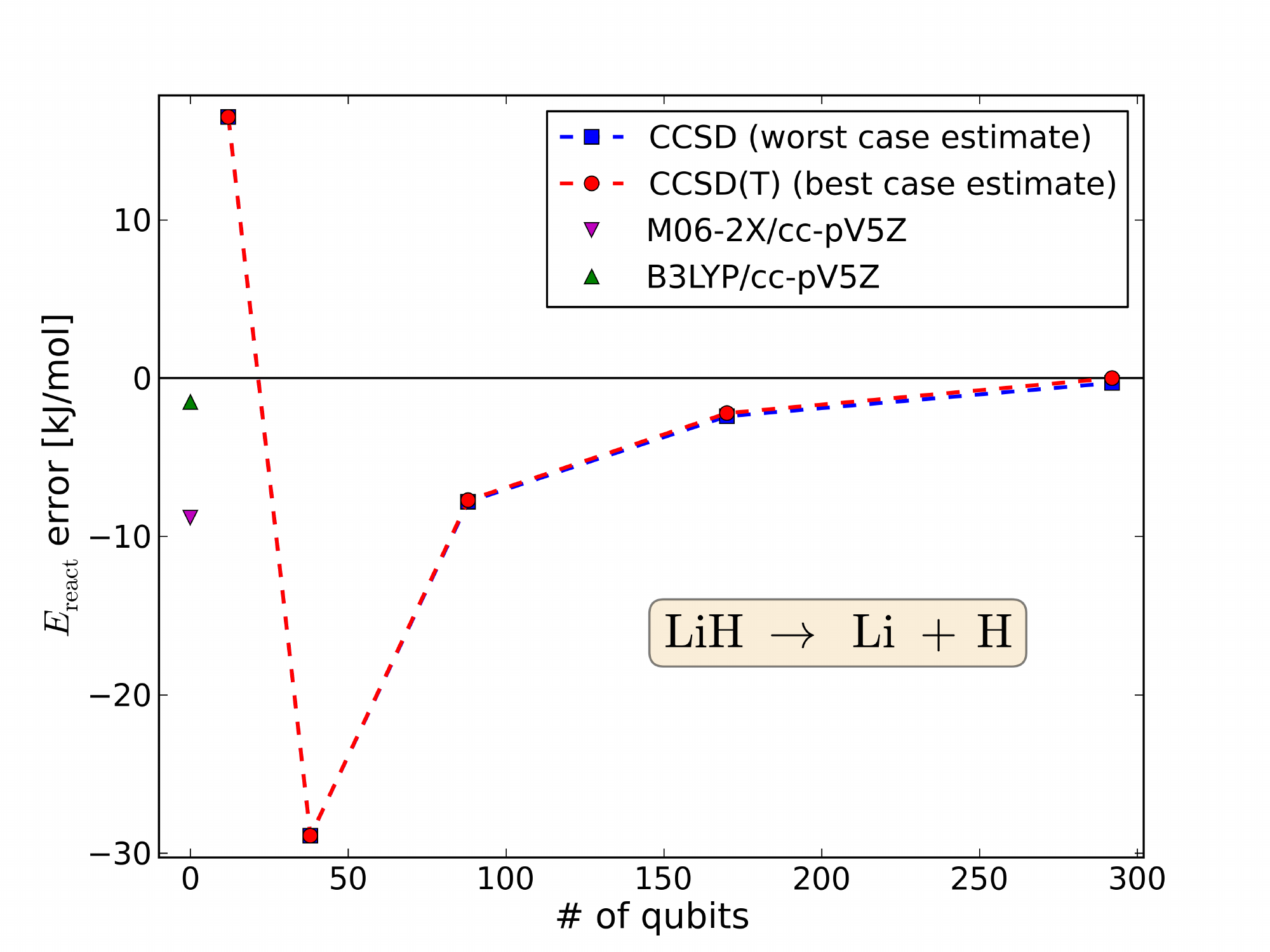}
}
\hspace{1cm}
\subfigure[]{
\includegraphics[width=7cm ,angle=0,clip=]{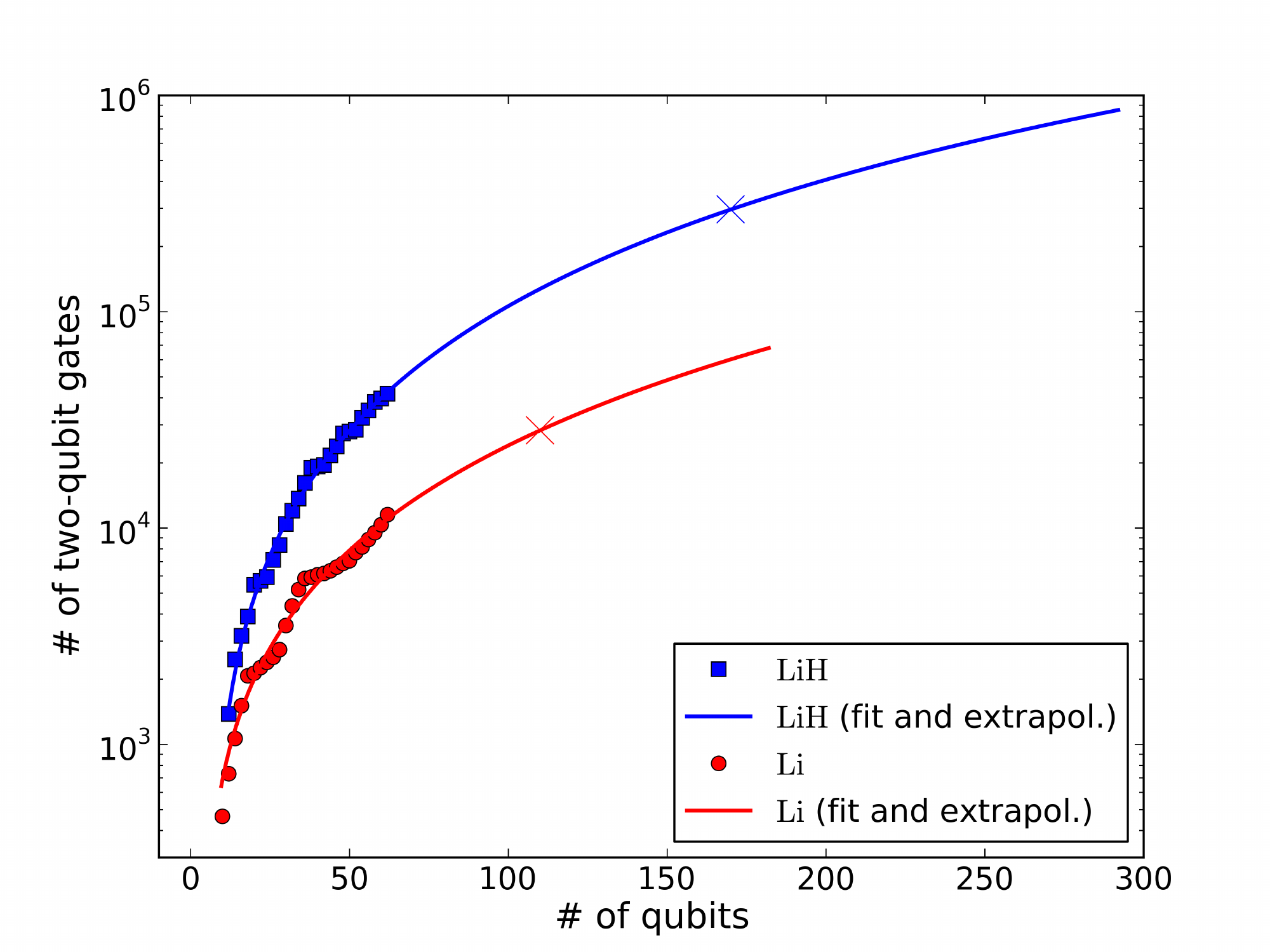}
}
\caption{
{\bf Estimation of the required quantum resources for obtaining reaction energies of certain accuracy for
the LiH-dissociation.} For further information see Table IV in the SI and Figure \ref{fig:appl:react1}.
\\
(a) Error in (extrapolated) UCCSD-VQE reaction energies with respect to the CCSD(T)/cc-pV5Z reaction energy (246.1 kJ/mol).
\\
\textcolor{black}{
(b) Estimated number of required two-qubit gates for an optimized UCCSD-VQE treatment. A quadratic polynomial ($an_{\text{qubits}}^2+bn_{\text{qubits}}+c$;
$a_{\rm LiH} = 9.7$, $b_{\rm LiH} = 103.5$, $c_{\rm LiH} = -1171.9$, $a_{\rm Li} = 1.6$, $b_{\rm Li} = 84.0$, $c_{\rm Li} = -312.0$)
is used for extrapolation to a larger number of qubits.
For further information see also Figure \ref{fig:appl:gates}.
}
}
\label{fig:appl:react2}
\end{figure*}

For the LiH-dissociation, see Figure \ref{fig:appl:react2},
both the absolute errors as well as the reference reaction
energy of $\rm 246.1\, kJ/mol$ (which is in very good agreement with the experimental result of $\rm 242\, kJ/mol$\cite{Chan86}) are smaller than for the $\rm H_{2}O$-dissociation leading to similar
relative errors.
Also, we find a comparably   
small error when applying the minimal basis set which $-$ in contrast to the larger cc-pV$X$Z basis sets $-$ overestimates the reference reaction energy. 
We note that the slow basis set convergence of LiH and Li molecular energies 
(i.e. comparably large errors when applying cc-pVTZ and cc-pVQZ) is not reflected in the LiH-dissociation reaction energy due to error compensation.
Thus, as for the $\rm H_{2}O$-dissociation applying the cc-pVQZ basis set leads to an error below
a few kJ/mol requiring 170 qubits and $\rm 3.0\cdot 10^{5}$ two-qubit gates 
with our worst- and best-case estimates for the UCCSD-VQE reaction energy being virtually identical.
We note that in case of the LiH-dissociation
we have additionally calculated actual UCCSD-VQE reaction energies (as done in case of the minimal basis set; not by extrapolation)
for the SV, DZ and TZ basis sets with all findings being in line with the previous discussion,
see Table III in the SI. 

\begin{figure*}[htb]
\subfigure[]{
\includegraphics[width=7cm ,angle=0,clip=]{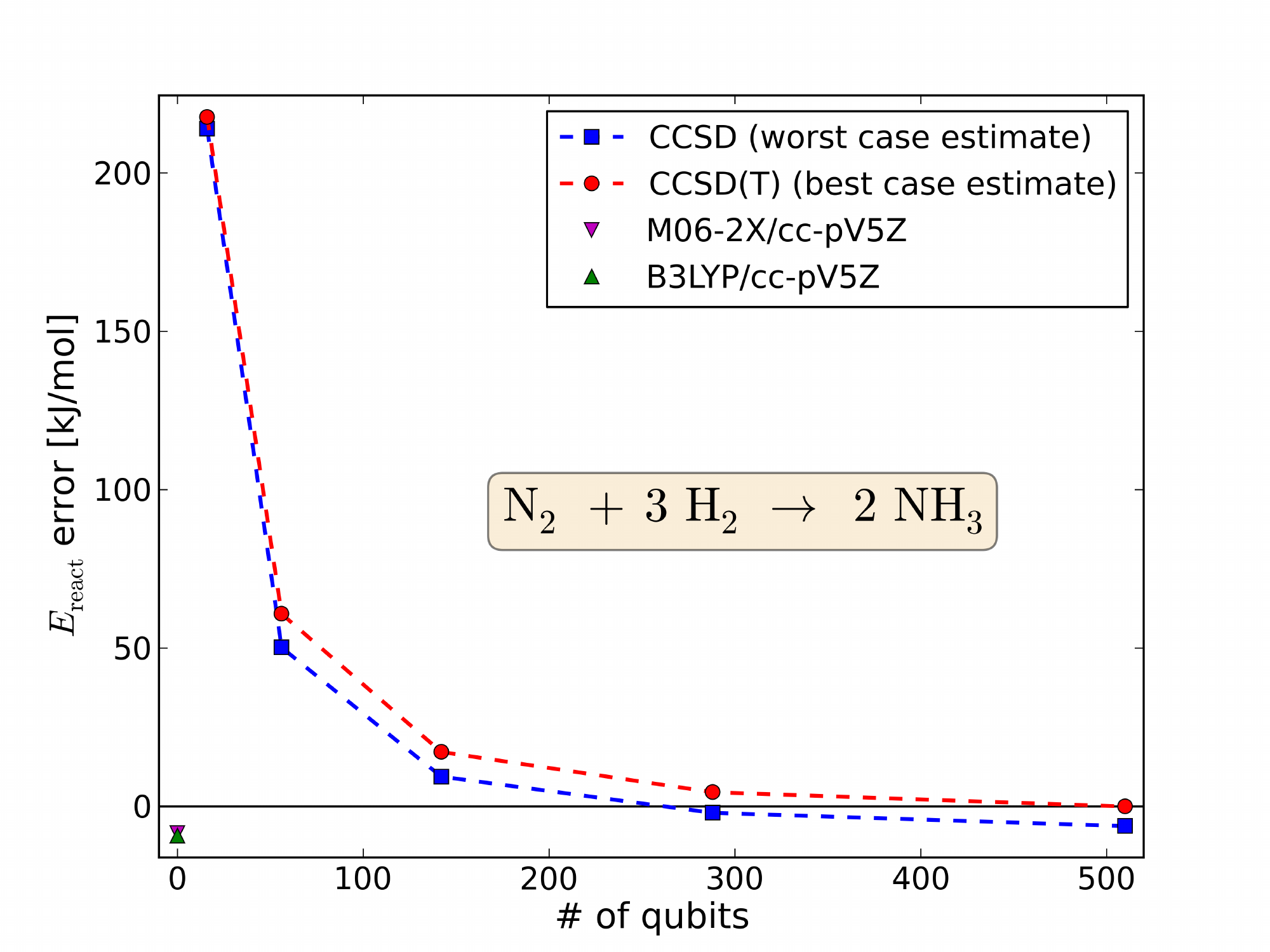}
}
\hspace{1cm}
\subfigure[]{
\includegraphics[width=7cm ,angle=0,clip=]{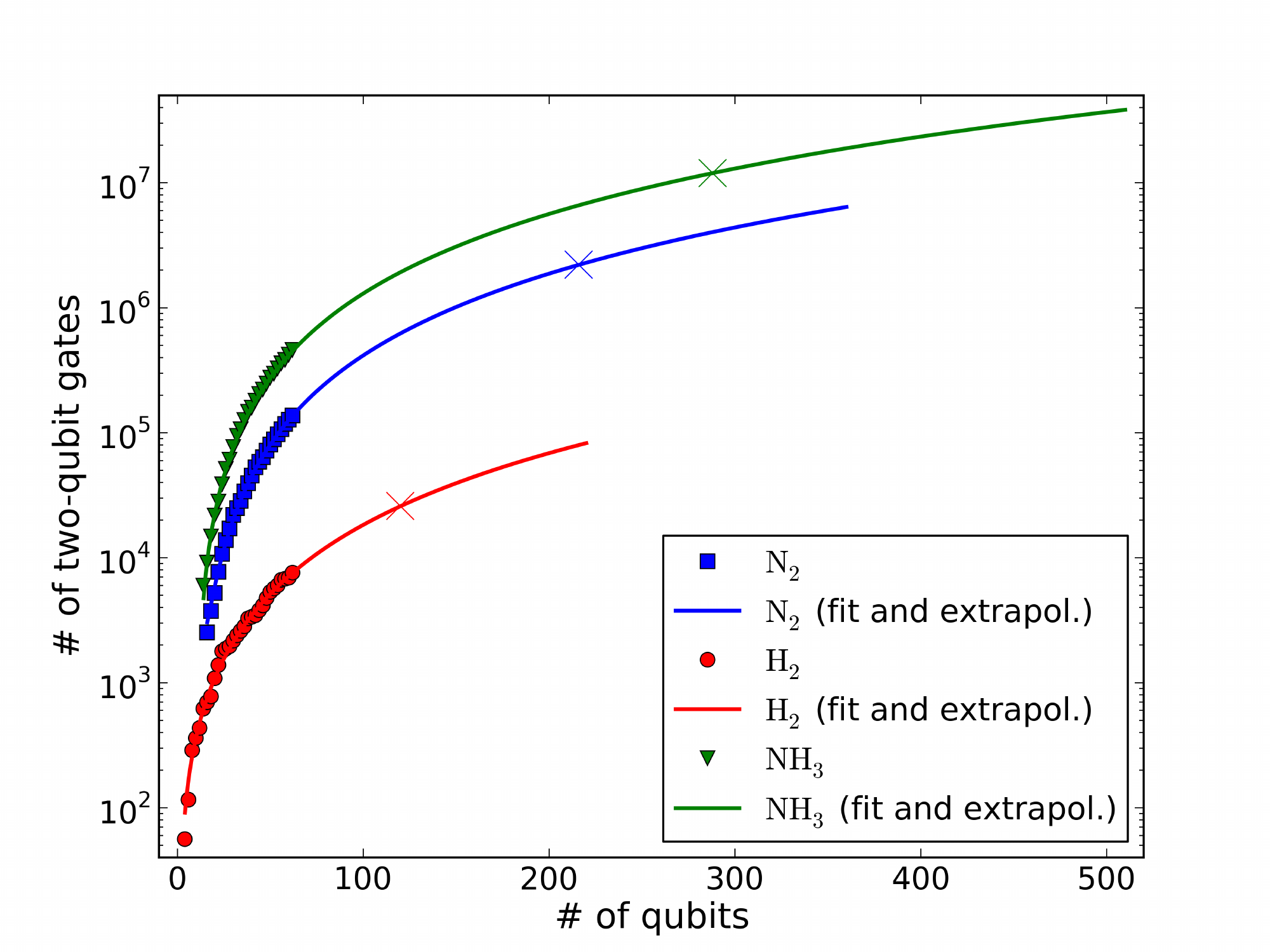}
}
\caption{
{\bf Estimation of the required quantum resources for obtaining reaction energies of certain accuracy for
the Haber-Bosch process.}
For further information see Table V in the SI and Figure \ref{fig:appl:react1}.
\\
(a) Error in (extrapolated) UCCSD-VQE reaction energies with respect to the CCSD(T)/cc-pV5Z reaction energy (-163.3 kJ/mol).
\\
\textcolor{black}{
(b) Estimated number of required two-qubit gates for an optimized UCCSD-VQE treatment. A quadratic polynomial ($an_{\text{qubits}}^2+bn_{\text{qubits}}+c$;
$a_{\rm N_2} = 52.6$, $b_{\rm N_2} = -1185.4$, $c_{\rm N_2} = 8542.9$, $a_{\rm H_2} = 1.6$, $b_{\rm H_2} = 23.7$, $c_{\rm H_2} = -29.1$,
$a_{\rm NH_3} = 151.7$, $b_{\rm NH_3} = -2215.0$, $c_{\rm NH_3} = 6026.5$)
is used for extrapolation to a larger number of qubits.
For further information see also Figure \ref{fig:appl:gates}.
}}
\label{fig:appl:react3}
\end{figure*}

For the Haber-Bosch process, see Figure \ref{fig:appl:react3},
absolute errors are somewhat larger than in case of the $\rm H_{2}O$-dissociation (relative errors more than twice as large)
and a slightly slower basis set convergence is observed.
Thus, at least the cc-pVQZ basis set needs to be applied
to achieve an error within a few kJ/mol
of the reference reaction energy requiring 288 qubits and $\rm 1.2\cdot 10^{7}$ two-qubit gates. 
At this basis set level the difference between our worst- and best-case estimates for the UCCSD-VQE reaction energy is also of similar size ($\rm 6\, kJ/mol$).
Remarkably, when 
using the minimal basis set a very large basis set incompleteness error of around $\rm +220\, kJ/mol$ is observed 
which even leads to a change in the sign of the reaction energy with respect to the cc-pV5Z reference result.
We also note that 
there is a sign change of the 
correlation energy contribution ($E_{\text{react,\,corr}}(\rm CCSD)$, see Table V in the SI)
when increasing the basis set from cc-pVDZ to cc-pVTZ, underlining 
again the importance of applying sufficiently large basis sets.

\begin{figure*}[htb]
\subfigure[]{
\includegraphics[width=7cm ,angle=0,clip=]{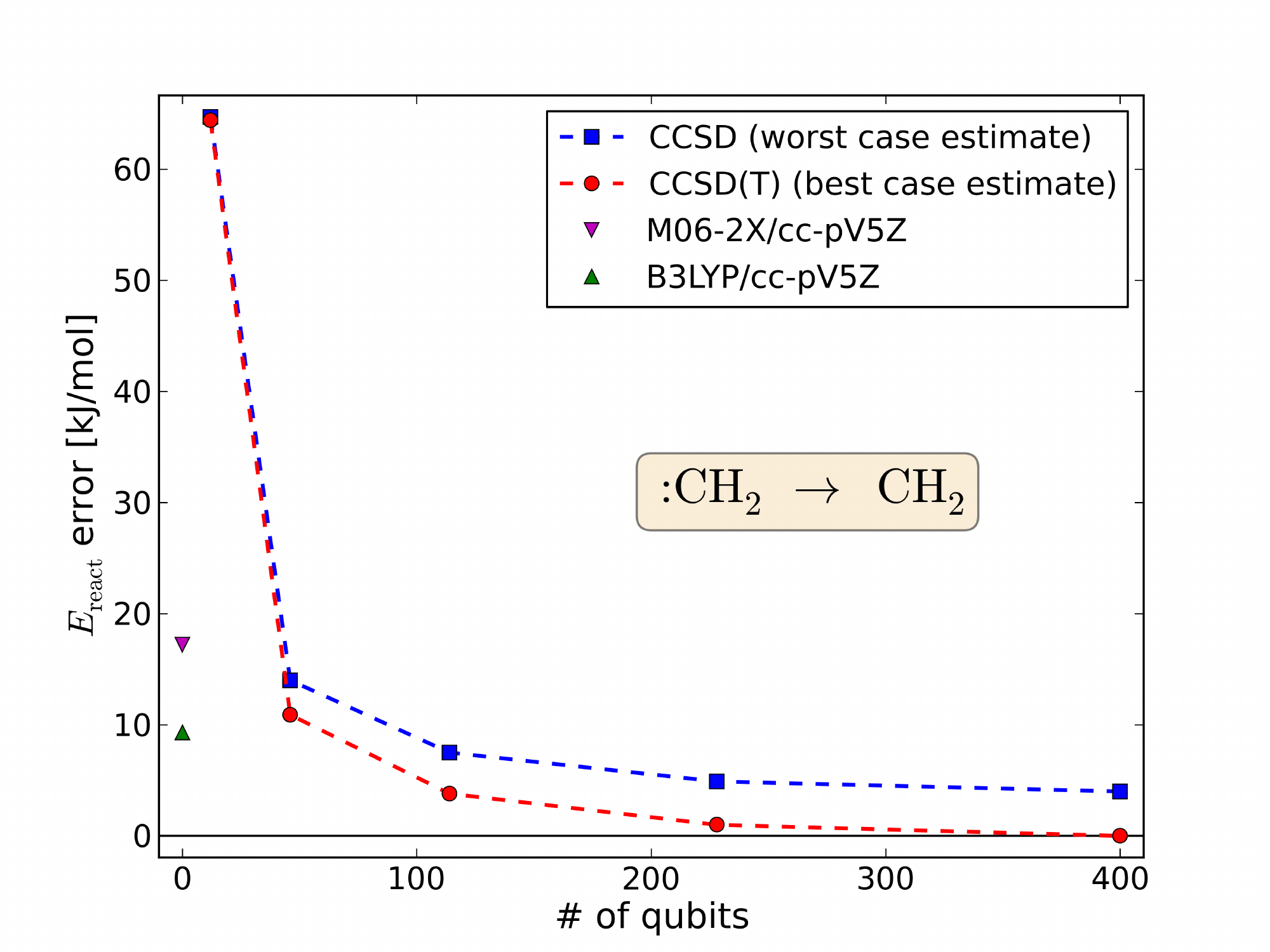}
}
\hspace{1cm}
\subfigure[]{
\includegraphics[width=7cm ,angle=0,clip=]{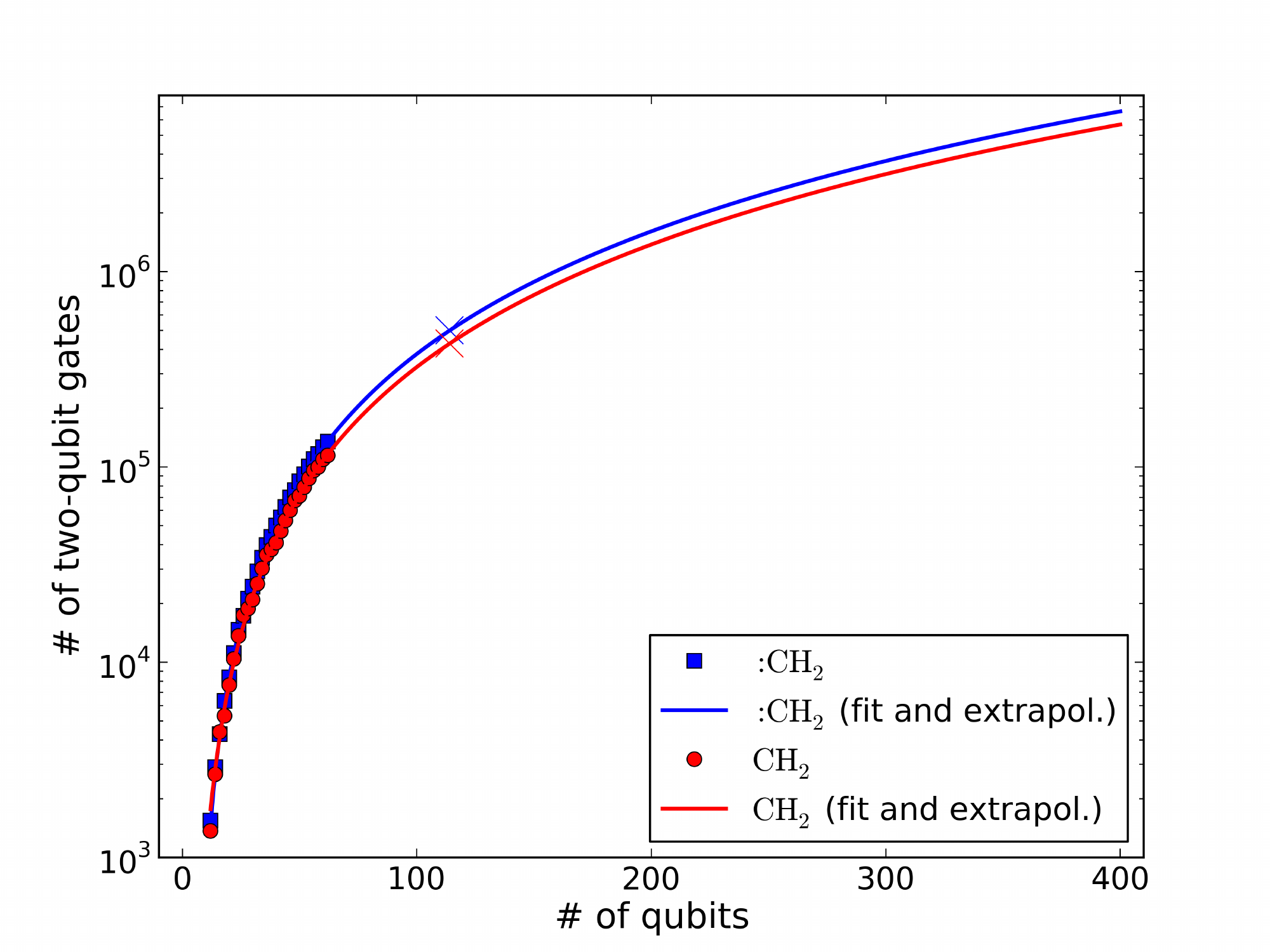}
}
\caption{
{\bf Estimation of the required quantum resources for obtaining transition energies of certain accuracy for
the triplet-singlet transition in $\rm CH_{2}$.}
For further information see Table VI in the SI and Figure \ref{fig:appl:react1}.
\\
(a) Error in (extrapolated) UCCSD-VQE transition energies with respect to the CCSD(T)/cc-pV5Z transition energy (39.4 kJ/mol).
\\
\textcolor{black}{
(b) Estimated number of required two-qubit gates for an optimized UCCSD-VQE treatment. A quadratic polynomial ($an_{\text{qubits}}^2+bn_{\text{qubits}}+c$;
$a_{\rm {:}CH_2} = 42.6$, $b_{\rm {:}CH_2} = -476.1$, $c_{\rm {:}CH_2} = 1015.7$, $a_{\rm CH_2} = 36.5$, $b_{\rm CH_2} = -413.1$, $c_{\rm CH_2} = 1468.2$)
is used for extrapolation to a larger number of qubits.
For orientation, the required quantum resources for a calculation
using the cc-pVTZ basis set (needed to obtain a reaction energy within or close to ``chemical accuracy'' of the CCSD(T)/cc-pV5Z reference)
are visualized by an ``x'' for each molecule.
For further information see also Figure \ref{fig:appl:gates}.
}
}
\label{fig:appl:react4}
\end{figure*}

For the 
triplet-singlet transition in $\rm CH_{2}$, see Figure \ref{fig:appl:react4}, 
absolute errors are by far smallest and basis set convergence is fastest (however comparably large relative errors due to a small reference transition energy of $\rm 39.4 kJ/mol$ that compares very well to the experimental value of 
$\rm 38\, kJ/mol$\cite{McKellar83}). This 
might partially be traced back to good error compensation
already for smaller basis sets 
since 
``only'' the two energetically lowest electronic states of one and the
same molecule need to be addressed instead of the formation and breaking of chemical bonds 
(one broken bond in case of the $\rm H_{2}O$- and LiH-dissociation and a complete rearrangement of atoms in case of the Haber-Bosch process).
However, we note that correctly computing the gap between (or even the order of) states with different spin multiplicity is in general not trivial, especially for transition metal compounds.
For the triplet-singlet transition in $\rm CH_{2}$ we find that already the cc-pVTZ basis set is sufficiently large to achieve an 
error within
a few kJ/mol of the reference transition energy requiring 114 qubits and $\rm 5.0\cdot 10^{5}$ two-qubit gates. 
At this basis set level the difference between our worst- and best-case estimates for the UCCSD-VQE 
transition energy is again of similar size ($\rm 4\, kJ/mol$).

\subsubsection*{Comparison to DFT and HF}

In Table \ref{tab:appl:reactener_extrapol_H2O} and
Tables IV-VI in the SI extrapolated UCCSD-VQE reaction energies are also compared to 
DFT and HF reaction energies ($E_{\text{react,\,total}}$). 
It is not very surprising for a reader who is acquainted with the characteristics of quantum-chemical methods, that
both DFT (where electron correlation is included via an exchange-correlation functional instead of systematic excitation of electrons to unoccupied orbitals) and HF (no electron correlation considered) show a faster basis set convergence 
than correlated methods such as CC due 
to a faster convergence of molecular energies. E.g. in case of the molecular energies of $\rm H_{2}O$ and OH
when applying cc-pVTZ the HF contribution ($E_{\text{total}}$(HF)) to the CCSD total molecular energy is converged within $\rm 0.01\, \%$ or $\rm +20\,kJ/mol$ 
whereas the CCSD correlation contribution ($E_{\text{corr}}$(CCSD)) still shows an error of about $\rm 9\, \%$ or $\rm +60\,kJ/mol$ 
with respect to the corresponding reference values in the cc-pV5Z basis set. We note that the highly different relative errors are also due to the fact that the HF contribution is by more than two orders of magnitude larger than the CCSD correlation contribution.
Thus, HF and DFT reaction energies within $\rm 4\, kJ/mol$ (threshold defined above) of the respective cc-pV5Z
result are usually already obtained with 
smaller basis sets 
than needed in case of correlated methods
(typically  
cc-pV($X$-1)Z is sufficient instead of cc-pV$X$Z as needed for CC): 
cc-pVTZ for the $\rm H_{2}O$- and LiH-dissociation, only cc-pVDZ for the triplet-singlet transition in $\rm CH_{2}$ 
and $-$ as an exception $-$ still cc-pVQZ in case of the Haber-Bosch process.

Comparing the DFT reaction energies to the respective CCSD(T)/cc-pV5Z reference reaction energies,
absolute errors amount to $\rm 1-11\, kJ/mol$ when applying the B3LYP functional,
$\rm 4-17\, kJ/mol$ in case M06-2X is used and $\rm 2-26\, kJ/mol$ for BP86.
Thus, for the simple exemplary reactions studied in this work already DFT in combination with B3LYP could allow for ``useful'' predictions that on average are of similar (in some cases slightly worse) accuracy 
as CCSD, CCSD(T) or UCCSD-VQE combined with sufficiently large basis sets.
However, the larger discrepancies for M06-2X and BP86 suggest that DFT is not in general suited to obtain reaction energies close to ``chemical accuracy''. 
It is worth mentioning that DFT calculations in general only come with a computational cost which is similar to that of HF
despite approximately accounting for electron correlation (however in a rather unsystematic fashion), i.e. no 
computationally expensive CC- or CI-like correlation expansion is necessary and thus DFT calculations can be 
run efficiently on classical hardware. 
Finally, we note that 
the HF method shows by far the largest errors 
with respect to the CCSD(T)/cc-pV5Z reference reaction energies
ranging up to
$\rm 158\, kJ/mol$
due to the total neglect of electron correlation.
This again demonstrates that the correlation
energy contribution to the reaction energy typically is of similar size as the reaction energy itself and once again underlines the importance of accurately accounting for electron correlation 
(e.g. by correlated methods in combination with sufficiently large basis sets or DFT).

\subsubsection*{Summary}

In summary, by using CCSD and CCSD(T) as a worst- and best-case estimates for 
UCCSD-VQE reaction energies we expect that results within or $-$ in cases where UCCSD-VQE is closer to CCSD than CCSD(T) $-$ close to ``chemical accuracy''
of the respective CCSD(T)/cc-pV5Z reference reaction energies 
are typically obtained when applying cc-pVQZ basis sets (cc-pVTZ for the triplet-singlet transition in $\rm CH_{2}$ is sufficient).
The errors due to the basis set incompleteness (within $\rm 4\, kJ/mol$) are then estimated to be 
of similar size 
as the ``uncertainty'' of our extrapolation, namely the difference between CCSD and CCSD(T) (within $\rm 7\, kJ/mol$). 
As expected, absolute errors in reaction energies are found to be significantly smaller than absolute errors in molecular energies due to partial error compensation. 
We find that reaction energies obtained with DFT 
in combination with the B3LYP functional exhibit a faster basis set convergence and are on average
of similar or in some cases slightly worse accuracy (errors between 1 and $\rm 11\, kJ/mol$; for M06-2X and BP86 somewhat larger errors) than CCSD, CCSD(T) and UCCSD-VQE reaction energies
while requiring significantly less computational resources (an efficient simulation on classical hardware is possible). 
Overall, the findings in this section are in line with our previous estimates for molecular energies (see Section \ref{sec:appl:molener:basis}), that
in order to obtain ``useful'' UCCSD-VQE results for practical applications 
(e.g. simulating chemical reactions of small molecules) basis sets of at least valence triple-$\zeta$ or 
quadruple-$\zeta$ quality should be used roughly requiring between 100 and 300 qubits
as well as between $\rm 10^{5}$ and $\rm 10^{7}$ two-qubit gates.

\section{Conclusion}
\label{sec:appl:conclusion}

We presented a detailed study estimating the accuracy and required quantum hardware resources
when running small quantum chemistry applications on a quantum computer. 
Using our implementation of the UCCSD-VQE approach in combination  
with reduced gate counts (exploiting gate cancellations as well as pre-screening based on MP2)
we computed the molecular energies of 
nine small atomic/molecular systems including open-shell species ($\rm H_{2}O$, OH, LiH, Li, $\rm N_2$, $\rm H_2$, $\rm NH_3$, triplet-$\rm CH_2$ and singlet-$\rm CH_2$) 
by simulating a ``noise-free'' quantum computer with up to 20 qubits. 
From these molecular energies we also calculated reaction energies $-$ in general relevant for studying and predicting chemical reactivity $-$
for the $\rm H_{2}O$- and LiH-dissociation, the Haber-Bosch process as well as the triplet-singlet transition in $\rm CH_2$.

Absolute errors of UCCSD-VQE molecular and reaction energies with respect to the exact (FCI) results are found to 
be very small (below $\rm 1 \,kJ/mol$) when using the minimal basis set. Reaction energies are always closer to the respective exact results
than molecular energies due to partial error compensation.
In case of $\rm N_2$/Haber-Bosch process which is known to be challenging for approximate electronic structure methods errors
range up to $\rm 5 \,kJ/mol$.
\textcolor{black}{UCCSD-VQE results are usually between or close to results obtained with CCSD, the classical non-variational counterpart, and CCSD(T), 
the quantum-chemical ``gold standard''\cite{Cooper10,Evangelista11,Harsha18}
indicating that CCSD might be used as a worst-case 
and CCSD(T) as a best-case estimate for UCCSD-VQE.} 
Thus no clear advantage in applying UCCSD-VQE instead of CCSD(T) was observed.
Within our optimized implementation the UCCSD-VQE treatment of the above-mentioned small systems in the minimal basis set 
typically comes with a computational cost of $\mathcal{O}(10)$ qubits and $\mathcal{O}(10^{3})$ two-qubit gates
with the latter being expected to be more critical. 

Since in general correlated approaches such as CC methods
require the usage of much larger basis sets to obtain ``useful'' results we also considered
correlation-consistent basis sets of up to valence quintuple-$\zeta$ quality which should provide results 
close enough to the basis set limit for practical purposes like reaction energies.
Due to the exponential scaling of UCCSD-VQE on classical hardware we had to rely on
extrapolation to access results and gate counts for those larger basis sets, which was done by
simply using the CCSD energy as a worst-case and the CCSD(T) energy as a best-case estimate in agreement with all previous findings. 
We showed that UCCSD-VQE is expected to provide ``useful'' reaction energies within or close to ``chemical accuracy'' ($\rm 4 \, kJ/mol$)
of our CCSD(T)/valence quintuple-$\zeta$ reference results, 
when using valence triple- or quadruple-$\zeta$ basis sets. This roughly requires 
between 100 and 300 qubits as well as between $10^{5}$ and $10^{7}$ two-qubit gates for small quantum chemistry applications
with the ``uncertainty'' of our extrapolation also being in the order of a few kJ/mol (within $\rm 7 \, kJ/mol$).  

For all investigated reactions DFT in combination with the B3LYP functional
provides results of similar (or in some cases slightly worse) accuracy as extrapolated
UCCSD-VQE 
(errors between 1 and $\rm 11\, kJ/mol$)
while running efficiently on classical hardware in contrast to UCCSD.
However, 
benchmark studies on much larger and diverse sets of chemical reactions 
as well as the fact that in our study discrepancies are larger for other density functionals
suggest that DFT is in general not suited to obtain reaction energies close to ``chemical accuracy''.\cite{Peverati11,Perdew05}

From our extrapolations we can further estimate that a medium-sized organic molecule 
like naphthalene ($\rm C_{10}H_{8}$, 68 electrons) would roughly require around
800/1500 qubits and $\mathcal{O}(10^{7})$/$\mathcal{O}(10^{8})$ two-qubit gates for a UCCSD-VQE treatment on a quantum computer when applying valence triple-/quadruple-$\zeta$ basis sets. 
Overall this means that for practically relevant UCCSD-VQE simulations of small- to
medium-sized systems already a device with a few hundred (logical) qubits and error correction is probably needed. 
However, it should be kept in mind that our resource estimations are not strict and certainly not best-case estimates. 
They should rather be regarded as rough guidelines since there is definitely room for further improvement $-$ in particular for reducing the number of required two-qubit gates, e.g. by strategies borrowed from ``classical'' quantum chemistry\cite{Motta18} or 
\textcolor{black}{more hardware-efficient techniques like hardware-efficient VQE,\cite{Kandala17,Barkoutsos18} the Hamiltonian variational approach,\cite{Wecker15} a low-depth circuit ansatz\cite{Dallaire-Demers18} and most recently qubit coupled-cluster.\cite{Ryabinkin18} }
On the other hand, for running simulations on actual quantum hardware additionally the effect of noise on the accuracy of the results as well as the connectivity of qubits on the chip has to be considered, which might further increase the number of required gates.
Our estimations clearly exceed the capabilities of today's quantum hardware, which $-$ in case of superconducting chips or trapped ion systems
with $\mathcal{O}(10)$ qubits $-$ roughly allow for $\mathcal{O}(10^{2}-10^{3})$ two-qubit gates on average.\cite{Barends14,ibm,Wright19} 
However, fast progress is made and when a scaled-up error-corrected quantum hardware finally becomes available other approaches besides UCCSD-VQE such as the phase estimation algorithm will come 
into play providing the exact result in large basis sets for comparably large systems and thus outperforming all classically feasible methods (and also UCCSD-VQE).
Until then, quantum algorithms will also have improved reducing the overall hardware requirements.  

\section*{Acknowledgments}
S.Z. and M.M. thank Prof. Alexander Shnirman and Prof. Frank Wilhelm-Mauch for their support.

\bibliography{literature}

\clearpage



\begin{figure*}
\subfigure[]{
\includegraphics[width=7.0cm ,angle=0,clip=]{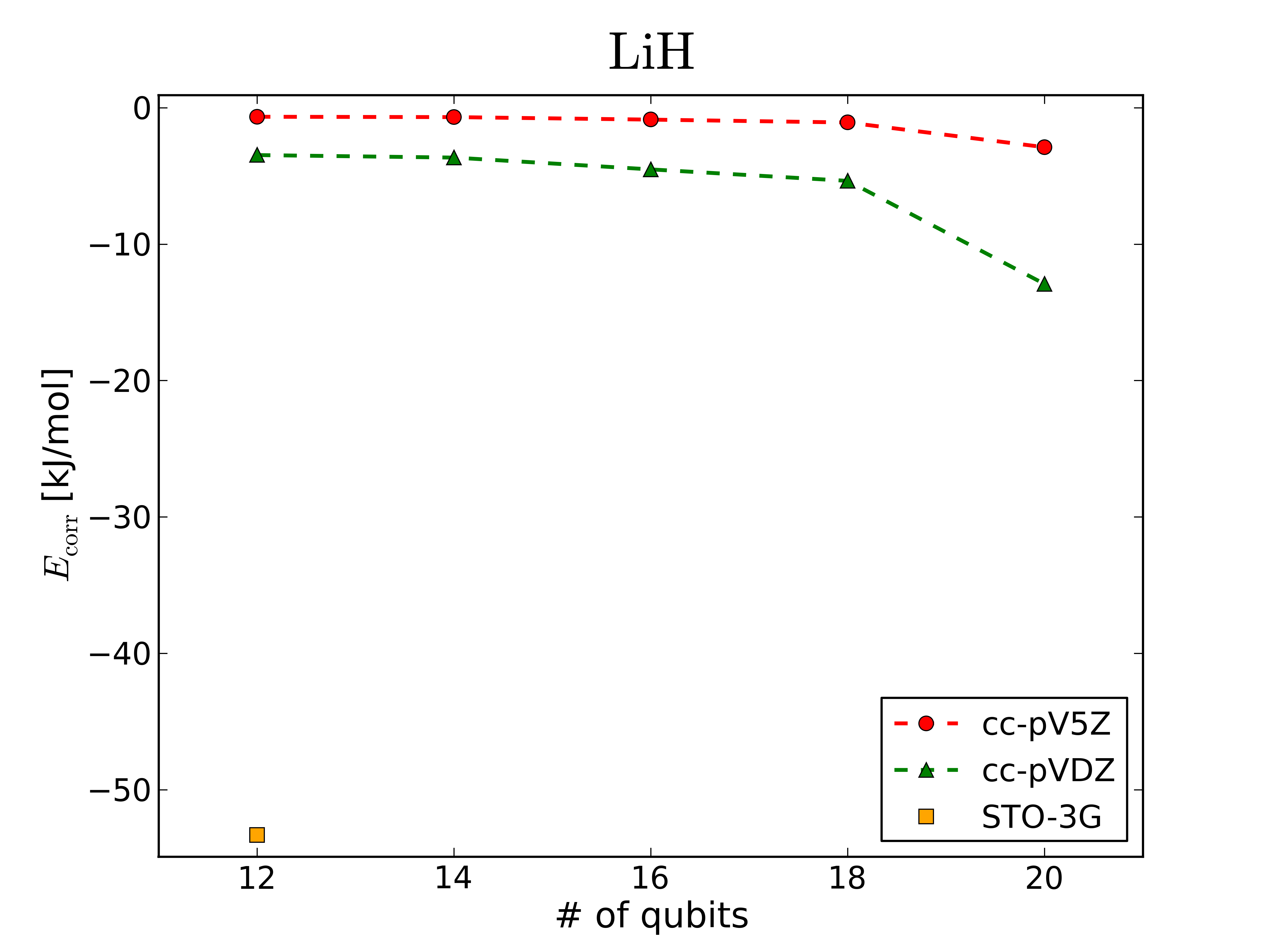}
}
\hspace{1cm}
\subfigure[]{
\includegraphics[width=7.0cm ,angle=0,clip=]{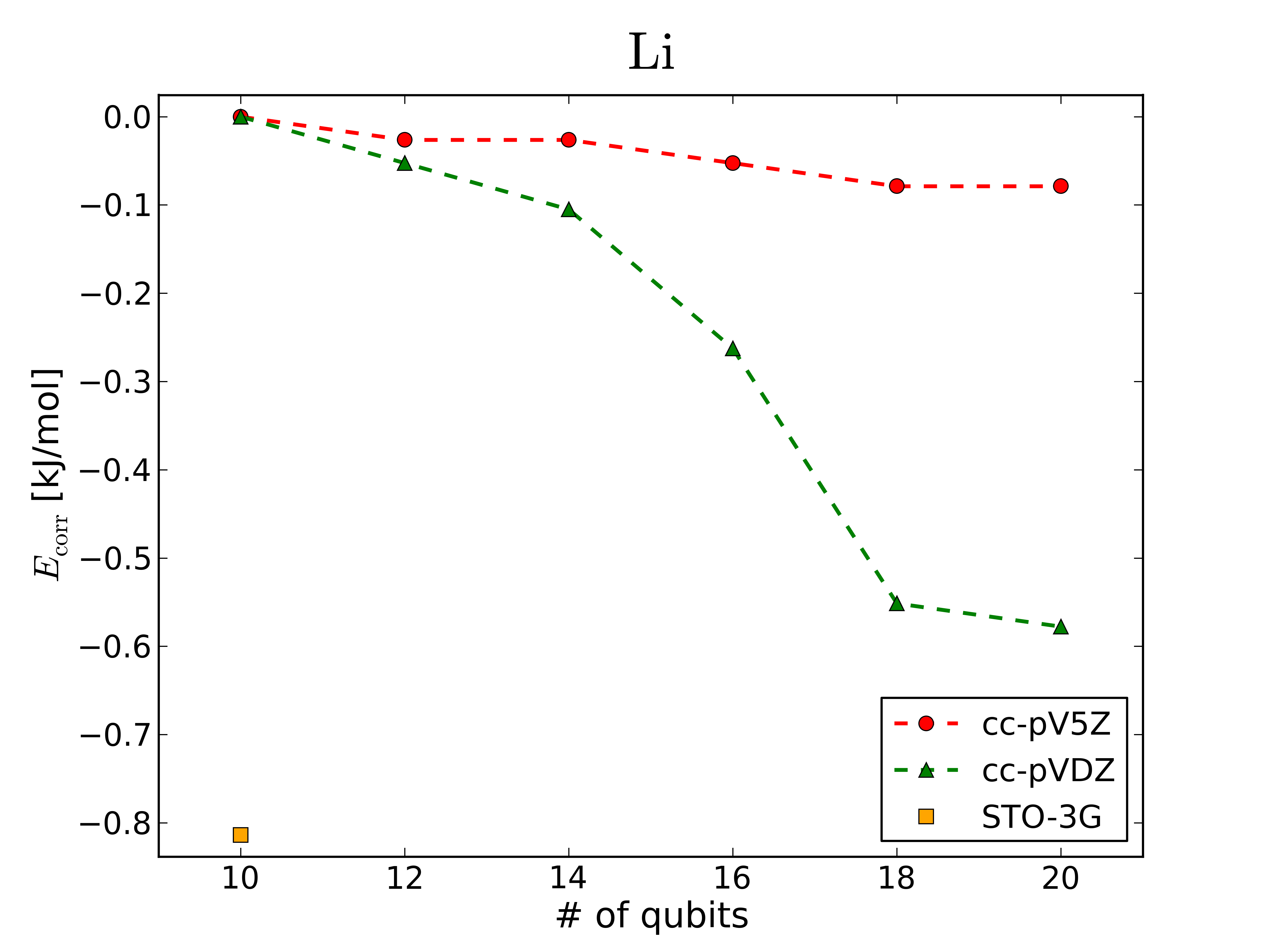}
} \\
\subfigure[]{
\includegraphics[width=7.0cm ,angle=0,clip=]{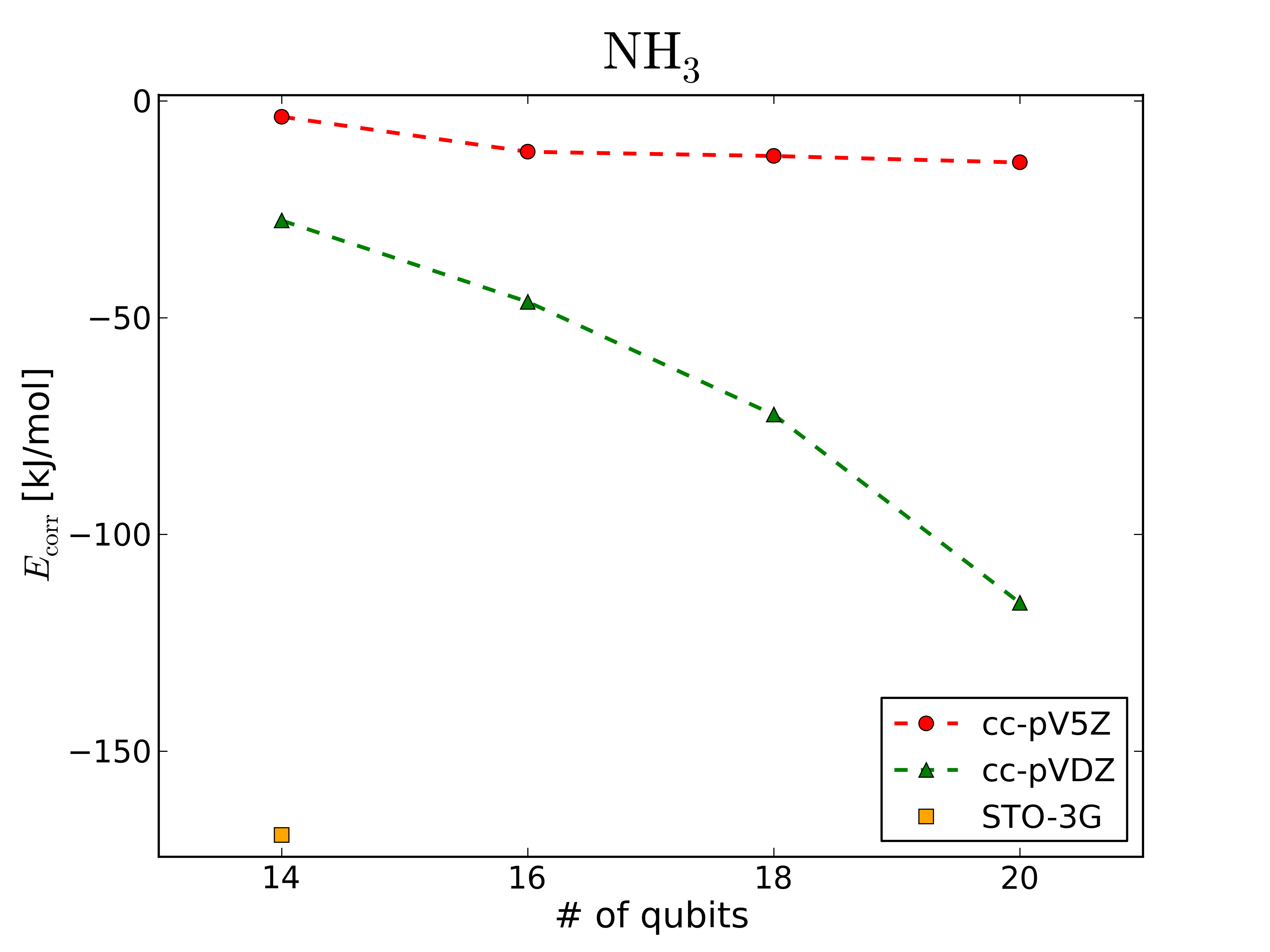}
}
\hspace{1cm}
\subfigure[]{
\includegraphics[width=7.0cm ,angle=0,clip=]{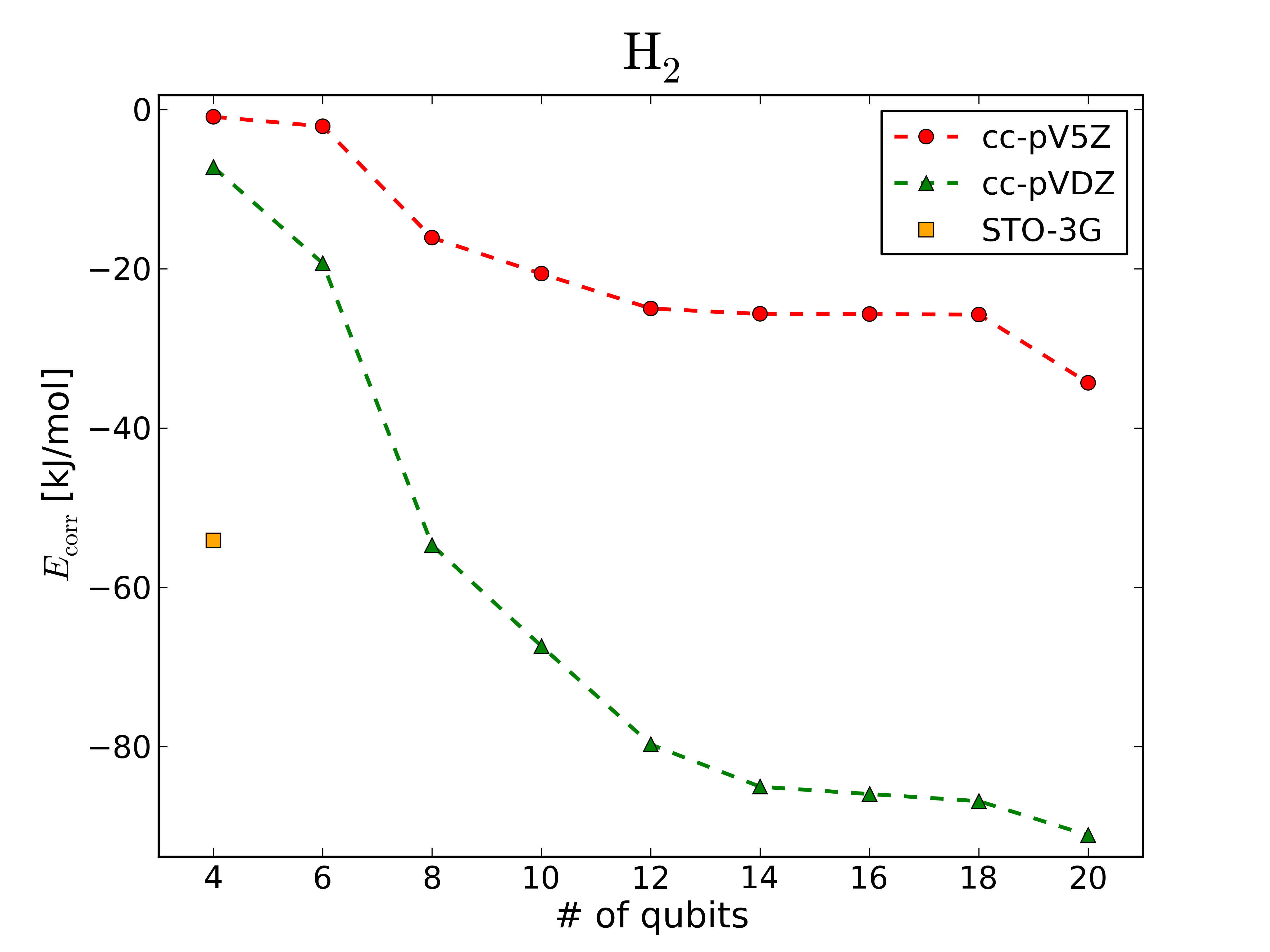}
}
\\
\subfigure[]{
\includegraphics[width=7.0cm ,angle=0,clip=]{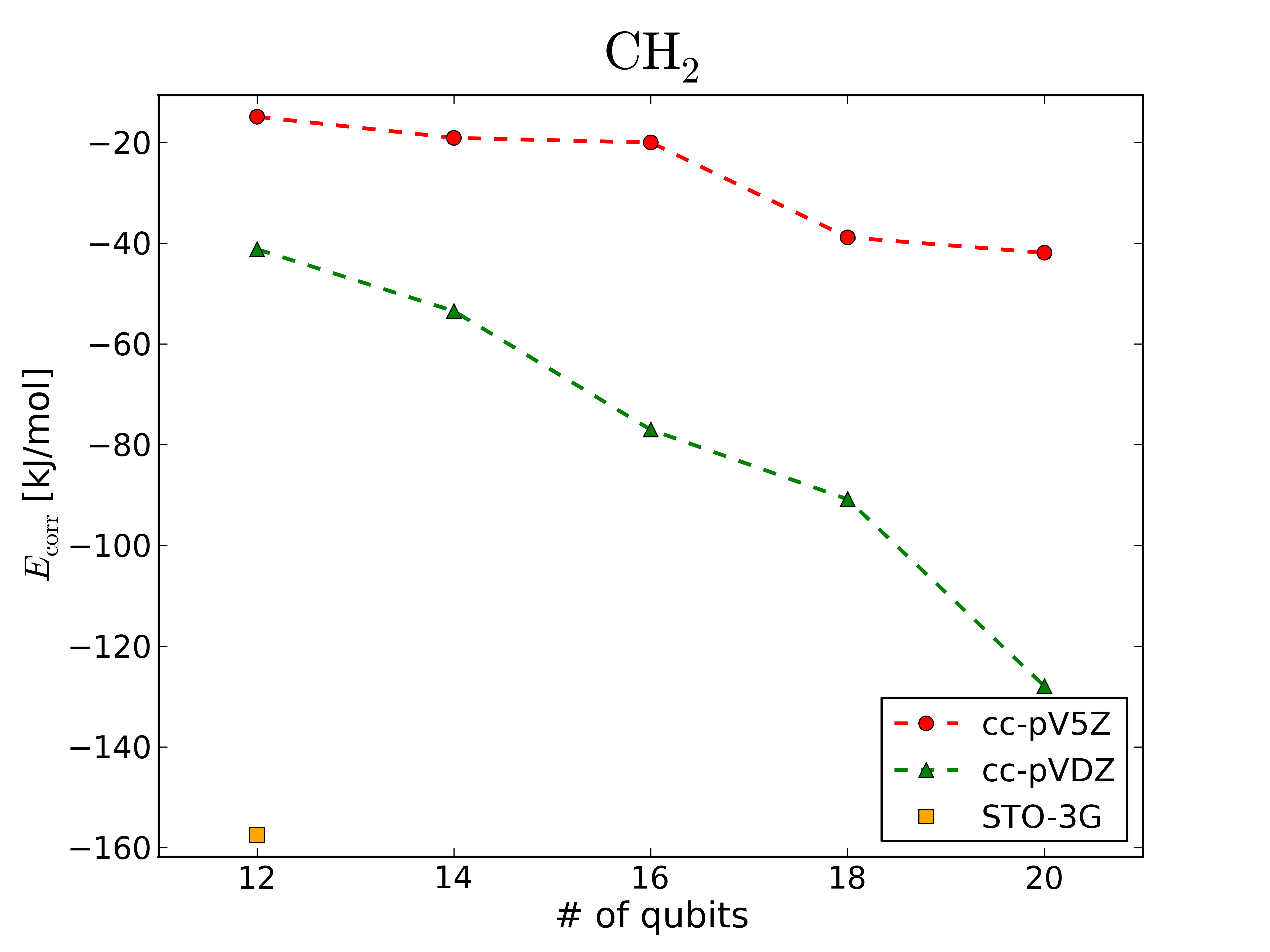}
}
\renewcommand\thefigure{SI 1}
\caption{
UCCSD-VQE correlation energies for up to 20 simulated qubits.
Results are shown for molecules ${\rm LiH}$ (a), ${\rm Li}$ (b), ${\rm NH_3}$ (c), ${\rm H_2}$ (d) and ${\rm CH_2}$ (e) in combination with  
different basis sets (STO-3G, cc-pVDZ and cc-pV5Z).
}
\label{fig:appendix:1}
\end{figure*}


\begin{figure*}
\subfigure[]{
\includegraphics[width=7cm ,angle=0,clip=]{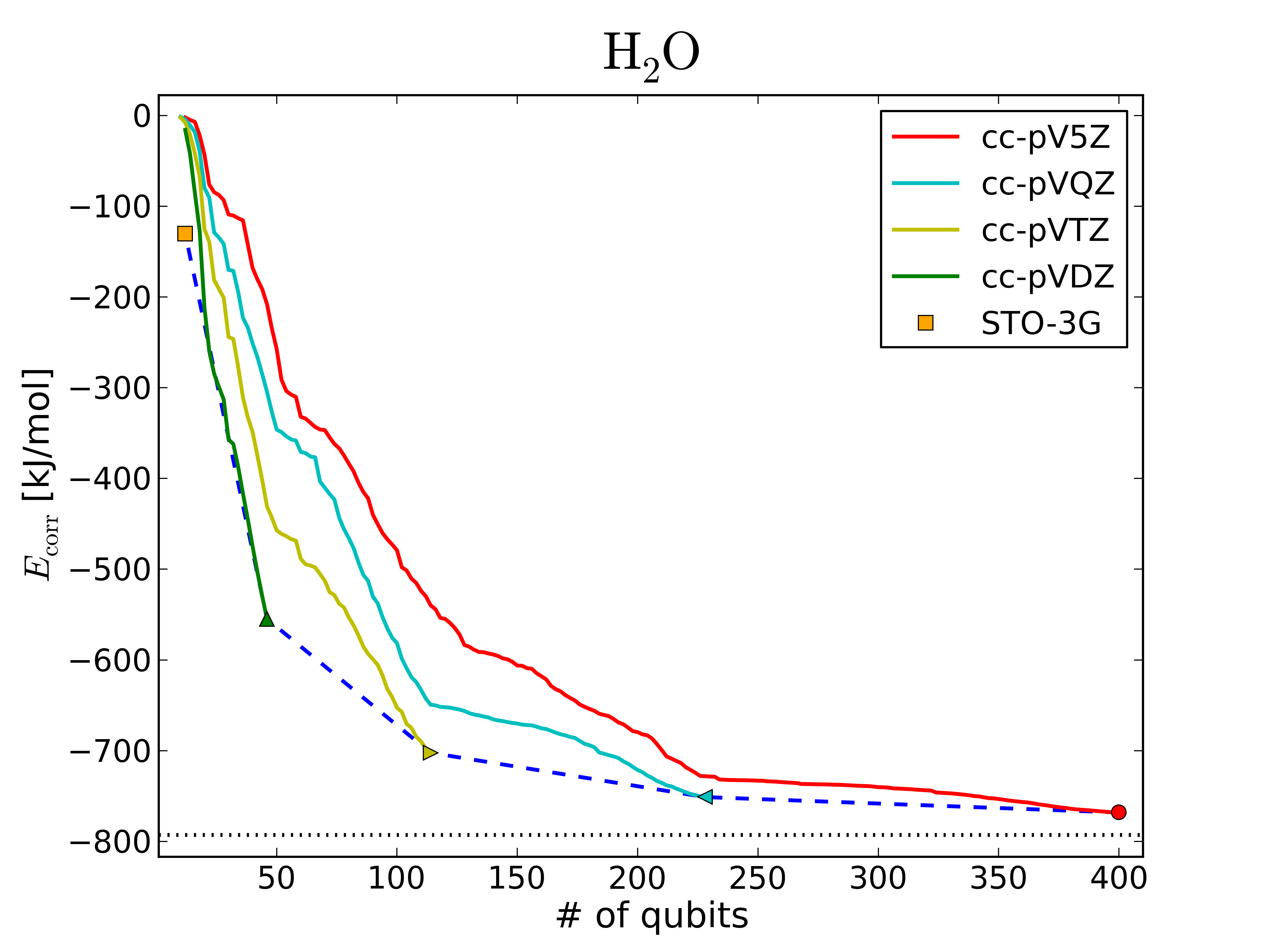}
}
\hspace{1cm}
\subfigure[]{
\includegraphics[width=7cm ,angle=0,clip=]{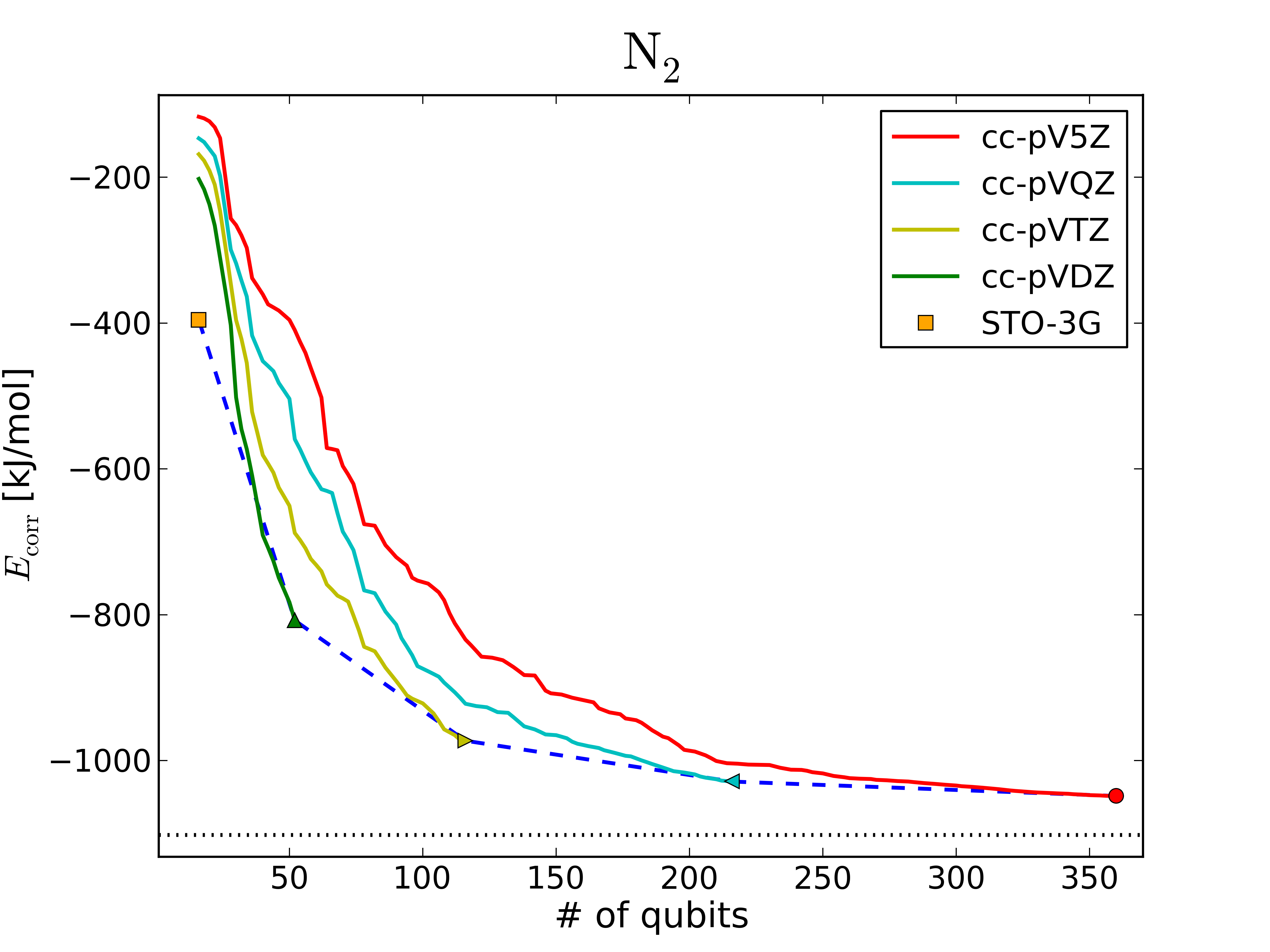}
} \\
\subfigure[]{
\includegraphics[width=7cm ,angle=0,clip=]{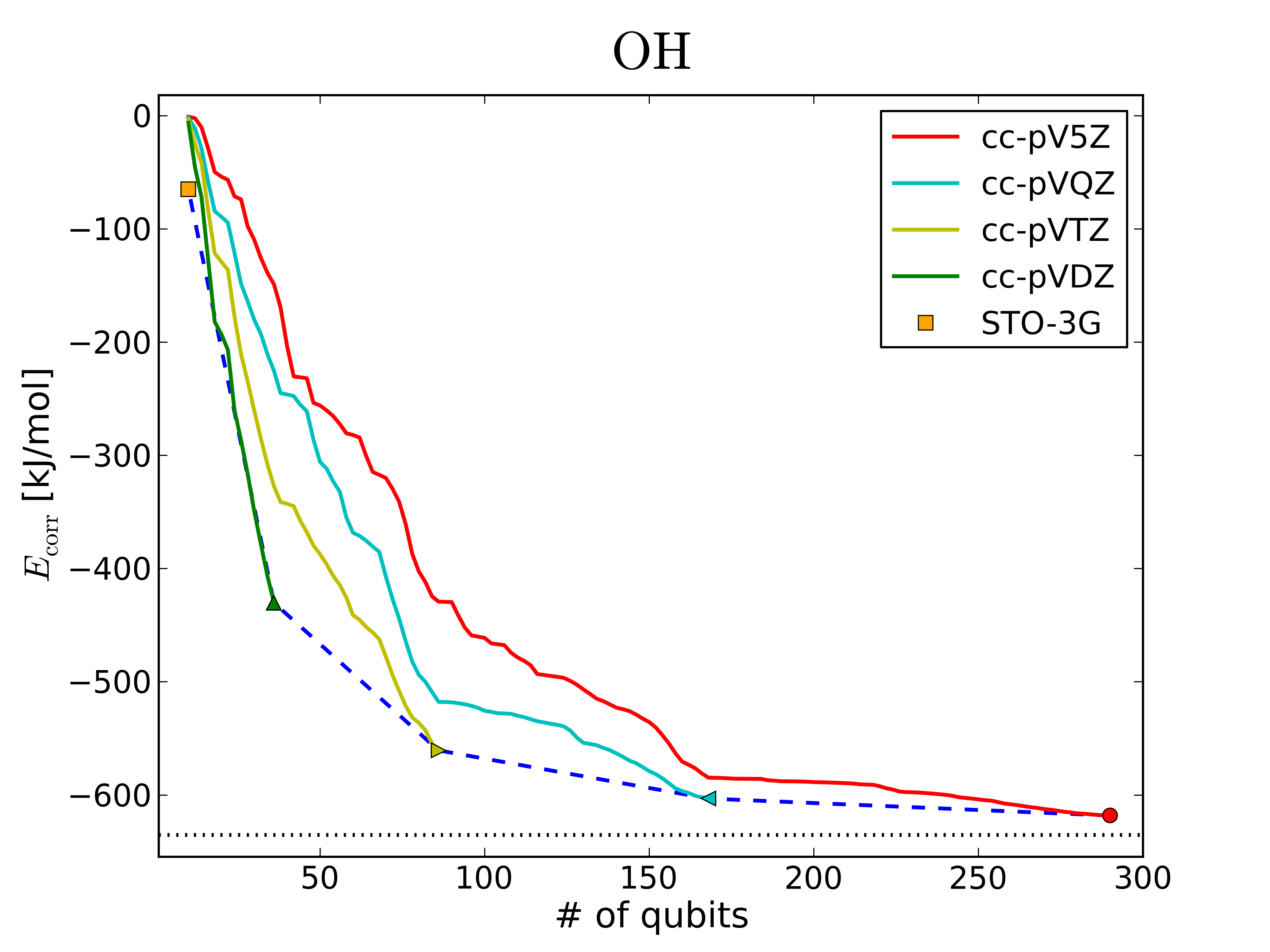}
}
\hspace{1cm}
\subfigure[]{
\includegraphics[width=7cm ,angle=0,clip=]{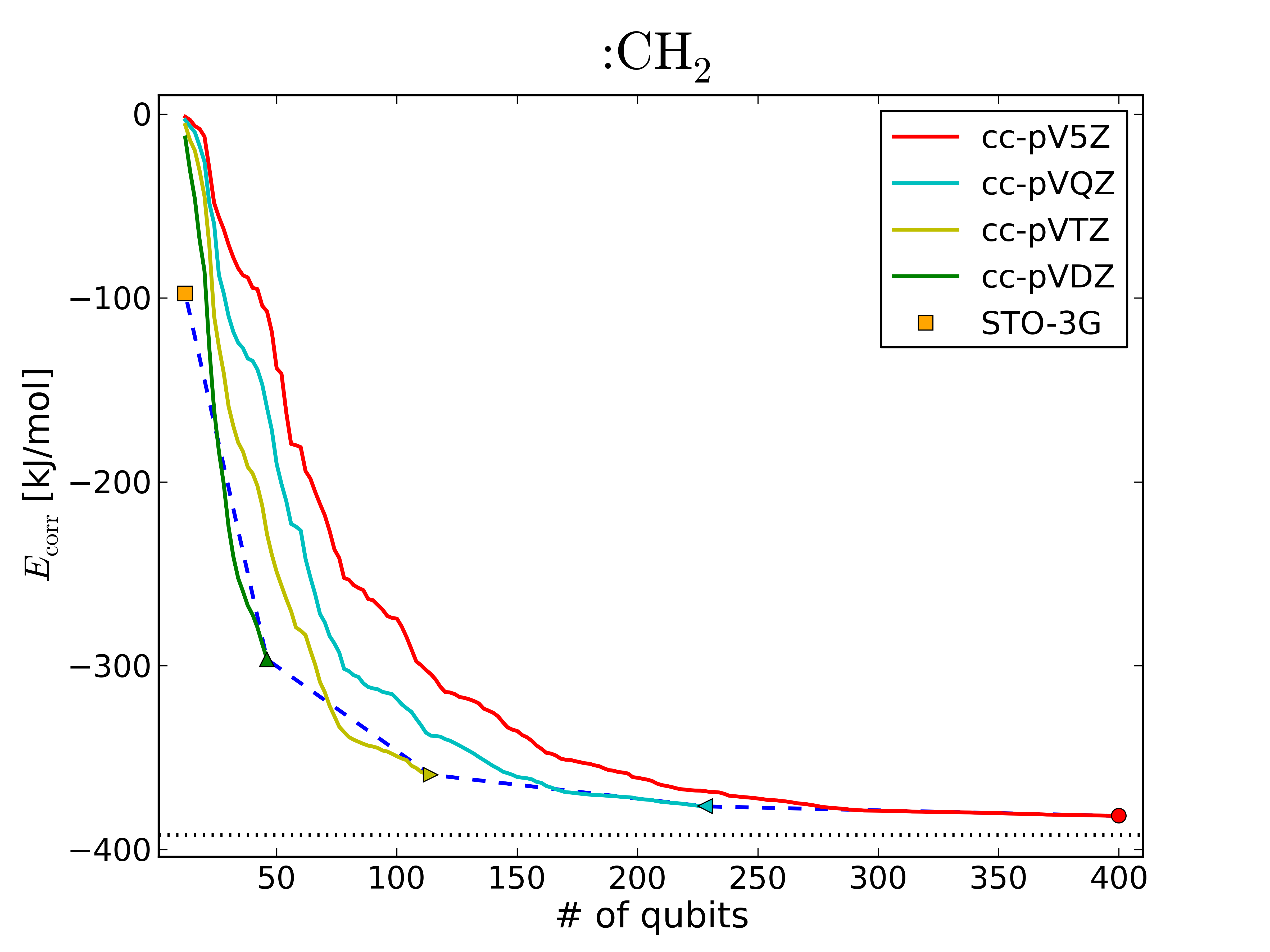}
}
\renewcommand\thefigure{SI 2}
\caption{
Extrapolated UCCSD-VQE correlation energies for a larger number of qubits.
The extrapolation is done by using CCSD correlation energies (worst-case estimates) in accordance with the results shown in Table I in the main paper.
Results are depicted for molecules ${\rm H_{2}O}$ (a), ${\rm N_{2}}$ (b), ${\rm OH}$ (c) and ${\rm {:}CH_2}$ (d) in combination with
basis sets of increasing size (STO-3G, cc-pVDZ, cc-pVTZ, cc-pVQZ and cc-pV5Z).
The dotted horizontal line represents the CCSD(T)/cc-pV5Z result which is chosen to be the target/reference energy.
For each basis set
we gradually increased the number of correlated spin orbitals (= number of qubits) by including more and more of the energetically lowest-lying virtual orbitals.
The blue dashed line connects the optimal CCSD correlation energy for each basis set (all virtual orbitals included in the correlation treatment).
}
\label{fig:appendix:2}
\end{figure*}


\begin{figure*}
\subfigure[]{
\includegraphics[width=7.0cm ,angle=0,clip=]{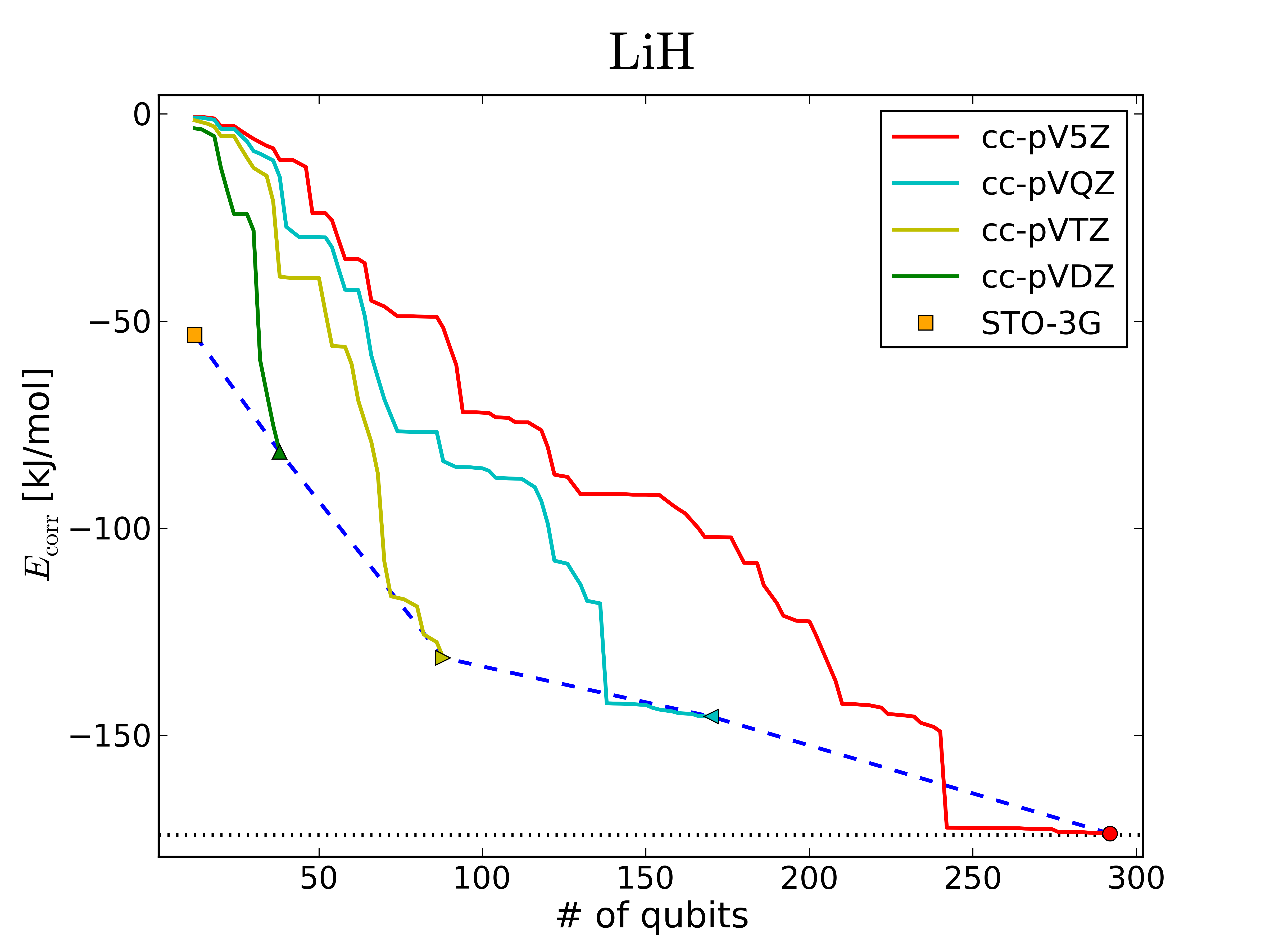}
}
\hspace{1cm}
\subfigure[]{
\includegraphics[width=7.0cm ,angle=0,clip=]{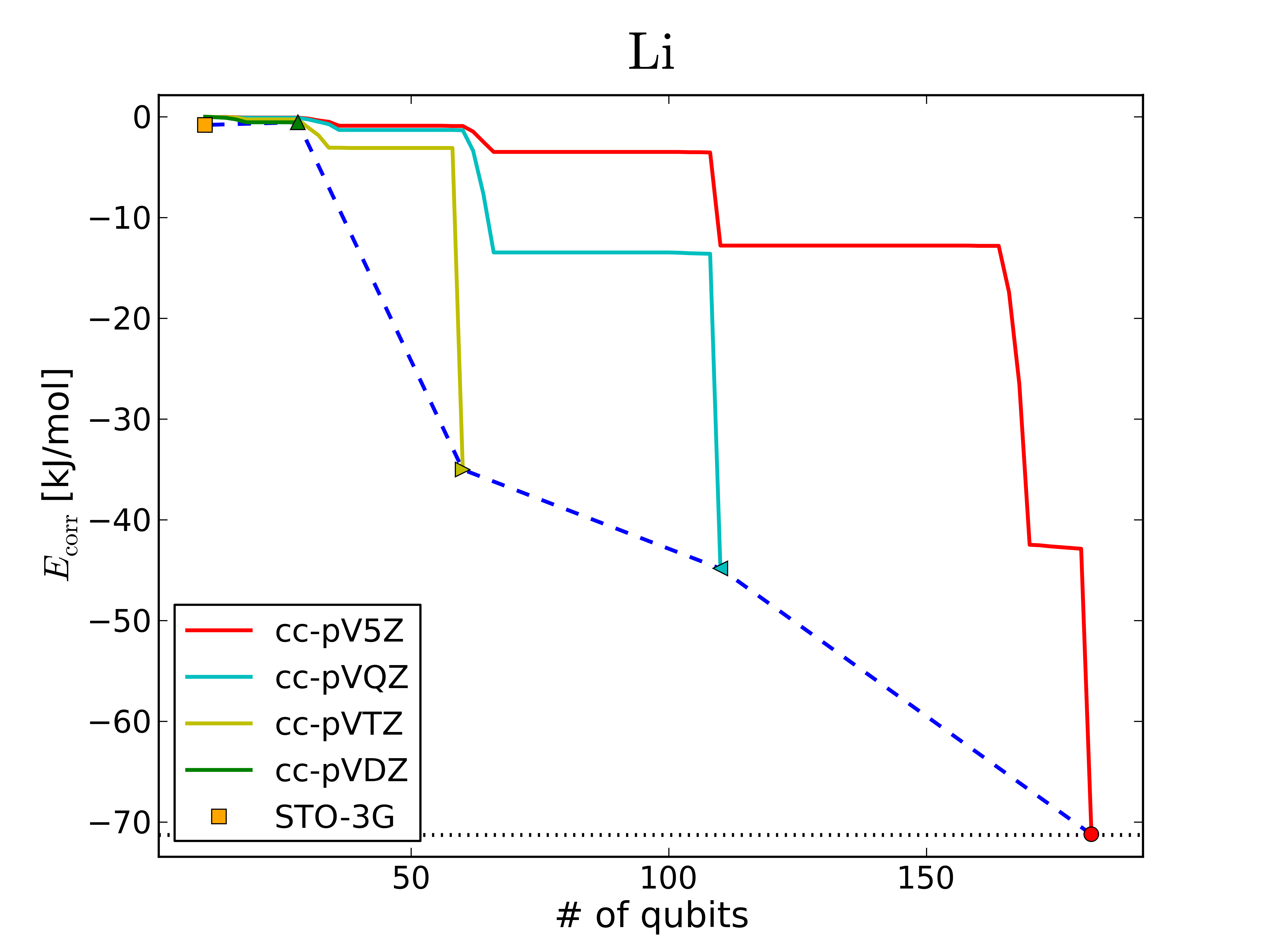}
} \\
\subfigure[]{
\includegraphics[width=7.0cm ,angle=0,clip=]{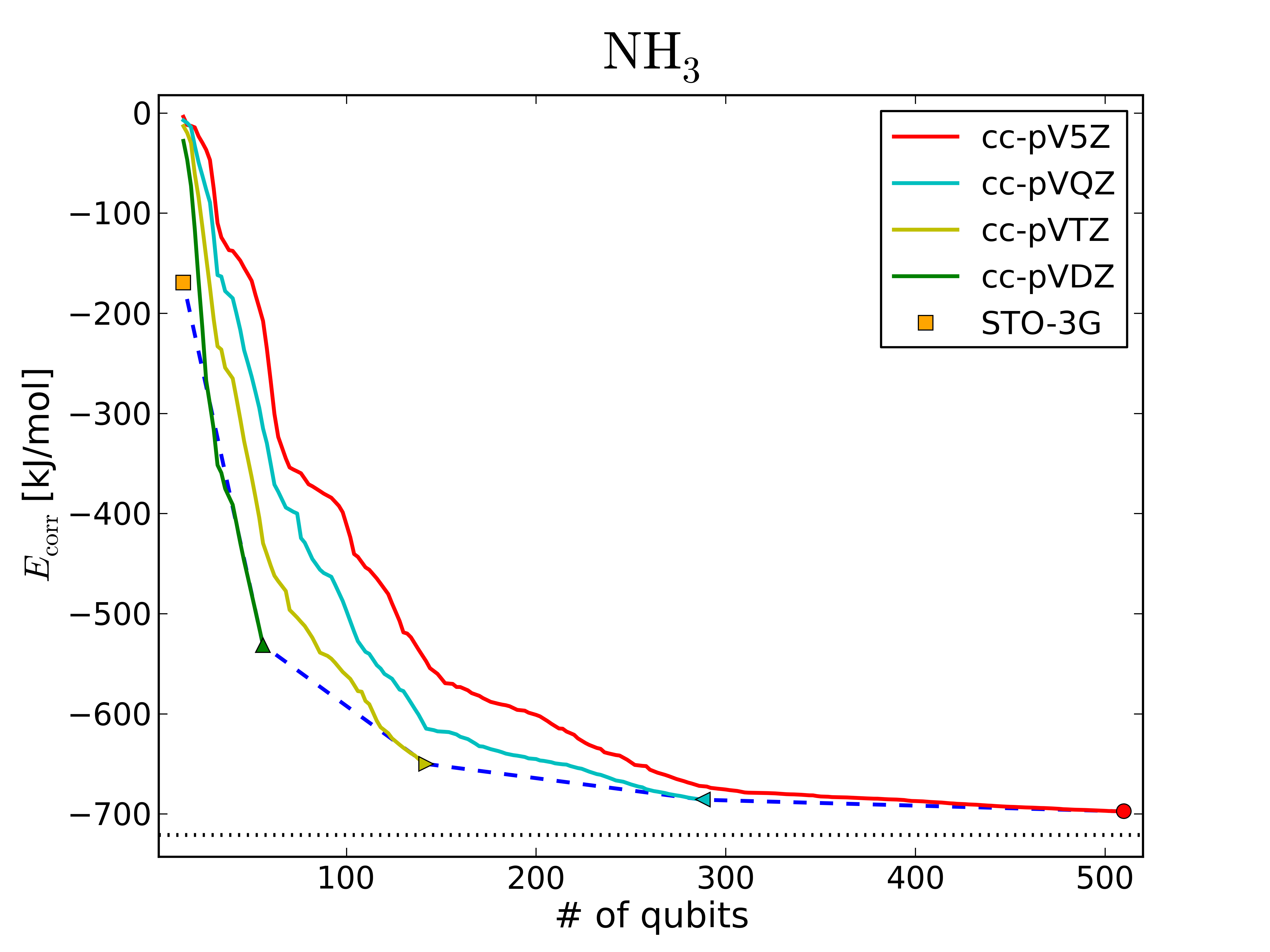}
}
\hspace{1cm}
\subfigure[]{
\includegraphics[width=7.0cm ,angle=0,clip=]{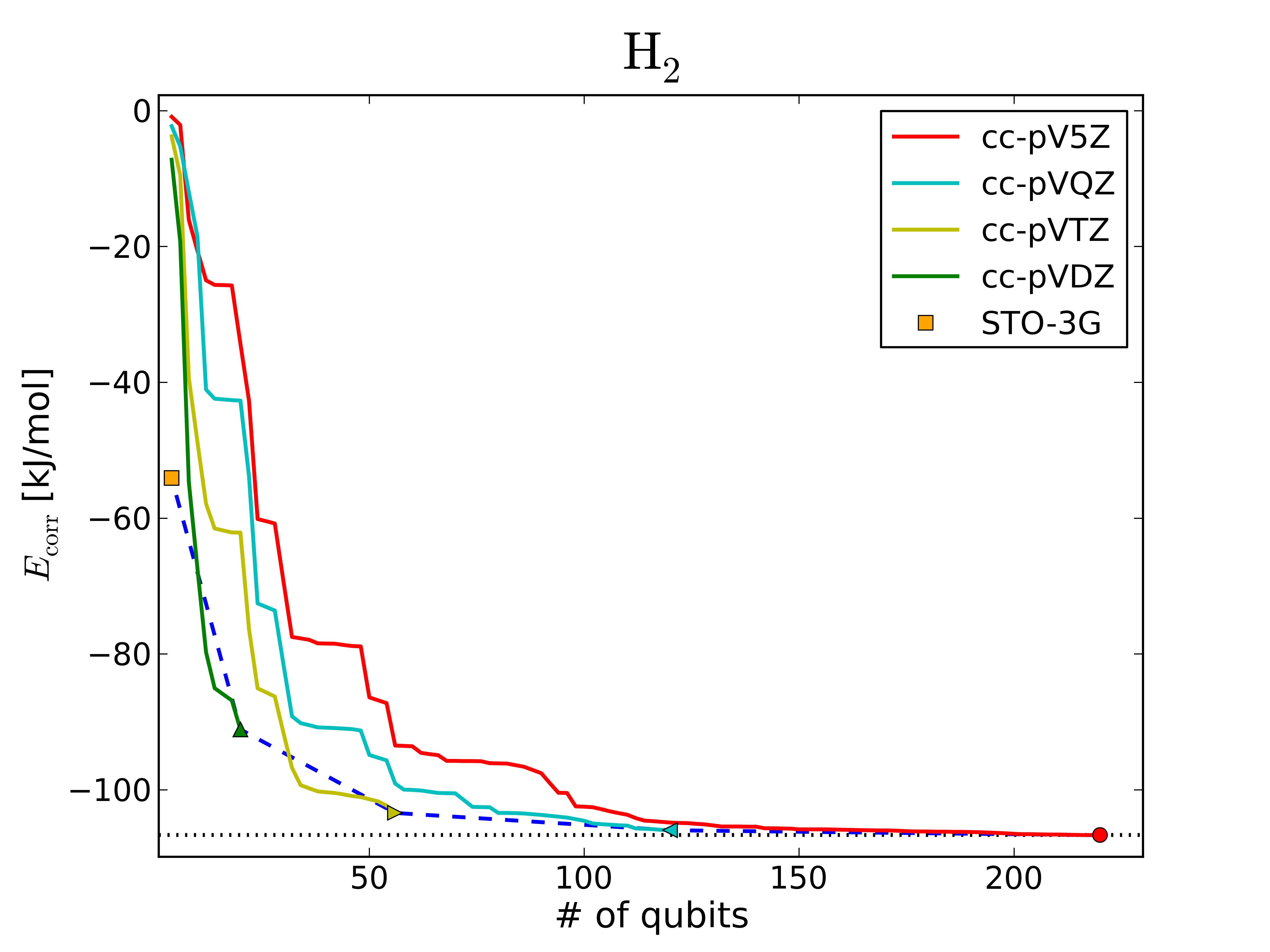}
}
\\
\subfigure[]{
\includegraphics[width=7.0cm ,angle=0,clip=]{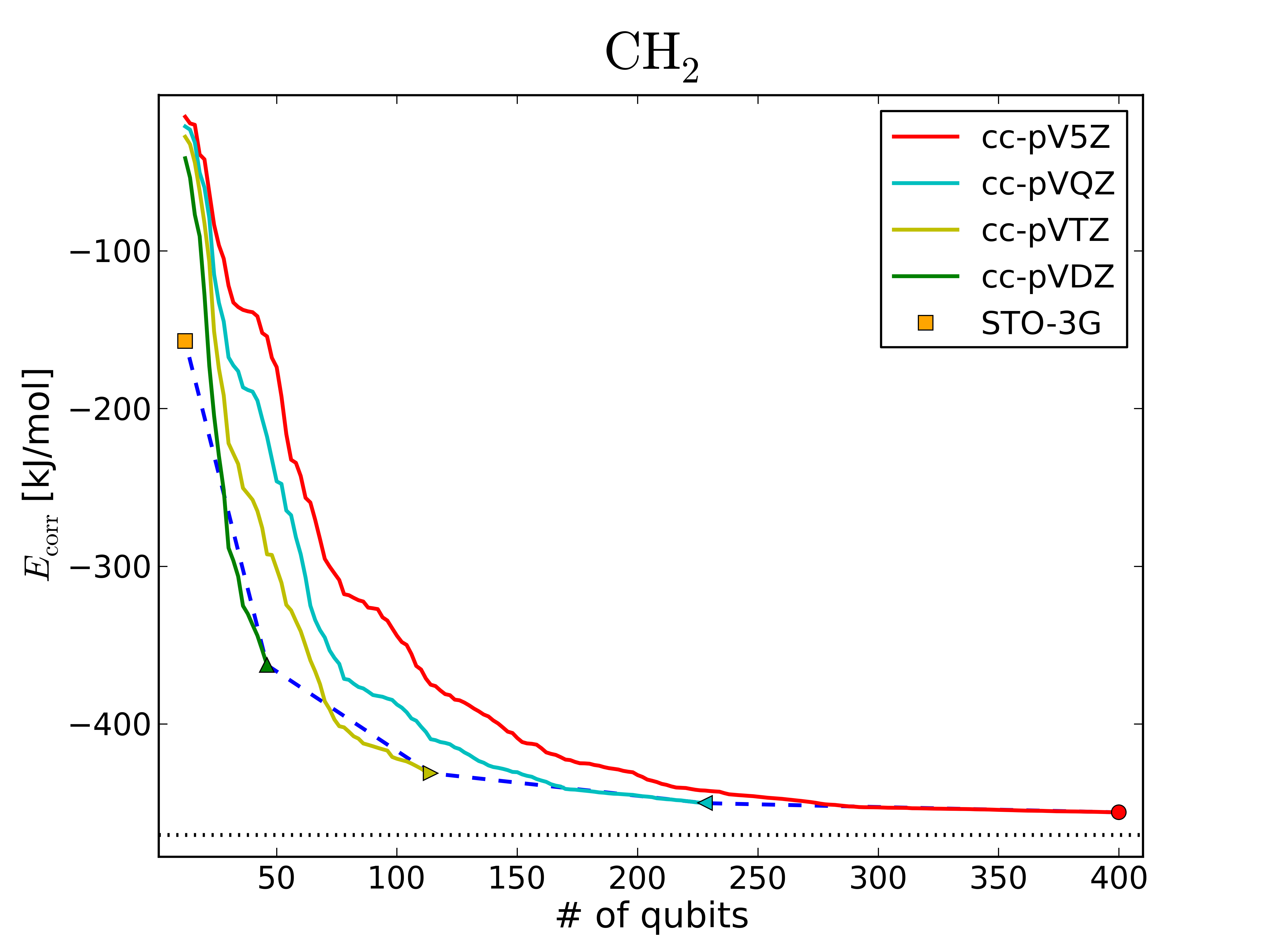}
}

\renewcommand\thefigure{SI 3}
\caption{
Extrapolated UCCSD-VQE correlation energies for a larger number of qubits. 
The extrapolation is done by using CCSD correlation energies (worst-case estimates) in accordance with the results shown in Table I in the SI.
Results are depicted for molecules ${\rm LiH}$ (a), ${\rm Li}$ (b), ${\rm NH_3}$ (c), ${\rm H_2}$ (d) and ${\rm CH_2}$ (e) in combination with
basis sets of increasing size (STO-3G, cc-pVDZ, cc-pVTZ, cc-pVQZ and cc-pV5Z).
For further information see Figure 2 in the SI.
}
\label{fig:appendix:3}
\end{figure*}


\begin{figure*}
\subfigure[]{
\includegraphics[width=7.0cm ,angle=0,clip=]{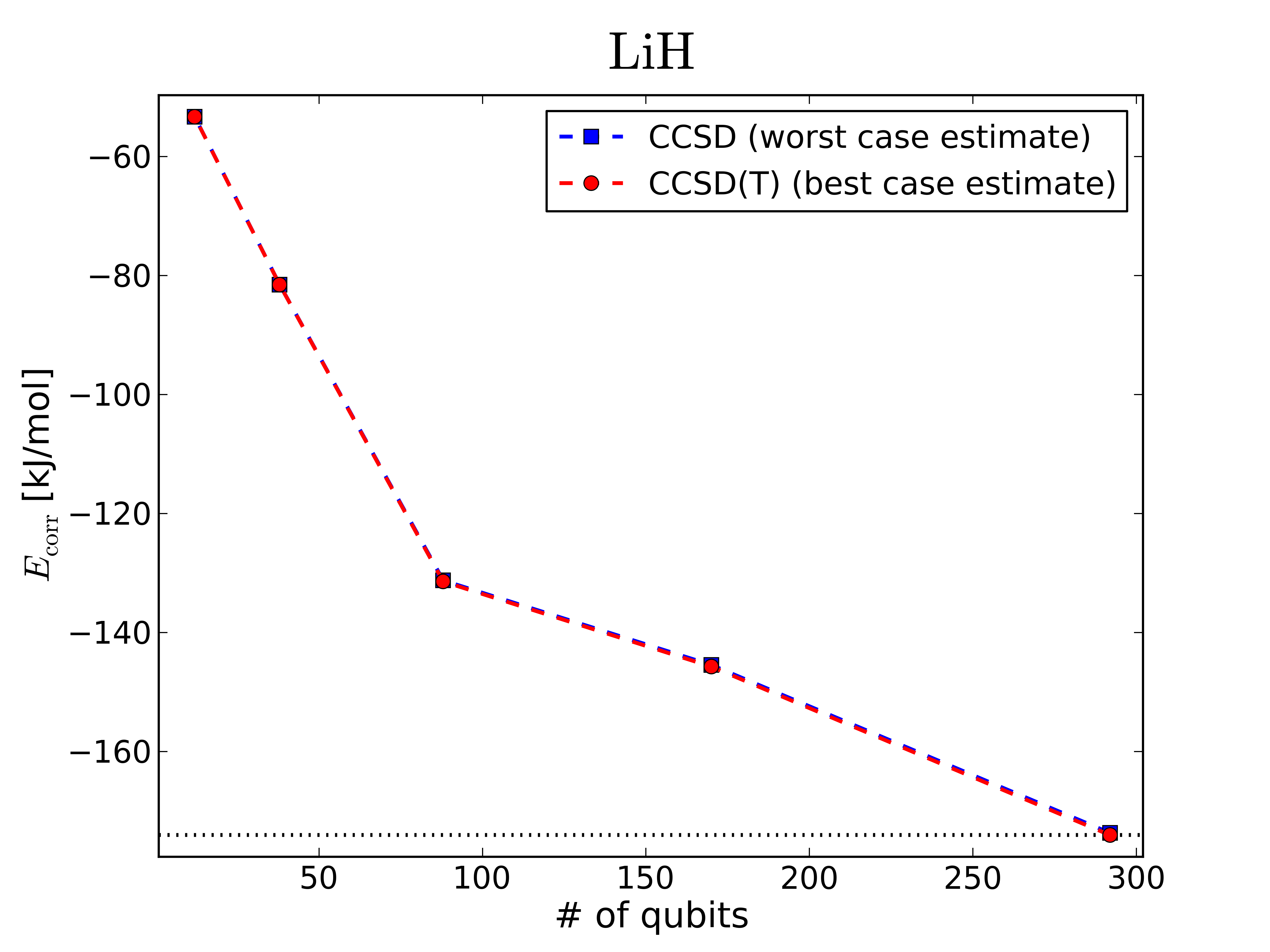}
}
\hspace{1cm}
\subfigure[]{
\includegraphics[width=7.0cm ,angle=0,clip=]{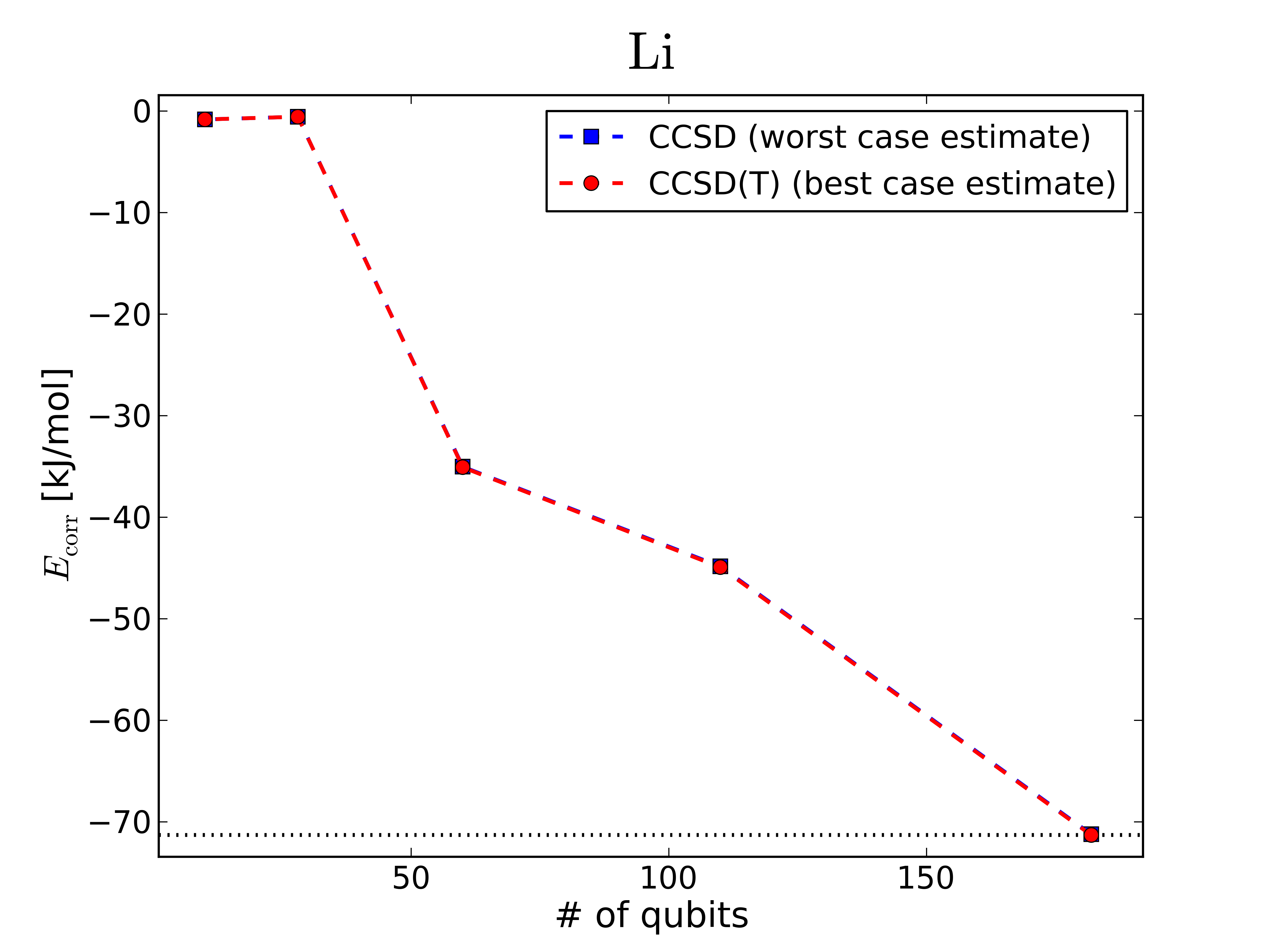}
} \\
\subfigure[]{
\includegraphics[width=7.0cm ,angle=0,clip=]{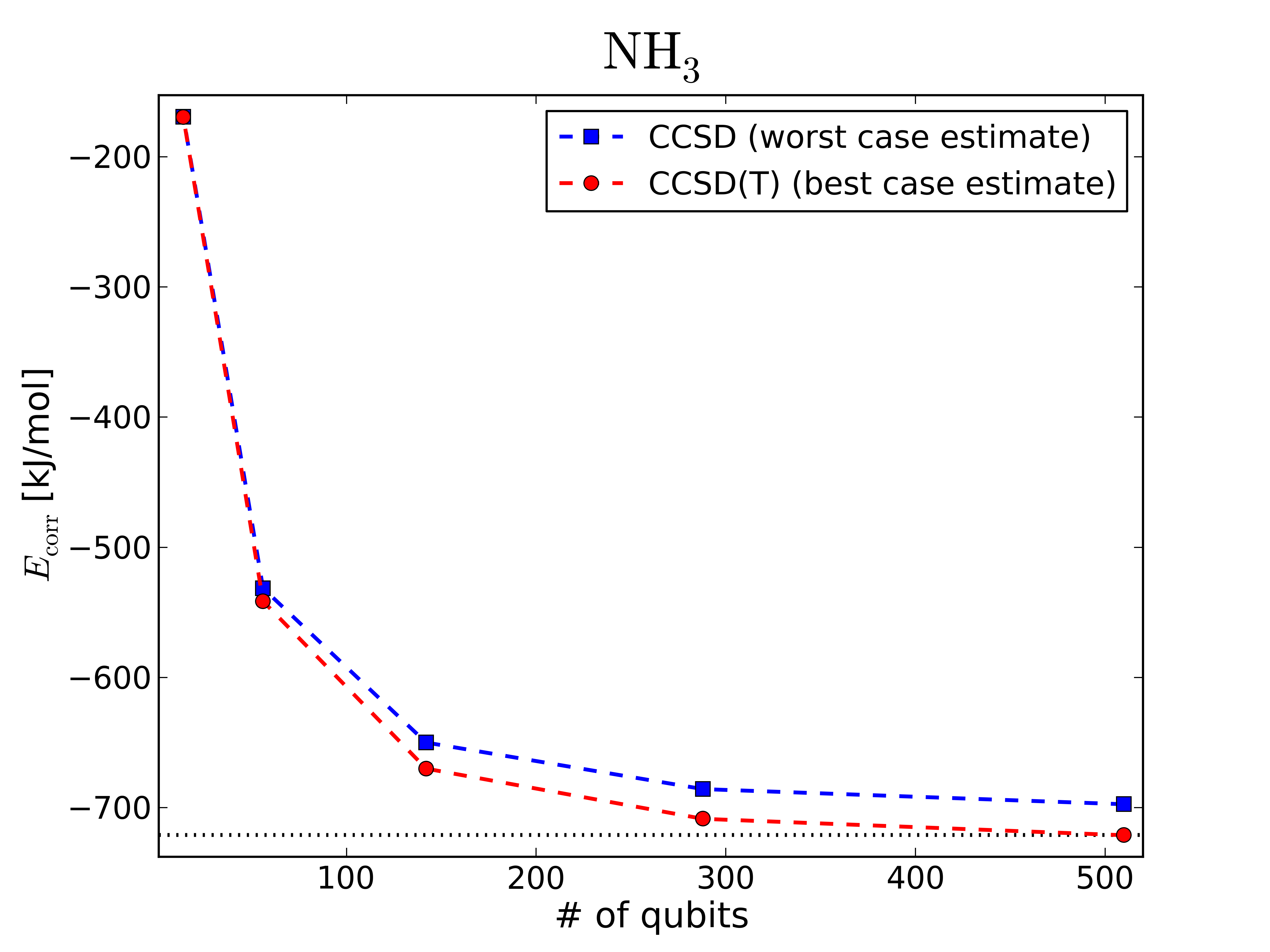}
}
\hspace{1cm}
\subfigure[]{
\includegraphics[width=7.0cm ,angle=0,clip=]{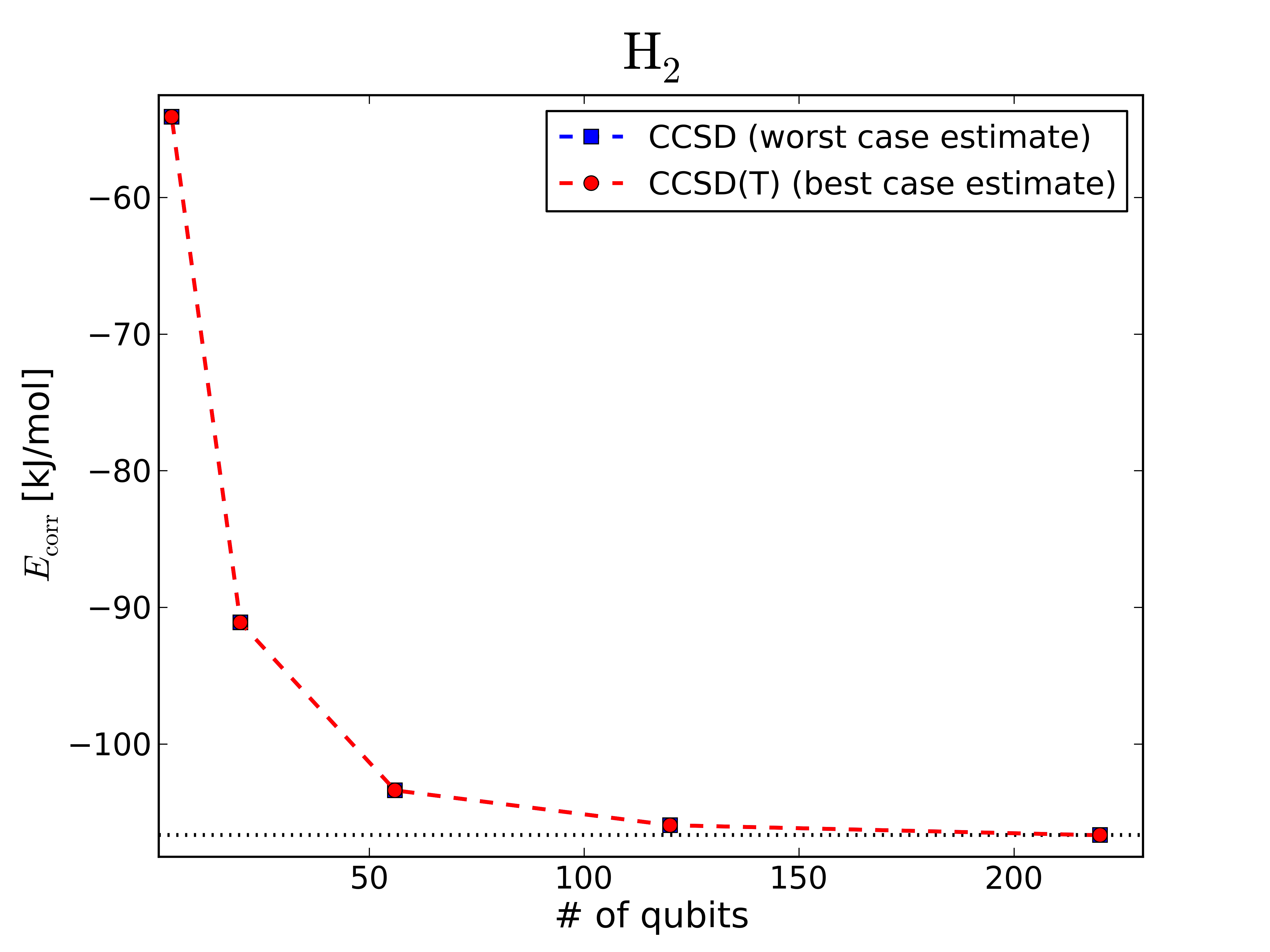}
}
\\
\subfigure[]{
\includegraphics[width=7.0cm ,angle=0,clip=]{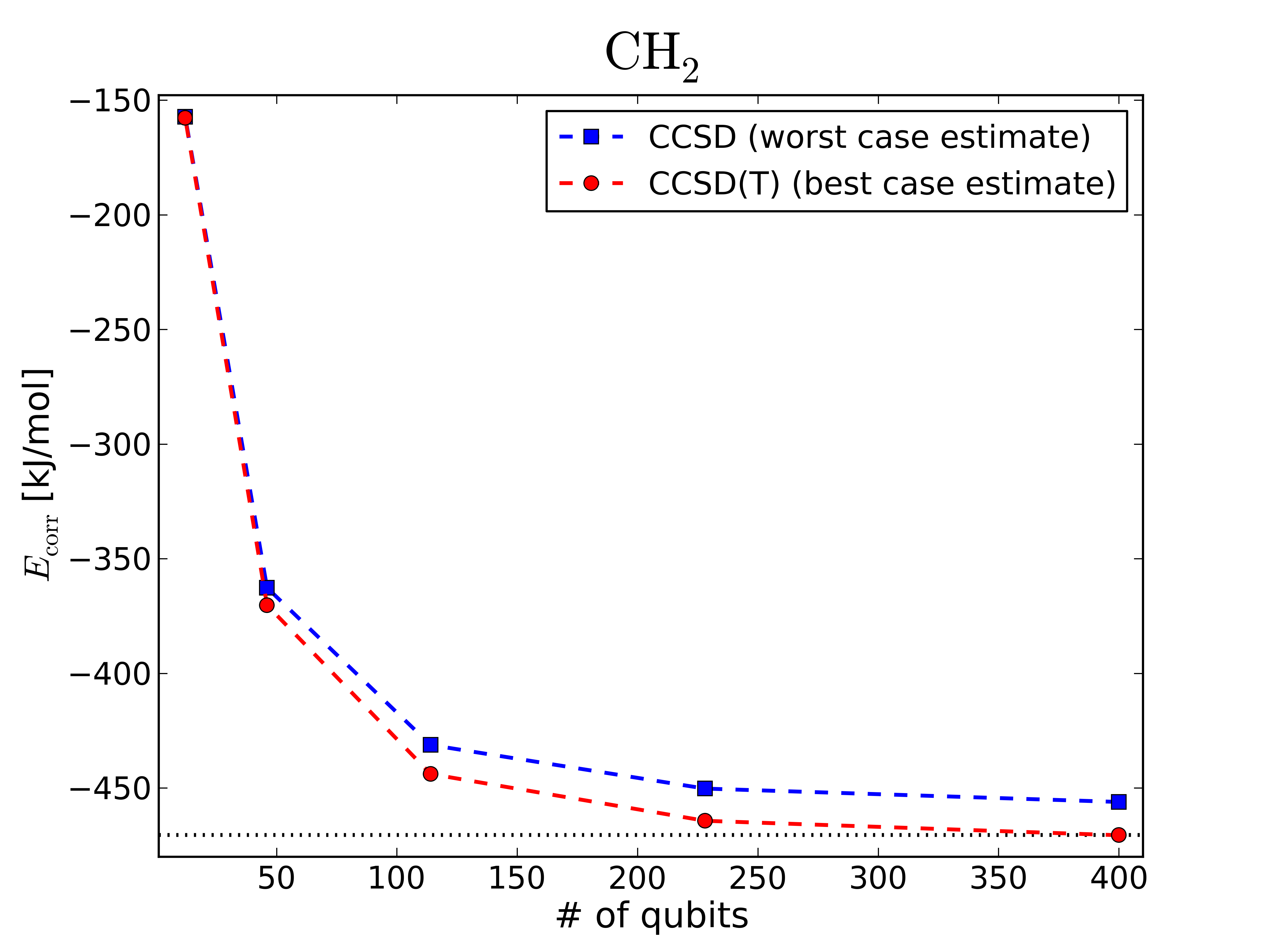}
}

\renewcommand\thefigure{SI 4}
\caption{
Extrapolated UCCSD-VQE correlation energies for a larger number of qubits. The extrapolation is done by using CCSD (worst-case estimate) and CCSD(T) (best-case estimate) correlation energies in accordance with the results shown in Table I in the SI.
For further information see Figure 3 in the SI. 
}
\label{fig:appendix:4}
\end{figure*}


\newpage

\begin{table*}[htbp]
\centering
\renewcommand\thetable{SI I}
\caption[]{ 
Accuracy of molecular energies obtained with UCCSD-VQE for ${\rm LiH}$, ${\rm Li}$, ${\rm NH_3}$, ${\rm H_2}$ and ${\rm CH_{2}}$.
The HF total energy as well as the CCSD, CCSD(T), FCI and UCCSD-VQE correlation energies are given together with the respective differences to the FCI energy ($\Delta$FCI). 
Additionally, the number of required qubits and two-qubit gates is shown.
All results were obtained using the minimal basis set STO-3G. Energies are in kJ/mol.
 }
\label{tab:appendix:1}
\tiny
\begin{ruledtabular}
\begin{tabular}{lccccccccccccccc}
 & \phantom{x} & \multicolumn{2}{c}{\bf{$\boldsymbol{\rm LiH}$}} &  \phantom{x}      &  \multicolumn{2}{c}{\bf{$\boldsymbol{\rm Li}$}} &  \phantom{x} & \multicolumn{2}{c}{\bf{$\boldsymbol{\rm NH_{3}}$}} & \phantom{x} & \multicolumn{2}{c}{\bf{$\boldsymbol{\rm H_{2}}$}} & \phantom{x} & \multicolumn{2}{c}{\bf{$\boldsymbol{\rm CH_{2}}$}} \\
 &              &  & $\Delta$FCI                                             &               &      & $\Delta$FCI                           &              &       & $\Delta$FCI                      &             &      & $\Delta$FCI   &             &      & $\Delta$FCI  \\
\hline
$E_{\text{total}}$({\text{HF}}) & &               -20642.0&$-$&&-19206.9&$-$&&       -145594.2&$-$&&      -2931.8&$-$&&      -100745.6	&	$-$     \\
\hline                                                                                                        		
$E_{\text{corr}}$(CCSD) & &        -53.320&0.028&&-0.815&0.000&&      -169.299&0.567&&     -54.085&0.000&&    -157.212	&	1.206     \\
$E_{\text{corr}}$(CCSD(T)) & &     -53.342&0.006&&-0.815&0.000&&      -169.618&0.248&&     -54.085&0.000&&    -157.759	&	0.658     \\
$E_{\text{corr}}$(FCI) & &         -53.348&0&&-0.815&0&&              -169.866&0&&         -54.085&0&&        -158.417	&	0        \\
\hline                                                                                                        		
$E_{\text{corr}}$(UCCSD-VQE) & &   -53.320&0.028&&-0.815&0.000&&      -169.400&0.466&&     -54.085&0.000&&    -157.491	&	0.927     \\
\# qubits & &                      12&$-$&&10&$-$&&                   14&$-$&&             4&$-$&&            12	&	$-$              \\
\# two-qubit gates & &             1382&$-$&&464&$-$&            &    5976&$-$&        &   56&$-$&         &  1366	&	$-$               \\

\end{tabular}
\end{ruledtabular}
\end{table*}


\begin{table*}[htbp]
\centering
\renewcommand\thetable{SI II}
\caption[]{
Accuracy of ${\rm LiH}$ molecular energies obtained with UCCSD-VQE for basis sets of increasing size (SV, DZ, TZ).
For further information see Table I in the SI.
Energies are in kJ/mol.
 }
\label{tab:appendix:2}
\tiny
\begin{ruledtabular}
\begin{tabular}{lccccccccc}
 & \phantom{x} & \multicolumn{2}{c}{\bf{$\boldsymbol{\rm SV}$}} &  \phantom{x}      &  \multicolumn{2}{c}{\bf{$\boldsymbol{\rm DZ}$}} &  \phantom{x} & \multicolumn{2}{c}{\bf{$\boldsymbol{\rm TZ}$}} \\
 &              &  & $\Delta$FCI                                             &               &      & $\Delta$FCI                           &              &       & $\Delta$FCI \\
\hline
$E_{\text{total}}$({\text{HF}}) & &               -20905.7	&	$-$	&		&	-20919.7	&	$-$	&		&	-20926.4	&	$-$\\
\hline                             														
$E_{\text{corr}}$(CCSD) & &        -48.865	&	0.003	&		&	-89.208	&	0.083	&		&	-93.417	&	0.121  \\
$E_{\text{corr}}$(CCSD(T)) & &     -48.868	&	0.000	&		&	-89.278	&	0.013	&		&	-93.497	&	0.042  \\
$E_{\text{corr}}$(FCI) & &         -48.868	&	0	&		&	-89.292	&	0	&		&	-93.538	&	0      \\
\hline                             														
$E_{\text{corr}}$(UCCSD-VQE) & &   -48.865	&	0.003	&		&	-89.205	&	0.087	&		&	-93.413	&	0.125  \\
\# qubits & &                      10	&	$-$	&		&	12	&	$-$	&		&	18	&	$-$                       \\
\# two-qubit gates & &             1918	&	$-$	&		&	3096	&	$-$	&		&	9944	&	$-$                                         \\

\end{tabular}
\end{ruledtabular}
\end{table*}


\begin{table*}[htbp]
\centering
\renewcommand\thetable{SI III}
\caption[]{
Accuracy of reaction energies for the LiH-dissociation, ${\rm LiH \, \rightarrow \, \, Li \, + \, H}$, obtained with UCCSD-VQE for basis sets of increasing size (SV, DZ, TZ).
The HF, CCSD, CCSD(T), FCI and UCCSD-VQE reaction energies are given together with the respective differences to the FCI reaction energy ($\Delta$FCI).
All values are in kJ/mol.
 }
\label{tab:appendix:3}
\tiny
\begin{ruledtabular}
\begin{tabular}{lccccccccc}
 & \phantom{x} & \multicolumn{2}{c}{\bf{$\boldsymbol{\rm SV}$}} &  \phantom{x}      &  \multicolumn{2}{c}{\bf{$\boldsymbol{\rm DZ}$}} &  \phantom{x} & \multicolumn{2}{c}{\bf{$\boldsymbol{\rm TZ}$}} \\
 &              &  & $\Delta$FCI                                             &               &      & $\Delta$FCI                           &              &       & $\Delta$FCI \\
\hline
$E_{\text{react}}$({\text{HF}}) & &               100.347     &       -48.691     &               &       98.673        &       -49.399     &               &       99.778        &       -52.443\\
\hline                                                                                                                          
$E_{\text{react}}$(CCSD) & &        149.034      &       -0.003   &               &       147.989 &       -0.083   &               &       152.101 &       -0.120  \\
$E_{\text{react}}$(CCSD(T)) & &     149.038      &       0.000   &               &       148.059 &       -0.013   &               &       152.180 &       -0.040  \\
$E_{\text{react}}$(FCI) & &         149.038      &       0       &               &       148.072 &       0       &               &       152.221 &       0      \\
\hline                                                                                                                          
$E_{\text{react}}$(UCCSD-VQE) & &   149.035      &       -0.003   &               &      147.985 &      -0.086   &               &       152.096 &       -0.124  \\
\end{tabular}
\end{ruledtabular}
\end{table*}


\begin{table*}[htbp]
\centering
\renewcommand\thetable{SI IV}
\caption[]{
Extrapolated UCCSD-VQE reaction energies for the LiH-dissociation, ${\rm LiH \, \rightarrow \, \, Li \, + \, H}$, along with corresponding individual molecular energies for basis sets of increasing size (STO-3G, cc-pVDZ, cc-pVTZ, cc-pVQZ and cc-pV5Z). 
The extrapolation is carried out by using CCSD energies (worst-case estimates) in accordance with the results shown in Table I in the SI, as also done in Figures 3 and 4 in the SI.
The obtained CCSD total energies are splitted into a HF total energy contribution and a CCSD correlation energy contribution. For comparison, DFT reaction energies using the functionals BP86, B3LYP and M06-2X are also shown.
Additionally, the number of required qubits as well as the number of two-qubit gates are estimated for the computationally most demanding system at a certain basis set quality (here ${\rm LiH}$), 
see \textcolor{black}{also the extrapolation in Figure 9 in the main paper}.
All energies are in kJ/mol.
 }
\label{tab:appendix:4}
\tiny
\begin{ruledtabular}
\begin{tabular}{llcccccc}
 &   & \phantom{x} & STO-3G & cc-pVDZ & cc-pVTZ & cc-pVQZ & cc-pV5Z  \\
\hline
                                    & $E_{\text{total}}$(HF)                     &&   	-20642.0	&		 	 	 	-20960.9	&	-20968.9	 	&	 	-20970.3	&	-20970.7            \\
{\bf{$\boldsymbol{\rm LiH}$}}    & $E_{\text{corr}}$(CCSD)                    &&   	-53.3	&		 	 	 	-81.5	&	-131.2	 	&	 	-145.4	&	-173.7                    \\
                                    & $E_{\text{total}}$(CCSD)                   &&   	-20695.4	&		 	 	 	-21042.5	&	-21100.1	 	&	 	-21115.7	&	-21144.4        \\
\hline                                                                                
                                    & $E_{\text{total}}$(HF)                     &&   	-19206.9	&		 	 	 	-19513.8	&	-19514.6	 	&	 	-19514.6	&	-19514.7         \\
{\bf{$\boldsymbol{\rm Li}$}}        & $E_{\text{corr}}$(CCSD)                    &&   	-0.8	&		 	 	 	-0.6	&	-35.0	 	&	 	-44.8	&	-71.2                          \\
                                    & $E_{\text{total}}$(CCSD)                   &&   	-19207.7	&		 	 	 	-19514.4	&	-19549.6	 	&	 	-19559.4	&	-19585.9          \\
\hline                                                                                
{\bf{$\boldsymbol{\rm H}$}}         & $E_{\text{total}}$                         &&   	-1225.0	&					-1310.9	&	-1312.3		&		-1312.6	&	-1312.7                    \\
\hline
                                    & $E_{\text{react,\,total}}$(HF)              &&  	210.1	&					136.2	&	142.1		&		143.1	&	143.3                              \\
                                    & $E_{\text{react,\,corr}}$(CCSD)             &&  	52.5	&					81.0	&	96.2		&		100.6	&	102.5                                  \\
\bf Reaction                            & $E_{\text{react,\,total}}$(CCSD)            &&  	262.6	&					217.2	&	238.3		&		243.7	&	\bf 245.8                               \\
                                    & \# qubits                                  &&   	12	&					38	&	88		&		170	&	292                                            \\
                                    & \textcolor{black}{\# two-qubit gates}        &&   	$1.5 \cdot 10^{3}$	&					$1.7 \cdot 10^{4}$	&	$8.3 \cdot 10^{4}$		&		$3.0 \cdot 10^{5}$	&	$8.5 \cdot 10^{5}$                                                   \\
\hline                                                                                
                                    & $E_{\text{react,\,total}}$(BP86)            &&  	327.6	&					237.8	&	243.3		&		244.4	&	244.3                                \\
\bf Reaction (DFT)                      & $E_{\text{react,\,total}}$(B3LYP)           &&  	321.2	&					238.1	&	243.6		&		244.8	&	244.6                                \\
                                    & $E_{\text{react,\,total}}$(M06-2X)          &&  	311.2	&					230.9	&	238.4		&		238.1	&	237.3                                         \\
\end{tabular}
\end{ruledtabular}
\end{table*}


\begin{table*}[htbp]
\centering
\renewcommand\thetable{SI V}
\caption[]{
Extrapolated UCCSD-VQE reaction energies for 
the Haber-Bosch process, ${\rm N_2 \, + \, 3\,H_2 \,  \rightarrow \, \, 2\, NH_3}$, along with corresponding individual molecular energies for basis sets of increasing size (STO-3G, cc-pVDZ, cc-pVTZ, cc-pVQZ and cc-pV5Z). 
For further information see Table IV in the SI. 
The number of required qubits as well as the number of two-qubit gates are estimated for the computationally most demanding system at a certain basis set quality (here ${\rm NH_3}$; in case of STO-3G more qubits are required for ${\rm N_2}$ which is the only exception),
see \textcolor{black}{also the extrapolation in Figure 10 in the main paper}.
All energies are in kJ/mol.
 }
\label{tab:appendix:5}
\tiny
\begin{ruledtabular}
\begin{tabular}{llcccccc}
 &   & \phantom{x} & STO-3G & cc-pVDZ & cc-pVTZ & cc-pVQZ & cc-pV5Z  \\
\hline
                                    & $E_{\text{total}}$(HF)                     &&       	-282224.5	&		 	 	 	-286061.3	&	-286139.5	 	&	 	-286159.7	&	-286164.1      \\
{\bf{$\boldsymbol{\rm N_{2}}$}}    & $E_{\text{corr}}$(CCSD)                    &&        	-395.5	&		 	 	 	-807.7	&	-972.7	 	&	 	-1028.6	&	-1048.4        \\
                                    & $E_{\text{total}}$(CCSD)                   &&       	-282620.0	&		 	 	 	-286869.0	&	-287112.2	 	&	 	-287188.3	&	-287212.5  \\
\hline                                                                                    
                                    & $E_{\text{total}}$(HF)                     &&       	-2931.8	&		 	 	 	-2963.4	&	-2974.6	 	&	 	-2975.9	&	-2976.3       \\
{\bf{$\boldsymbol{\rm H_{2}}$}}        & $E_{\text{corr}}$(CCSD)                    &&    	-54.1	&		 	 	 	-91.1	&	-103.4	 	&	 	-105.9	&	-106.7                 \\
                                    & $E_{\text{total}}$(CCSD)                   &&       	-2985.9	&		 	 	 	-3054.5	&	-3078.0	 	&	 	-3081.8	&	-3082.9        \\
\hline                                                                                    
                                    & $E_{\text{total}}$(HF)                     &&       	-145594.2	&		 	 	 	-147541.4	&	-147600.0	 	&	 	-147613.7	&	-147618.0   \\
{\bf{$\boldsymbol{\rm NH_{3}}$}}         & $E_{\text{total}}$                         &&  	-169.3	&		 	 	 	-531.5	&	-650.0	 	&	 	-685.8	&	-697.4                  \\
                                    & $E_{\text{total}}$(CCSD)                   &&       	-145763.5	&		 	 	 	-148072.9	&	-148249.9	 	&	 	-148299.5	&	-148315.4    \\
\hline
                                    & $E_{\text{react,\,total}}$(HF)              &&      	-168.5	&					-131.1	&	-136.7		&		-140.2	&	-143.2                  \\
                                    & $E_{\text{react,\,corr}}$(CCSD)             &&      	219.2	&					18.0	&	-17.1		&		-25.1	&	-26.4                         \\
\bf Reaction                            & $E_{\text{react,\,total}}$(CCSD)            &&      	50.7	&					-113.1	&	-153.9		&		-165.3	&	\bf -169.5                     \\
                                    & \# qubits                                  &&       	16	&					56	&	142		&		288	&	510                                    \\
                                    & \textcolor{black}{\# two-qubit gates}        &&       	$4.7 \cdot 10^{3}$	&					$3.6 \cdot 10^{5}$	&	$2.7 \cdot 10^{6}$		&		$1.2 \cdot 10^{7}$	&	$3.8 \cdot 10^{7}$                                            \\
\hline                                                                                    
                                    & $E_{\text{react,\,total}}$(BP86)            &&      	-83.6	&					-145.7	&	-166.8		&		-175.0	&	-180.4                     \\
\bf Reaction (DFT)                      & $E_{\text{react,\,total}}$(B3LYP)           &&      	-108.9	&					-141.6	&	-158.6		&		-167.1	&	-172.6                    \\
                                    & $E_{\text{react,\,total}}$(M06-2X)          &&      	-118.4	&					-129.0	&	-156.9		&		-163.7	&	-171.7                                             \\
\end{tabular}
\end{ruledtabular}
\end{table*}


\begin{table*}[htbp]
\centering
\renewcommand\thetable{SI VI}
\caption[]{
Extrapolated UCCSD-VQE transition energies for 
the triplet-singlet transition in $\rm CH_{2}$, ${\rm {:}CH_2 \, \rightarrow \, \, CH_2}$, along with corresponding individual molecular energies for basis sets of increasing size (STO-3G, cc-pVDZ, cc-pVTZ, cc-pVQZ and cc-pV5Z). 
For further information see Table IV in the SI. 
The number of required qubits as well as the number of two-qubit gates are estimated for ${\rm {:}CH_2}$, 
see \textcolor{black}{also the extrapolation in Figure 11 in the main paper}.
All energies are in kJ/mol.
 }
\label{tab:appendix:6}
\tiny
\begin{ruledtabular}
\begin{tabular}{llcccccc}
 &   & \phantom{x} & STO-3G & cc-pVDZ & cc-pVTZ & cc-pVQZ & cc-pV5Z  \\
\hline
                                    & $E_{\text{total}}$(HF)                     &&       	-100909.4	&		 	 	 	-102201.8	&	-102230.7	 	&	 	-102237.2	&	-102238.8      \\
{\bf{$\boldsymbol{\rm {:}CH_{2}}$}}    & $E_{\text{corr}}$(CCSD)                    &&    	-97.6	&		 	 	 	-296.5	&	-359.2	 	&	 	-376.4	&	-381.6           \\
                                    & $E_{\text{total}}$(CCSD)                   &&       	-101007.0	&		 	 	 	-102498.3	&	-102589.9	 	&	 	-102613.6	&	-102620.4  \\
\hline                                                                                    
                                    & $E_{\text{total}}$(HF)                     &&       	-100745.6	&		 	 	 	-102082.3	&	-102111.8	 	&	 	-102119.0	&	-102121.0   \\
{\bf{$\boldsymbol{\rm CH_{2}}$}}        & $E_{\text{corr}}$(CCSD)                    &&   	-157.2	&		 	 	 	-362.6	&	-431.2	 	&	 	-450.2	&	-456.0            \\
                                    & $E_{\text{total}}$(CCSD)                   &&       	-100902.8	&		 	 	 	-102444.9	&	-102543.0	 	&	 	-102569.2	&	-102577.0    \\
\hline
                                    & $E_{\text{react,\,total}}$(HF)              &&      	163.7	&					119.5	&	118.9		&		118.1	&	117.8                   \\
                                    & $E_{\text{react,\,corr}}$(CCSD)             &&      	-59.6	&					-66.2	&	-72.0		&		-73.8	&	-74.4                    \\
\bf Reaction                            & $E_{\text{react,\,total}}$(CCSD)            &&      	104.1	&					53.4	&	46.9		&		44.3	&	\bf 43.4                        \\
                                    & \# qubits                                  &&       	12	&					46	&	114		&		228	&	400                                \\
                                    & \textcolor{black}{\# two-qubit gates}        &&       	$1.8 \cdot 10^{3}$	&					$6.9 \cdot 10^{4}$	&	$5.0 \cdot 10^{5}$		&		$2.1 \cdot 10^{6}$	&	$6.6 \cdot 10^{6}$                                        \\
\hline                                                                                    
                                    & $E_{\text{react,\,total}}$(BP86)            &&      	94.3	&					68.9	&	66.7		&		65.9	&	65.4                          \\
\bf Reaction (DFT)                      & $E_{\text{react,\,total}}$(B3LYP)           &&      	80.2	&					52.8	&	50.2		&		49.2	&	48.7                          \\
                                    & $E_{\text{react,\,total}}$(M06-2X)          &&      	88.6	&					58.8	&	56.5		&		55.8	&	56.6                                     \\
\end{tabular}
\end{ruledtabular}
\end{table*}

\begin{table*}[htbp]
\centering
\renewcommand\thetable{SI VII}
\caption[]{
Coordinates of $\rm H_{2}O$ obtained at the B3LYP/def2-QZVPP level.}
\begin{ruledtabular}
\begin{tabular}{rrrr}
O &    0.0044960 &   0.0057136 &   0.0000000\\
H &    0.9646713 &  -0.0073990 &   0.0000000\\
H &   -0.2330023 &   0.9361335 &   0.0000000 \\
\end{tabular}
\end{ruledtabular}
\end{table*}

\begin{table*}[htbp]
\centering
\renewcommand\thetable{SI VIII}
\caption[]{
Coordinates of $\rm OH$ obtained at the B3LYP/def2-QZVPP level.}
\begin{ruledtabular}
\begin{tabular}{rrrr}
O  &  -0.0062713 &   0.0000000  &  0.0000000 \\
H  &   0.9674223 &   0.0000000  &  0.0000000 \\
\end{tabular}
\end{ruledtabular}
\end{table*}

\begin{table*}[htbp]
\centering
\renewcommand\thetable{SI IX}
\caption[]{
Coordinates of $\rm LiH$ obtained at the B3LYP/def2-QZVPP level.}
\begin{ruledtabular}
\begin{tabular}{rrrr}
Li &  -3.2319355 &   1.1691641 &   0.0000000 \\
H  &  -1.6405907 &   1.1691641 &   0.0000000 \\
\end{tabular}
\end{ruledtabular}
\end{table*}

\begin{table*}[htbp]
\centering
\renewcommand\thetable{SI X}
\caption[]{
Coordinates of $\rm N_2$ obtained at the B3LYP/def2-QZVPP level.}
\begin{ruledtabular}
\begin{tabular}{rrrr}
N  &  -4.2451484 &   2.5760816  &  0.0000000 \\
N  &  -3.1862016  &  2.3168784 &   0.0000000 \\
\end{tabular}
\end{ruledtabular}
\end{table*}

\begin{table*}[htbp]
\centering
\renewcommand\thetable{SI XI}
\caption[]{
Coordinates of $\rm H_2$ obtained at the B3LYP/def2-QZVPP level.}
\begin{ruledtabular}
\begin{tabular}{rrrr}
H  &  -4.0759023  &  2.5346545 &   0.0000000 \\
H  &  -3.3554477  &  2.3583055 &   0.0000000 \\
\end{tabular}
\end{ruledtabular}
\end{table*}

\begin{table*}[htbp]
\centering
\renewcommand\thetable{SI XII}
\caption[]{
Coordinates of $\rm NH_3$ obtained at the B3LYP/def2-QZVPP level.}
\begin{ruledtabular}
\begin{tabular}{rrrr}
N  &  -0.6800848  &  1.2693644  &  0.0227335 \\
H  &   0.3314114  &  1.2772854 &  -0.0142918 \\
H  &  -0.9994515  &  0.5770216  & -0.6429998 \\
H  &  -0.9993852  &  2.1717686 &  -0.3063819 \\
\end{tabular}
\end{ruledtabular}
\end{table*}

\begin{table*}[htbp]
\centering
\renewcommand\thetable{SI XIII}
\caption[]{
Coordinates of $\rm {:}CH_2$ obtained at the B3LYP/def2-QZVPP level.}
\begin{ruledtabular}
\begin{tabular}{rrrr}
C  &  -2.8332513 &   0.6986016  &  0.0000000 \\
H  &  -1.8294369 &   1.0885324  &  0.0000000 \\
H  &  -3.8206119 &   1.1284960  &  0.0000000 \\
\end{tabular}
\end{ruledtabular}
\end{table*}

\begin{table*}[htbp]
\centering
\renewcommand\thetable{SI XIV}
\caption[]{
Coordinates of $\rm CH_2$ obtained at the B3LYP/def2-QZVPP level.}
\begin{ruledtabular}
\begin{tabular}{rrrr}
C &   -2.8371013  &  0.5067306 &   0.0000000 \\
H &   -1.9616221  &  1.1871592 &   0.0000000 \\
H &   -3.6845767  &  1.2217402 &   0.0000000 \\
\end{tabular}
\end{ruledtabular}
\end{table*}

\end{document}